\documentclass[preprint, authoryear]{elsarticle}

\usepackage{lineno,hyperref}

\usepackage{multicol}
\usepackage{lipsum} 
\usepackage{graphicx}
\usepackage{float}
\usepackage{fancyhdr}
\usepackage{caption}
\usepackage{adjustbox}
\usepackage{array}
\usepackage{rotating}
\usepackage{multirow}
\usepackage{hhline}
\usepackage{subfiles}
\usepackage{titlesec}
\usepackage{subcaption}
\usepackage[bottom]{footmisc}
\usepackage[misc]{ifsym}
\usepackage{natbib}

\RequirePackage{amsmath,amssymb,amsfonts,amsbsy,amsthm,booktabs,epsfig,graphicx,rotating}

\newcommand{\squeezeup}{\vspace{-4.5mm}}

\newcolumntype{L}[1]{>{\raggedright\let\newline\\\arraybackslash\hspace{0pt}}m{#1}}
\newcolumntype{C}[1]{>{\centering\let\newline\\\arraybackslash\hspace{0pt}}m{#1}}
\newcolumntype{R}[1]{>{\raggedleft\let\newline\\\arraybackslash\hspace{0pt}}m{#1}}

\RequirePackage{times}      
\RequirePackage{mathptmx}   

\usepackage{xcolor}
\usepackage{mdframed}
\definecolor{shadecolor}{RGB}{228,230,240}

\modulolinenumbers[5]

\journal{Computational Statistics \& Data Analysis}

\usepackage[nodots]{numcompress}\biboptions{authoryear}

\let\linenumbers\nolinenumbers\nolinenumbers
\begin{document}

\begin{frontmatter}

\title{Sample Size Considerations for Bayesian Multilevel Hidden Markov Models: A Simulation Study on Multivariate Continuous Data with highly overlapping Component Distributions based on Sleep Data}

\author{Jasper Ginn}
\author{Sebastian Mildiner Moraga}
\author{Emmeke Aarts\corref{mycorrespondingauthor}}
\address{Department of Methodology and Statistics, Faculty of Social and Behavioural Sciences, Utrecht University, the Netherlands}

\cortext[mycorrespondingauthor]{Corresponding author}
\ead{e.aarts@uu.nl}

\begin{abstract}
Spurred in part by the ever-growing number of sensors and web-based methods of collecting data, the use of Intensive Longitudinal Data (ILD) is becoming more common in the social and behavioural sciences. The ILD collected in this field are often hypothesised to be the result of latent states (e.g. behaviour, emotions), and the promise of ILD lies in its ability to capture the dynamics of these states as they unfold in time. In particular, by collecting data for multiple subjects, researchers can observe how such dynamics differ between subjects. The Bayesian Multilevel Hidden Markov Model (mHMM) is a relatively novel model that is suited to model the ILD of this kind while taking into account heterogeneity between subjects. While the mHMM has been applied in a variety of settings, large-scale studies that examine the required sample size for this model are lacking. In this paper, we address this research gap by conducting a simulation study to evaluate the effect of changing (1) the number of subjects, (2) the number of occasions, and (3) the between subjects variability on parameter estimates obtained by the mHMM. We frame this simulation study in the context of sleep research, which consists of multivariate continuous data that displays considerable overlap in the state dependent component distributions. In addition, we generate a set of baseline scenarios with more general data properties. Overall, the number of subjects has the largest effect on model performance. However, the number of occasions is important to adequately model latent state transitions. We discuss how the characteristics of the data influence parameter estimation and provide recommendations to researchers seeking to apply the mHMM to their own data. 
\end{abstract}

\begin{keyword}
Multilevel hidden Markov model \sep Random effects model \sep Monte Carlo simulation \sep Bayesian estimation \sep Intensive longitudinal data 
\end{keyword}

\end{frontmatter}

\linenumbers

\section{Introduction}

The use of longitudinal data is commonplace in social and behavioral research. Longitudinal data tracks variables across time and across subjects, allowing researchers to observe differences between subjects as well as differences over time within a subject. Increasingly, researchers in the social and behavioral sciences have access to intensive longitudinal data (ILD). Although ILD tends to be considerably larger than traditional longitudinal data in terms of the number of occasions that are collected for each participant across time, what really sets ILD apart are the kind of hypotheses it can address, as well as the complexity that arises in modeling such data \citep{schafer2006models}. 

For many researchers, the promise of ILD lies in its ability to capture processes as they unfold in time, and its rising importance has spurred the development of new models or adapt existing ones that can analyze such data. That is, in addition to violating important assumptions that underpin many statistical models (e.g. independence), conventional models are simply not able to optimally exploit the information contained within ILD. To address such issues, researchers increasingly borrow techniques from other academic fields, such as econometrics and engineering, in which time-series modeling is common practice. For example, the Dynamic Structural Equation Model (DSEM) combines time-series modeling with structural equation modeling by including autoregressive elements. \citep{asparouhov2018dynamic,hamaker2018frontiers}. Another example is the The Multilevel hidden Markov model (mHMM) \citep{altman2007mixed, maruotti2011mixed, aartsusing2019}, which will be examined in this paper. However, without thoroughly investigating model performance of these models in relation to design factors (e.g., sample size) and data complexity, we cannot advocate their widespread use. 

The mHMM combines the hidden Markov model (HMM; \citep{rabiner1989tutorial, zucchini2017hidden}) with the multilevel framework (see e.g. \cite{hox2017multilevel,snijders2011multilevel, gelman2006data}). Broadly speaking, an HMM is a model that estimates latent (or "unobserved") states based on observed time-series data for a single subject. In this context, two or more states are defined as a set of mutually exclusive categories, such as moods ("happy", "sad", "angry" etc.) or behavior ("moving", "resting", "eating" etc.). Hence, the HMM and mHMM are suited to a specific type of ILD in which observed data are uniquely determined by some finite set of latent underlying states. 

Multilevel modeling is often used in cases where the data is thought to display some hierarchical or nested structure. In the case of ILD, we think of this hierarchy or nesting as follows. We have $N$ subjects, each of which is associated with their own time-series data of length $N_T$. This forms the lowest level of the hierarchy. Next, we think of the subjects as belonging to a common group, which forms the second level. As such, the multilevel framework allows us to fit a model at the group level, as well as a model for each subject individually. 

By combining the HMM with the multilevel modeling framework in the mHMM, we have at our disposal a tool to analyze ILD of multiple subjects. Recently, the mHMM has been implemented using Bayesian estimation, which has been shown to work well for both regular HMMs as well as mHMMs \citep{scott2002bayesian, de2017use}. However, not much is currently known about what constitutes an "acceptable" sample size for this model in terms of the number of subjects $N$ or the number of occasions $N_T$. Without this information, applied researchers should be cautious when interpreting parameter estimates obtained by fitting the mHMM. 

In this article, we study the effect of subject and occasion sample sizes and between-subject variability on the quality of the parameter estimates obtained from the Bayesian mHMM on Gaussian multivariate data with highly overlapping component distributions. In particular, we vary the following quantities:

\begin{enumerate}
    \item{The number of subjects $N$.}
    \item{The number of occasions $N_T$ for each subject.}
    \item{The variability in subject-level parameters.}
\end{enumerate}

To our knowledge, no major investigation into the effects of sample sizes and between-subject variances has previously been conducted on the parameter estimates obtained by the Bayesian mHMM on (multivariate) continuous data. \cite{altman2007mixed} reports that, in the context of a Frequentist mHMM and using Poisson-distributed outcome data, $60$ subjects is generally sufficient. However, she does not systematically vary the sample size and occasion size. \cite{rueda2013bayesian} conduct a small simulation study on Bayesian mHMMs using Gaussian outcome data. Their study is also limited, however, given that they focus primarily on investigating the ability of the model to accurately estimate random effects as well as its ability to intuit the number of latent states automatically.  In addition, a recent publication within the field of Ecology by \cite{mcclintock_worth_2021} also sheds some light on this matter, however evaluation of model performance is geared towards correct state assignment and dynamics of the latent states (and not the observational part of the model). Results relate to datasets common to animal movement behaviour biotelemetry studies, fitting a model in which random effects relate to the dynamics of the latent states only, and is applied to univariate, gamma distributed observations of moderate length (30 to 250 observations per animal).  Hence, model performance for data more typical of social behavioral science (e.g, multivariate data, more observations per subject, and allowing for subject variation in the observational part of the model) is still lacking.

Based on the extensive body of literature on multilevel models, it is reasonable to expect that the number of subjects is the most important determinant of parameter quality. Previous research into sample sizes in multilevel models suggest that this is particularly true for the random effects that measure variance between subjects (see e.g.  \cite{theall2011impact, hox2014analyzing,laszkiewicz2013sample, landau2013sample} and \cite{smid2020bayesian} for a comparison between Bayesian and Frequentist multilevel models). More specifically for ILD, \cite{schultzberg2018number} conduct a study to examine the required sample size for a range of DSEM models. They vary the complexity of their models, the number of subjects $N$ and the number of occasions $N_T$ per subject. Their main conclusion is that, in terms of getting better parameter estimates, increasing the number of subjects is always superior to collecting more occasions for each subject. However, we expect that, in the particular case of the mHMM, the occasion sample size will be of particular importance to estimate parameters related to the dynamics of the latent states. Of particular interest is the effect of changing the variance between subjects because this quantity can vary widely depending on the problem context. As such, it is important to understand the extent to which heterogeneity between persons affects the quality of the parameter estimates obtained by the model. 

The rest of this paper is structured as follows. In section two, we describe the HMM and mHMM in more detail. In turn, we elaborate on the design of the simulation study in section three. Section four reports on the results of the simulation study. In section five, we apply the mHMM on real EEG and EOG data of multiple subjects to detect sleep states. Finally, we discuss the results and offer some recommendations to researchers seeking to apply the mHMM on their own data in section six.


\section{Single-level and multilevel hidden markov models}

In this section, we first discuss the regular, single-level HMM and introduce the necessary notation. We then turn to a short explanation of the multilevel framework and the mHMM, and discuss ways in which one can estimate their parameters. 

\subsection{Hidden markov models} \label{sec:HMM}

The HMM attempts to estimate a set of discrete, unobserved states from observed time-series data \citep{zucchini2017hidden}.  The HMM is characterized by the following features \citep{visser2011seven, zucchini2017hidden}:

\begin{enumerate}
    \item{The observed data is a mixture distribution which consists of several \textit{component distributions}. The component distributions can have any shape (e.g. Poisson, Gaussian, Binomial etc.), but in this paper we only discuss component distributions that are normally distributed. The existence of component distributions implies that there are two or more "hidden" or \textit{latent} states that generate the values of the observed data.}
    \item{The latent states are not independently and identically distributed across the occasions, but rather follow a Markov process. Hence, the latent states (and therefore the observed data) are dependent rather than independent draws from the individual distributions. The observed data shows (severe) auto-correlation, which disappears when accounting for the latent states.}
\end{enumerate}

Consider the data shown in table \ref{tab:heart-rates} below. This data forms a time-series sequence of length $(1, 2, \dots, N_T)$ for a single subject. At any point in time $t$, we observe some value of the variable \textit{heart rate}.

\begin{table}[H]
\centering
\begin{tabular}{|l|l|}
\hline
Occasion & Heart Rate \\ \hline
$t=1$        & $86$         \\ \hline
$t=2$        & $72$         \\ \hline
...      & ...        \\ \hline
$t=N_T$        & $68$         \\ \hline
\end{tabular}
\caption{Time-series data of observed heart rate for a single subject}
\label{tab:heart-rates}
\end{table}

Our hypothesis is that the values of the heart rate variable are uniquely determined by some underlying, latent state at each occasion $t$. Every latent state has its own component distribution with its own component parameters. Hence, at any occasion $t$ we observe one of $m$ distinct states $C$ such that:

\begin{equation}
C_t = i, i \in \{1,2, \dots, m\}
\label{eq:statespace}
\end{equation}

\begin{figure}[H]
\centering
\includegraphics[scale=0.25]{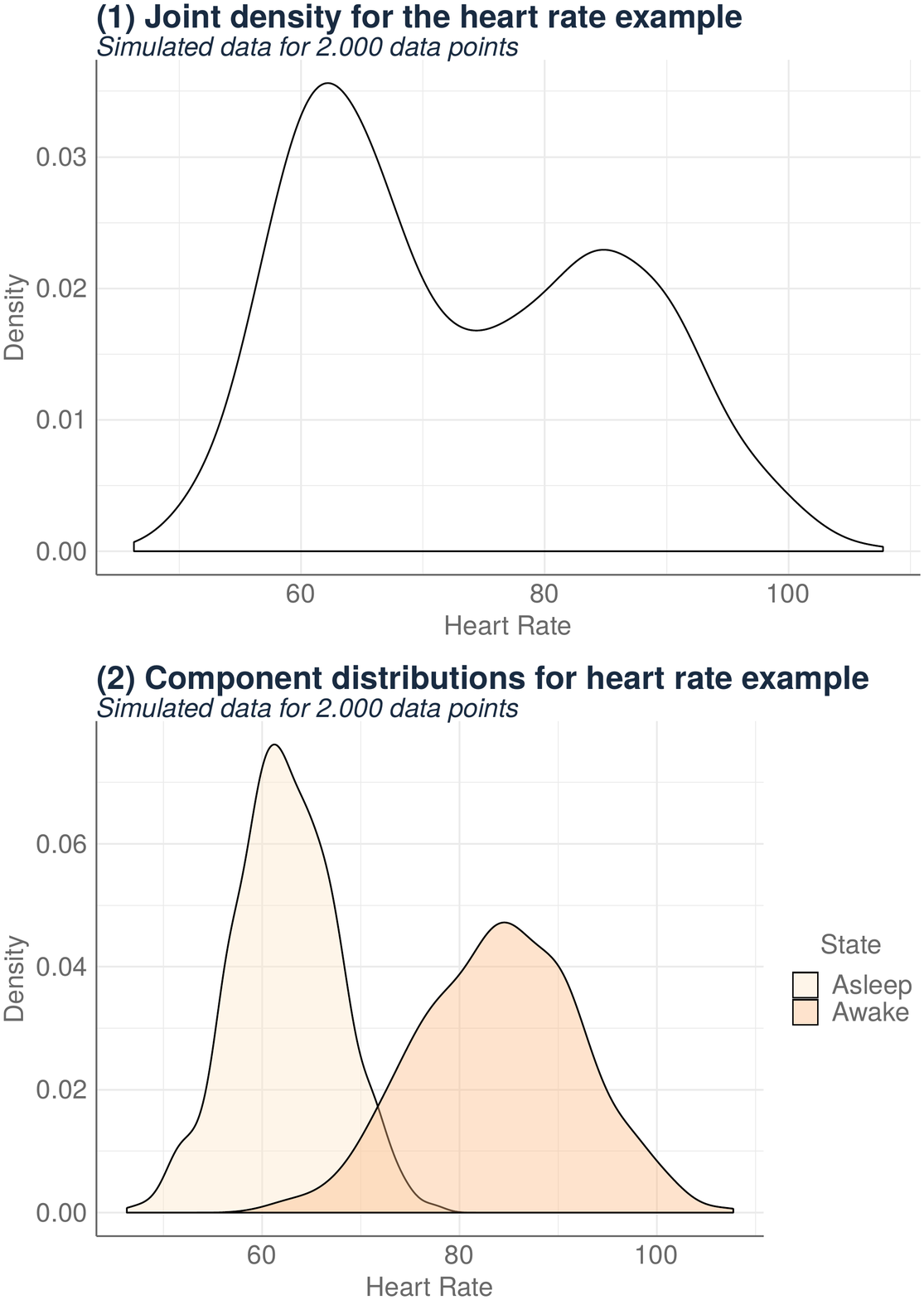}
\caption{Joint density of simulated heart rate data (panel 1). Component distributions of the heart rate data when we take into account the latent states "awake" and "asleep" (panel 2).}
\label{fig:component_densities}
\end{figure}

The meaning of the $m$ latent states depend on the problem context. For example, in figure \ref{fig:component_densities} we use the heart rate example to decompose the mixture distribution of $m=2$ independent normal distributions (panel 1) into the component distributions belonging to a state we call "awake" and a state we call "asleep" (panel 2).

Accordingly, the distribution of the heart rate variable $X_t$ is dependent on the state $C_t = i$ at occasion $t$. That is:

\begin{equation}
P(X_t = x_t) = P(X_t = x_t | C_t = i) 
\label{eq:joint_density}
\end{equation}

\noindent With respect to the heart rate example, equation \ref{eq:joint_density} yields the probability of observing a heart rate value (e.g. a heart rate of $83$) given the latent state. With reference to figure \ref{fig:component_densities}, it should be obvious that the probability of observing a value of $83$ is much higher for the state "awake" than the state "asleep".

Now consider the following. The distribution of states $C$ at occasion $t$ are not independently and identically distributed. Rather, the distribution of states follow a Markov process such that:

\begin{equation}
P(C_{t+1} = j) = P(C_{t+1} = j | C_{t} = i)
\label{eq:markov_chain}
\end{equation}

\begin{figure}[H]
\centering
\includegraphics[scale=0.5]{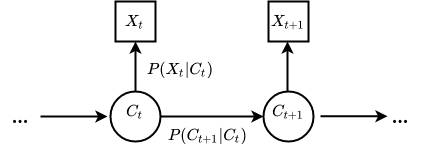}
\caption{An HMM as a directed graph. Each latent state, shown here as circles, depends only on the previous state (see equation \ref{eq:markov_chain}). The outcome data, shown here as squares, depends only on the value of the latent state at time $t$ (see equation \ref{eq:joint_density}).}
\label{fig:single_hmm_directed}
\end{figure}

Hence, the probability of observing the state $j$ at time $t+1$ depends only on the value of state $i$ observed at the previous occasion $t$ (see figure \ref{fig:single_hmm_directed}). 

To represent this dependency, the probabilities of transitioning from some state $i$ at occasion $t$ to another state $j$ at occasion $t+1$ are collected in the \textit{transition probability matrix} (TPM) $\pmb{\Gamma} \in \mathbb{R}^{m \times m}$. The position $\gamma_{ij}$ represents the probability of transitioning from state $i$ to state $j$ at occasion $t+1$. Each row in the TPM should sum to unity to ensure that it is a valid probability distribution.

Consider again the heartbeat example and the following TPM:

\begin{figure}[H]
\centering
\includegraphics[scale=0.5]{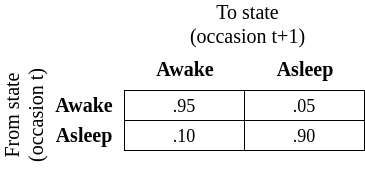}
\label{fig:HB-TBM}
\end{figure}
\squeezeup

The rows indicate the state $i$ at time $t$, and the columns indicate the state $j$ at time $t+1$. For example, the probability of transitioning to the state "awake" at time $t+1$ given that we observe the state "asleep" at time $t$ is $0.1$. The diagonal entries are called \textit{self-transition probabilities}, and denote the probability of observing the same state $C$ at both occasion $t$ and $t+1$. For example, this is the probability that a subject that is, say, awake at time $t$ is also awake at time $t+1$. 

The last set of parameters that needs to be established is the probability of starting in any of the $m$ states at occasion $t=1$. This probability is governed by the \textit{initial distribution} $\pmb{\delta} \in \mathbb{R}^{m \times 1}$. The initial distribution is usually either estimated from the likely sequence of states obtained by the model or from the TPM (see \cite{zucchini2017hidden} for details).  

Finally, we often make a simplifying assumption about the transition probabilities. Namely, we assume that they are stable across time. In such a case, we call the HMM \textit{time-homogenous} and the  TPM is exactly the same across all occasions $N_T$.

In total, the HMM estimates the following sets of parameters:

\begin{enumerate}
    \item{The transition probability matrix $\pmb{\Gamma}$}
    \item{The parameters of the component distributions, collectively represented in $\pmb{\theta}$. Given a Gaussian component distribution, $\pmb{\theta}$ contains the state-dependent means and standard deviations. }
\end{enumerate}

\noindent Note that we omit the $m$ initial probabilities collected in the vector $\pmb{\delta}$ from this list. As previously mentioned, we can find these probabilities implicitly by examining the likely state sequence.

\subsubsection{HMMs in the academic literature}

HMMs are used in a wide variety of contexts. One area in which HMMs have found a ready application is that of speech recognition \citep{rabiner1989tutorial, gales2008application} and Natural-Language Processing (NLP) applications such as Part-Of-Speech (POS) tagging \citep{jurafsky2008speech}. \cite{castellano2010bayesian} use HMMs on financial data to determine whether the US Dollar will increase or decrease in value. Here, the latent states refer to "appreciation" or "depreciation" of the US dollar. HMMs are also used to model animal behavior \citep{leos2018introduction, bode2018using, whoriskey2017hidden, vsabata2016modeling}. In this context, observed data is collected that is thought to be connected to behavioral states, such as "eating", "moving" and "resting". Similar studies attempt to extract similar latent behavioral states from human behavior. The objective of such studies ranges from recognizing emotional states \citep{yamato2002recognizing} to predicting future behavior \citep{mitterbauer2009behavior} or to extract early-warning signals with respect healthcare needs of elderly hospital patients \citep{chung2008daily}.

HMMs can be used with both univariate and multivariate outcome data. For example, \cite{flexerand2002automatic} use multivariate outcome data to automatically extract sleep states ("Awake", "REM" and "Non-REM"). In the context of climate change research, \cite{hughes1994incorporating} use multivariate climate-related data to predict "characteristic weather states". \cite{song2017hidden} use multivariate outcome data to investigate the prevention of cocaine use.

Another important extension is the possibility to introduce covariates that influence the TPM or component distribution parameters. This allows us to improve the estimates of these parameters, and to model time trends and seasonality. \citep{zucchini2017hidden, vermunt1999discrete, vermunt2010longitudinal}.

The wide application of HMMs serves to illustrate the flexibility of the model. In short, researchers need only collect time-series data that is believed to be influenced by unobserved, discrete states, each of which are associated with their own component distributions. In the next section, we turn to the case in which we want to estimate a model for multiple subjects.

\subsection{Multilevel hidden Markov models} \label{mHMM-theory}

An important limitation of the HMM is that it is only suited to model the time-series data of a single subject. If one has multiple subjects, then the researcher either needs to fit one model and assume that it holds for all subjects, or they must decide that each subject merits their own model and hence assume that they are very dissimilar. The former approach is difficult to justify because even small differences between subjects can lead to a poor summary of the data. The latter approach is unsatisfactory because fitting a model for each subject individually is time-consuming, not parsimonious, and not efficient.

Consider again the sleep state example based on observed heart rate data. Assume that we collect heart rate data for $100$ subjects who are similar on many traits (e.g. gender, age and so on). Even though these subjects are very similar, we still want to account for between-subject variance in their sleep state patterns because we expect each person to exhibit \textit{some} different behavior. A parsimonious and computationally friendly way to do this is to use the multilevel framework (see e.g. \cite{hox2017multilevel,snijders2011multilevel,gelman2006data}). In a multilevel model, we assume that there is some hierarchical structure in the data such that it can be modeled by defining a cluster (e.g., group) level  and a within-cluster (e.g., within-group, hence subject-specific) level. Doing so allows us to represent the overall group model while accommodating differences that occur between subjects. 

The group-level parameters describe the overal mean in and variation that can be found between subjects, and are most frequently modeled using normal distributions. In the example of heart rate, the group-level mean describes the average heart beat between subjects for the states "awake" and "asleep", and the group-level variance describes the variation between subjects in the average heart beat value for the states "awake" and "asleep". For the TPM, we also model the between-subject variation in such a way, although doing so is more complicated (see section \ref{sec:model-tpm}). The subject-specific parameters are realizations of the group-level parameters. 

\subsubsection{Multilevel HMMs in the academic literature} \label{sec:mHMM-in-lit}

\cite{altman2007mixed} provides the first formalization of the mHMM, in which she uses a combination of random effects and covariates to model between-subject differences. 
She applies her model to a data set of lesion counts in multiple sclerosis patients and shows that it is able to adequately model between-subject differences. This version of the mHMM is also used by \cite{schliehe2012application} and \cite{mckellar2014using} to model the behavior of multiple animals. 

Altman's version of the mHMM estimates the model parameters by directly optimizing the likelihood of the model \citep{altman2007mixed,turner2008direct}. This places a severe constraint on the number of parameters allowed to vary over subjects as it is computationally burdensome and time-consuming to perform the necessary numerical integration to fit the model \citep{altman2007mixed, schliehe2012application, mckellar2014using}.
To this end, some authors have proposed a variety of adaptations. \cite{maruotti2011mixed} adapts Altman's model by using the Expectation-Maximization (EM) method \citep{baum1970maximization} instead of direct numerical optimization. \cite{maruotti2011mixed} further investigates the use of various non-parametric group-level distributions, which he shows are easier to estimate by Maximum Likelihood (ML) methods than parametric group-level distributions. He uses this approach to model multivariate legislative count data at the regional and national level in Italy \citep{lagona2015multilevel}. \cite{jackson2015two} use the EM algorithm in conjunction with numerical integration to model risky teenage driving behavior. They further tweak the model to relax the assumption of conditional independence for a subject's time-series data. \cite{dedieu2014mixed} propose an adapted version of the EM algorithm to speed up computations. 

A flexible alternative to fit mHMMs is to use Bayesian estimation. This approach was first developed in the context of regular HMMs as an alternative to the EM method \citep{fruhwirth2001markov, scott2002bayesian, ryden2008versus, leos2018introduction}. In a comparison of both methods, \cite{ryden2008versus} finds that, for more complex models, Bayesian estimation is superior in terms of computation time. Additionally, Bayesian methods yield many additional metrics, such as coverage, that are much harder to extract using Frequentist methods (see e.g. \cite{zucchini2017hidden}, chapter 3). The observation that Bayesian methods are faster than their Frequentist alternatives also holds in the case of mHMMs \citep{zhang2014bayesian}. One issue in using Bayesian estimation for (m)HMMs is that of \textit{label switching}. This term describes the situation in which, during successive draws in the MCMC sampling algorithm, the labels of two or more states are "switched" around, even though the complete-data likelihood is exactly the same as when label switching does not occur \citep{scott2002bayesian}. Typically, one can observe this in the posterior distributions, which display sudden "jumps" if label switching occurs. Both \cite{scott2002bayesian} and \cite{shirley2010hidden} recommend to choose good starting values when using (m)HMMs to minimize issues related to label switching.

Several authors have used the Bayesian mHMM. For example, \cite{shirley2010hidden} use it in the context of a clinical study investigating the treatment of alcoholism. \cite{rueda2013bayesian} show that the Bayesian approach can also be used to estimate the number of hidden states, and show by means of a simulation study that this model is more accurate than a single-level HMM. De Haan (2017) argue that the Bayesian approach to fitting mHMMs yields additional advantages compared to Frequentist methods. For example, they argue that Bayesian methods are appropriate when dealing with small samples, although this may be only the case when using strong informative priors (see \cite{smid2020bayesian} for a more general discussion on this topic). Furthermore, \citeauthor{de2017use} argue that Bayesian methods are robust to missing data and that they are computationally friendly. They also apply the Bayesian mHMM on several data sets that measure psychological processes, such as negative and positive affect and aggression during therapeutic sessions.


In summary, it is advantageous to use the multilevel hidden Markov model when modeling ILD of multiple subjects. By using random effects, we can fit a parsimonious and relatively efficient model using Bayesian methods. 
Given the potential of this model in the social and behavioral sciences and beyond, it is necessary to examine its ability to adequately estimate model parameters for different sample sizes at the level of subjects $N$ and number of occasions $N_T$. We next turn to the methods section, in which we describe the procedure of the simulation study.


\section{Methods} \label{sec:methods}

The purpose of the simulation study was to empirically assess the performance of the MHMM on multivariate data with continuous outcomes with varying sample sizes and amount of heterogeneity between subjects. The simulated datasets in this study are inspired by electroencephalogram (EEG) and electrooculography (EOG) recordings used to infer sleep stages (here, sleep states), a detailed description of which is given in section \ref{sec:emp-application}. For the purposes of this analysis, a "sleep state" is one of three categories: (1) "Awake", (2) "REM" (Rapid Eye Movement) sleep and (3) "NREM" (Non-REM) sleep. 

The sleep dataset displays two typical features not necessarily present in other data, challenging model performance. The first feature is the strong overlap between the state-dependent component distributions. Based on a preliminary analysis, we select three variables that display the least amount of overlap, and in which the states overlap in different ways. Nonetheless, the overlap in the state-dependent component distributions can be considered to be on the extreme end of the spectrum (see figure \ref{fig:emiss-densities-densplot} in appendix \ref{sec:appendix1}). The second feature is that of very high ($> 0.95$) self-transition probabilities. This is not an uncommon feature of ILD, especially when it concerns data on behavioral states (see section \ref{sec:simproc}). However, there are also many settings in which the self-transition probabilities are much lower.

To examine model performance in less extreme scenarios and to allow for generalizabilty to a broader set of datasets, we also run a set of $10$ baseline scenarios in which we use less extreme values for the component distributions and transition probabilities. For the remainder of this paper, we refer to the results that are based on the sleep dataset as the "sleep data simulation results". We refer to the results that are based on the baseline scenarios as the "baseline simulation results".

\subsection{Model parameter settings} \label{sec:model}

The following notation will aid us in defining these parameters such that they suit the multilevel framework. We have $n \in \{1, 2, \dots, N\}$ subjects, for which we each have outcome data on $t \in \{1, 2, \dots, N_T\}$ occasions. Given that the observed data is multivariate, each subject has observed data on $k = 3$ outcome variables. In total, we therefore have $\pmb{y}_{11}, \pmb{y}_{12}, \dots, \pmb{y}_{1N_T}, \dots, \pmb{y}_{N1}, \dots, \pmb{y}_{NN_T}$ observations, where $\pmb{y}_{NN_T} \in \mathbb{R}^{3 \times 1}$. Finally, we have $m=3$ latent states. 

\subsubsection{Component distributions} \label{sec:model-components}

We assume a normal distribution for each of the component distributions, for which the mean is allowed to vary over subjects and the variance is assumed to be fixed over subjects. The subject specific deviations to the group level mean follow a zero mean normal distribution. As such, we model three parameters for the component distributions: 

\begin{enumerate}
    \item{The group-level mean $\beta_{00km}$ for component distribution $m$ of dependent variable $k$. We refer to the group-level mean $\beta_{00km}$ as the \textit{component group-level mean}.}
    \item{The random variance $\sigma^2_{u_0,km}$ which models the subject-specific deviation term $u_{0nkm}$, where $u_{0nkm}$ represents the subject $n$ specific deviation from the group-level mean on component distribution $m$ of dependent variable $k$. The subject-specific deviation term $u_{0nkm}$ is assumed to follow a zero mean normal distribution with variance $\sigma^2_{u_0,km}$, $u_{0nkm} \sim N(0, \sigma^2_{u_0,km})$. We refer to the variance term $\sigma^2_{u_0,km}$ as the \textit{component distribution random effect}.}
  \item{The residual variance $\sigma^2_{\epsilon,km}$ which models the subject-level residual error term $\epsilon_{nkmt}$, where $\epsilon_{nkmt}$ captures the residual error between occasion $t$ for person $n$ on component distribution $m$ of dependent variable $k$, and the component and subject specific mean given by $\beta_{00km}  + u_{0nkm} $.  The subject-level residual error term $\epsilon_{nkmt}$ is assumed to follow a fixed over subjects zero mean normal distribution with variance $\sigma^2_{\epsilon,km}$,  $\epsilon_{km} \sim N(0, \sigma^2_{\epsilon,km})$. } 
\end{enumerate}

The used population values of each of the $m$ component distribution group-level means for each outcome variable are shown in table \ref{tab:popval-means}. In our study, we set the residual variance term $\sigma^2_{\epsilon,km}$  to $0.1$. Note that we model each of the outcome variables as independent normal distributions. That is, the covariance between any two outcome variables is assumed to be zero after conditioning on the latent state. 

\begin{table}[t]
\centering
\begin{adjustbox}{width=0.45\textwidth}
\begin{tabular}{ | L{10em} | R{4em} | R{4em} | R{4em} | }
\hline
\multicolumn{4}{| c |}{Sleep data simulations} \\
\hline
                              & \textbf{Awake} & \textbf{NREM} & \textbf{REM}                                     \\ \hline
\textbf{EEG mean beta} & -0.360        & -0.600        & 0.700 \\ \hline
\textbf{EOG median theta}     & 1.010          & -1.310        & -0.240                                           \\ \hline
\textbf{EOG min beta}         & 0.750          & -1.310        & 0.005                                            \\ \hline       
\multicolumn{4}{| c |}{Baseline simulations} \\
\hline
                              & \textbf{State 1} & \textbf{State 2} & \textbf{State 3}                                     \\ \hline
\textbf{Dependent variable 1} & -3.900        & -1.000        & 2.400 \\ \hline
\textbf{Dependent variable 2}     & 3.050          & -3.400        & -0.500                                           \\ \hline
\textbf{Dependent variable 3}         & 0.400          & 3.500        & -2.800                                            \\ \hline 
                      \end{tabular}
\end{adjustbox}
\caption{Population values of the group-level means for each of the $m$ component distributions on three dependent variables selected for the simulation study.}
\label{tab:popval-means}
\end{table}

\subsubsection{Transition probabilities} \label{sec:model-tpm}

To incorporate between-subject heterogeneity in the transition probabilities, a multinomial regression model with fixed and random effects for each row of the subject-specific TPM is used \citep{altman2007mixed, aartsusing2019, zucchini2017hidden}. This ensures that each row of the subject-specific TPM $\pmb{\Gamma}_n$ sums to unity. For each latent state $i \in \{1,2,3\}$ and $j \in \{2,3\}$, and each subject $n \in \{1,2,\dots, N\}$, we say that the transition probability of going from state $i$ to state $j$ for the $n^{\text{th}}$ person is given by: 

\begin{equation}
    \gamma_{nij} = \frac{\exp{(\alpha_{nij})}}{ 1 + \sum_{j \in {\{2, 3}\}} \exp{(\alpha_{nij})}} = \text{MNL}(\alpha_{nij})
    \label{eq:subj-tpm}
\end{equation}

And:

\begin{equation}
    \alpha_{nij} = \bar{\alpha}_{ij} + \psi_{nij} 
\end{equation}

As such, each row of the transition probability from state $i$ to state $j$ for person $n$ is modeled by a set of intercepts $\pmb{\alpha_{ni}} \in \{\alpha_{ni2}, \alpha_{ni3}\}$, where each intercept $\alpha_{nij}$ composed of a group-level mean intercept $\bar{\alpha}_{ij}$ and a subject-specific error term $\psi_{nij}$, where $\psi_{nij} \sim N(0, \sigma^2_{\psi, ij})$. Throughout the remainder of this text, we refer to the term $\bar{\alpha}_{ij}$ as the \textit{TPM group-level intercept} and we refer to the term $\sigma^2_{\psi, ij}$ as the \textit{TPM random effect}. We refer to the transition probabilities that are derived from the TPM group-level intercepts as group-level transition probabilities. For model identification purposes, the first category for each row in the subject-specific TPM $\pmb{\Gamma}_n$ is the baseline category, and is estimated by setting the numerator in equation \ref{eq:subj-tpm} equal to $1$. 


The population values for the group-level TPM are given in figure \ref{tab:popval-tpm}.

\begin{table}[t]
\centering
\begin{adjustbox}{width=0.45\textwidth}
\begin{tabular}{ C{1em} C{1em}  L{4em}  R{4em}  R{4em}  R{4em}  }
\hline
\multicolumn{6}{c}{Sleep data simulations} \\
&&& \multicolumn{3}{c }{To state} \\
&&& \multicolumn{3}{c }{\small (occassion $t$ + 1)} \\
\parbox[t]{2mm}{\multirow{4}{*}{\rotatebox[origin=c]{90}{From state}}} &
\parbox[t]{2mm}{\multirow{4}{*}{\rotatebox[origin=c]{90}{(occassion $t$)}}}                             
& & \textbf{Awake} & \textbf{NREM} & \textbf{REM}                                     \\ \cline{4-6}

&&  \textbf{Awake}     & 0.984          & 0.003        & 0.013 \\ \cline{4-6}
& & \textbf{NREM}     & 0.007          & 0.959        & 0.034                                           \\ \cline{4-6}
& & \textbf{REM}        & 0.012          & 0.021        & 0.967                                            \\ 
\hline
\multicolumn{6}{c}{Baseline simulations scenario 2-5} \\
&&& \multicolumn{3}{c }{To state} \\
&&& \multicolumn{3}{c }{\small (occassion $t$ + 1)} \\
\parbox[t]{2mm}{\multirow{3}{*}{\rotatebox[origin=c]{90}{From state}}}  &
\parbox[t]{2mm}{\multirow{3}{*}{\rotatebox[origin=c]{90}{(occassion $t$)}}} 
& &                             \textbf{State 1} & \textbf{State 2} & \textbf{State 3}                                     \\ \cline{4-6}

& &\textbf{State 1}     & 0.800          & 0.100        & 0.100 \\ \cline{4-6}
&& \textbf{State 2}     & 0.150          & 0.700        & 0.150                                           \\ \cline{4-6}
&& \textbf{State 3}        & 0.180          & 0.640        & 0.180                                            \\ \cline{4-6}

                      \end{tabular}
\end{adjustbox}
\caption{Population values of the group-level TPM.}
\label{tab:popval-tpm}
\end{table}

\subsection{Sample size and between-subject variance} \label{sec:simproc}

Recall that we vary three quantities in this study: (1) the number of subjects $N$, (2) the number of occasions $N_T$, and (3) the between-subject variance. We use the following values for the number of subjects and the number of occasions in the sleep data simulations:

\begin{enumerate}
    \item{\textbf{Number of subjects:} the number of subjects varies as $N = 10,\ 20,\ 40,\ 80$.}
    \item{\textbf{Number of occasions per subject:} the number of occasions per subject varies as $N_T = 400,\ 800,\ 1.600$.}
\end{enumerate}

Two considerations inform the choices for the sample sizes at the subject and occasion level. Firstly, even though the simulated data are based on EOG and EEG measurements to detect sleep states, it is important for this example to generalize to other settings. The literature on HMMs and mHMMs shows a large variety of different sample sizes with respect to the number of subjects and the number of occasions. The number of subjects can range anywhere from $3$ \citep{whoriskey2017hidden} to $7.000$ \citep{dedieu2014mixed}, and the number of occasions can vary from $4$ \citep{song2017hidden} to well over $10.000$ \citep{vsabata2016modeling}. This is often related to the type of measurement (e.g. sensor data, questionnaire) and the research field. In the social and behavioral sciences, it is rare to observe a large number of subjects in combination with a large number of occasions. However, this is more common in studies that model behavior and use a form of sensor data. For example, \cite{de2017use} use the results of a study using data on $141$ subjects and $539$ occasions. Moreover, many studies that model animal behavior collect very long sequences of occasions (often in excess of $10.000$). Usually, the number of animals that are tracked lies between $30$ and a $100$ individuals. Given that studies using fewer subjects exist in the literature, we also consider smaller subject-level sample sizes. Additionally, it is often claimed Bayesian methods are robust to small sample sizes. However, such claims must be investigated to prevent misapplication of the model \citep{smid2020bayesian, mcneish2019two}. 

Secondly, a previous (small) simulation study conducted by \cite{altman2007mixed} for Frequentist mHMMs used sample sizes of $30$ and $60$ subjects with a small occasion sample sizes ($20$ occasions). We note that she restricts her study to the estimation of a single random effect and her observed data follows a poisson distribution. Altman finds that $60$ subjects is a sufficiently large number to estimate the random effect in her study. 

In the case of our simulation study, we choose a minimum occasion sample size for practical reasons. As is common in this type of data in which the measurement time between occasions is short, the self-transition probabilities (transition probabilities on the diagonal entries of the TPM) are very high. This means that sufficiently long occasion sequences are required to ensure that all state transitions occur. Given that the model is computationally burdensome and given the large number of simulation iterations that will be executed, we set $1.600$ occasions as an upper limit.

To vary the between-subject variance, we need to take into account that this requires manipulation of both the component distribution and TPM random effects. These will be manipulated as follows within the sleep data simulations:

\begin{enumerate}
    \item{\textbf{Component distributions:} The between-subject deviations from the component distribution means are modeled as normal distributions with mean $0$ and variance $\sigma^2_{u_0,mk}$ (see section \ref{sec:model}). The variance term $\sigma^2_{u_0,mk}$ will be varied as $\zeta=0.25,0.5,1,2$.}
    \item{\textbf{TPM:} Varying the TPM random effect is less straightforward because of the multinomial regression models used to model the transition probabilities for each row of the subject-specific TPM. (see section \ref{sec:model-tpm}). We choose values that correspond to a small, medium and large amount of between-subject variance $\sigma^2_{\psi, ij}$, and vary $\sigma^2_{\psi, ij}$ as $Q = 0.1, 0.2, 0.4$.} 
\end{enumerate}

In the baseline scenario's, all component distributions for each of the $k$ dependent variables are well separated, see table \ref{tab:popval-means}, and  the TPM random effect is fixed at $\sigma^2_{\psi, ij}$ at  $Q = 0.1$. We vary the random variance $\sigma^2_{u_0,mk}$ of the between-subject deviations from the component distribution means to $\zeta = 0.25$ (scenarios A) and $\zeta = 0.50$ (scenarios B). The TPM, number of subjects $N$, and number of occasions per subject $N_T$ are varied as follows: 
\begin{enumerate}
    \item{\emph{Baseline scenario 1}: Sleep data simulation TPM (see table \ref{tab:popval-tpm}), $N = 40$, $N_T = 800$.}
    \item{\emph{Baseline scenario 2}: Baseline simulation TPM (see table \ref{tab:popval-tpm}), $N = 40$, $N_T = 800$.}
    \item{\emph{Baseline scenario 3}: Baseline simulation TPM (see table \ref{tab:popval-tpm}), $N = 80$, $N_T = 800$.}
    \item{\emph{Baseline scenario 4}: Baseline simulation TPM (see table \ref{tab:popval-tpm}), $N = 80$, $N_T = 3200$.}
    \item{\emph{Baseline scenario 5}: Baseline simulation TPM (see table \ref{tab:popval-tpm}), $N = 140$, $N_T = 800$.}
\end{enumerate}

We will evaluate the sleep data scenarios in full factorial design, leading to a total of $4 \times 3 \times 4 \times 3 = 144$ plus $2 \times 5 = 10$ baseline scenarios. In this study, we run $250$ iterations for each of the simulation scenarios. See appendix \ref{sec:mcmcse} for a short discussion on the number of simulation iterations.  All data sets are created using the R package \textit{mHMMbayes} \citep{aartsusing2019}. A precise description of the process by which the data is generated is given in appendix \ref{sec:appendix1-datasim}.

\subsubsection{Model settings and hyper-prior specification}

For the purposes of this paper, we use only uninformative hyper-prior distributions except in the case of the prior state-dependent component distributions, for which we use the sample means. For further details on the hyper-prior distributions used in the model, see appendix \ref{sec:appendixI-hyperpriors}.

For the component distributions, we construct starting values for each simulation iteration by adding random noise from a uniform distribution with lower limit $-0.2$ and upper limit $0.2$ to the population parameters. The diagonal entries of the TPM (the self-transition probabilities) are randomly generated from a uniform distribution with lower limit $0.5$ and upper limit $0.8$. The off-diagonal entries are equal to each other and are chosen such that the rows of the TPM sum to unity.

Based on a preliminary analysis, we set the number of MCMC iterations for each model to $3.250$ with a burn-in sample of $1.250$. These settings were shown to be adequate while also respecting the computational burden of the model. 

\subsubsection{Model convergence}

For each scenario, we store the complete model data for three simulation iterations to check the convergence of the models. To this end, we run a second chain for these models using different starting values. We analyze and report the number of models that do not converge, as well as the number of cases (if any) for which inadmissible estimates are found. 

\subsection{Evaluation metrics} \label{sec:eval-metrics}

The following metrics are used to evaluate the quality of parameter estimates for the Bayesian mHMM, and are taken from \cite{morris2019using}:

\begin{enumerate}
    \item{\textbf{Parameter bias:} this is the discrepancy between the average parameter estimate in a scenario and the true population value used to generate the data in scenario $r$, expressed either as a number or as a percentage relative to the value of the population parameter.}
    \item{\textbf{Empirical Standard Error (empirical SE):} this is a measure of precision of the estimator of the population parameter, and estimates the long-run standard deviation of the parameter estimates as the number of iterations in scenario $r$ grows large.}
    \item{\textbf{Average Model Standard Error (model SE):} the model SE represents the average standard error of a parameter estimate. It is closely related to the empirical SE; if the model SEs are estimated well, then the average model SE should be approximately equal to the empirical SE. A large or small model SE relative to the empirical SE indicates a bias in the estimation of the standard errors around a parameter estimate.}
    \item{\textbf{Mean-Squared Error (MSE):} the MSE is commonly known as the sum of squared bias an the variance of $\hat{\theta}$ (or the square of the empirical SE). Hence, it is a composite measure of both bias and variance of the estimator.}
    \item{\textbf{Coverage:} the percentage of generated datasets for which the population value of a model parameter lies within the estimated confidence interval in scenario $r$.}
    \item{\textbf{Bias-corrected coverage:} bias-corrected coverage is a metric introduced by \cite{morris2019using}. Instead of using the population parameter to compute the coverage, we use the average parameter estimate. This eliminates parameter bias as a source of poor coverage.}
\end{enumerate}

The primary evaluation metric is parameter bias. A bias of 5\% of the parameter value is considered acceptable. Furthermore, we consider a coverage of 92\% to 98\% acceptable. Finally, we note that, although we use all of the metrics listed above to evaluate the simulation study, we do not report the results of all metrics. All results, however, are available in the research archive.


\section{Results}

In this section, we present the results of the simulation study. We first examine the posterior distributions and report on model convergence. We then turn to the model performance with respect to each of the parameters.

\subsection{Posterior distributions and model convergence} \label{sec:check-convergence}

The mean and median \textit{Maximum a Posterior} (MAP) estimates for the parameters are very similar, indicating no skewness in the posterior distributions of the model parameters. No inadmissible values were found for any of the parameters. In some of the simulation scenarios, the distribution of parameter estimates across the $250$ simulation iterations is bi-modal, although these are not pronounced and hence not considered to be an issue.

\begin{figure}[!b]
\centering
\includegraphics[width=1\linewidth]{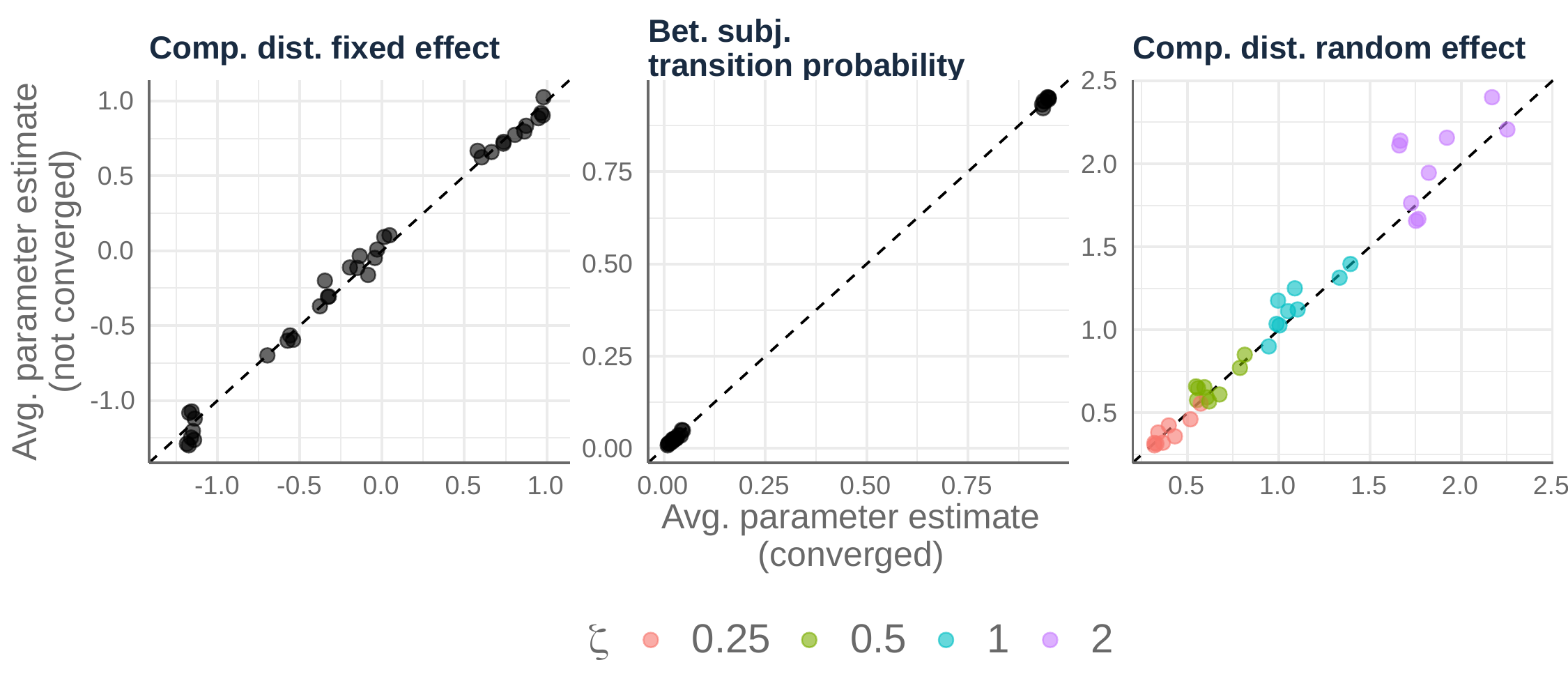}
\caption{Comparison of the average parameter estimates that converged (x-axis) versus the average parameter estimate of those cases where the parameter did not converge (y-axis). The dashed line indicates a one-to-one mapping of the values on both axes. The component distribution random effects are colored by the value of $\zeta$.}
\label{fig:convergence-parameter-estimates-diff}
\end{figure}

Model convergence was not an issue for any of the baseline scenarios. With respect to the models used in the simulation study, we observe that, on about 68\% of all models that we checked, at least one parameter did not converge. In 34\% of cases, more than $5$ parameters did not converge. The parameter that most often did not converge is the transition probability from NREM to REM sleep (30\%), followed by the transition probability from REM to NREM sleep (26\%) and the component distribution random effect of the REM state for the variable \textit{EOG min beta} (24\%). All other parameters converged at least in 77\% of all cases. Across the conditions, the group-level means ($>88\%$) and between-subject variances ($>78\%$) converged most often when the component distribution random effect $\zeta \leq 0.5$. When the component distribution random effect is large, the convergence rates are at their worst for the means of the REM state ($68\%$).

The convergence of the transition probabilities varies somewhat across the value of the TPM random effect $Q$, and performs best when $Q=0.1$. When $Q=0.4$, the lowest convergence rate is that of the transition probability from REM to NREM sleep (66\%). 
Finally, figure \ref{fig:convergence-parameter-estimates-diff} indicates that the parameter values that did not converge do not differ a lot from those parameter values that did converge except for the estimates of the variance term when the component distribution random effect $\zeta$ is large.

\subsection{Component distribution group-level means} \label{sec:results-comp-fixed-effects}

The results for the component distribution group level means $\beta_{00km}$ indicate that the subject sample size is the most important determinant of parameter quality across all conditions, whereas the occasion sample size plays little to no role in obtaining better parameter estimates. The quality of the estimates typically worsen as the component distribution random effects $\sigma^2_{u_0,mk}$ of the between-subject deviations from the component distribution means increase. 

The baseline scenarios indicate that, when the state-dependent component distributions barely overlap, we observe parameter bias on the component distribution group-level means when the number of subjects $N=40$, with percent bias rising to 16\%. When we increase the sample size to $N=80$, the bias becomes much less pronounced (percent bias $\leq 11$\%). Parameter bias falls within the acceptable 5\% mark across all outcome variables when $N=140$. See appendix \ref{sec:appendixII-baseline} for full results on the baseline scenario's. 

\begin{figure*}[!t]
\centering
\begin{subfigure}{.5\textwidth}
  \centering
  \includegraphics[width=.85\linewidth]{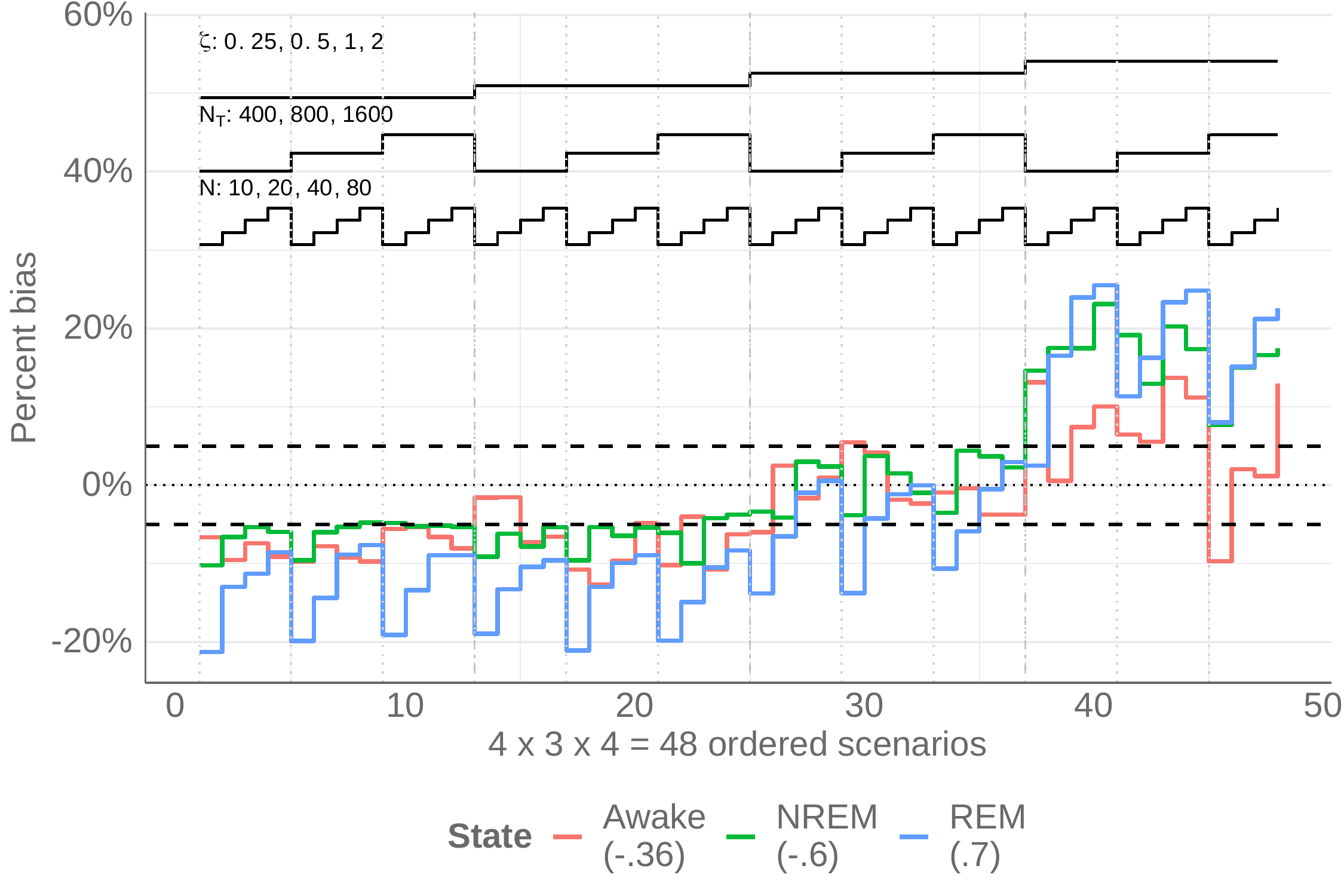}
  \caption{Nested loop plot (percent bias), EEG mean beta}
  \label{fig:pbias-emiss-1}
\end{subfigure}%
\begin{subfigure}{.5\textwidth}
  \centering
  \includegraphics[width=.85\linewidth]{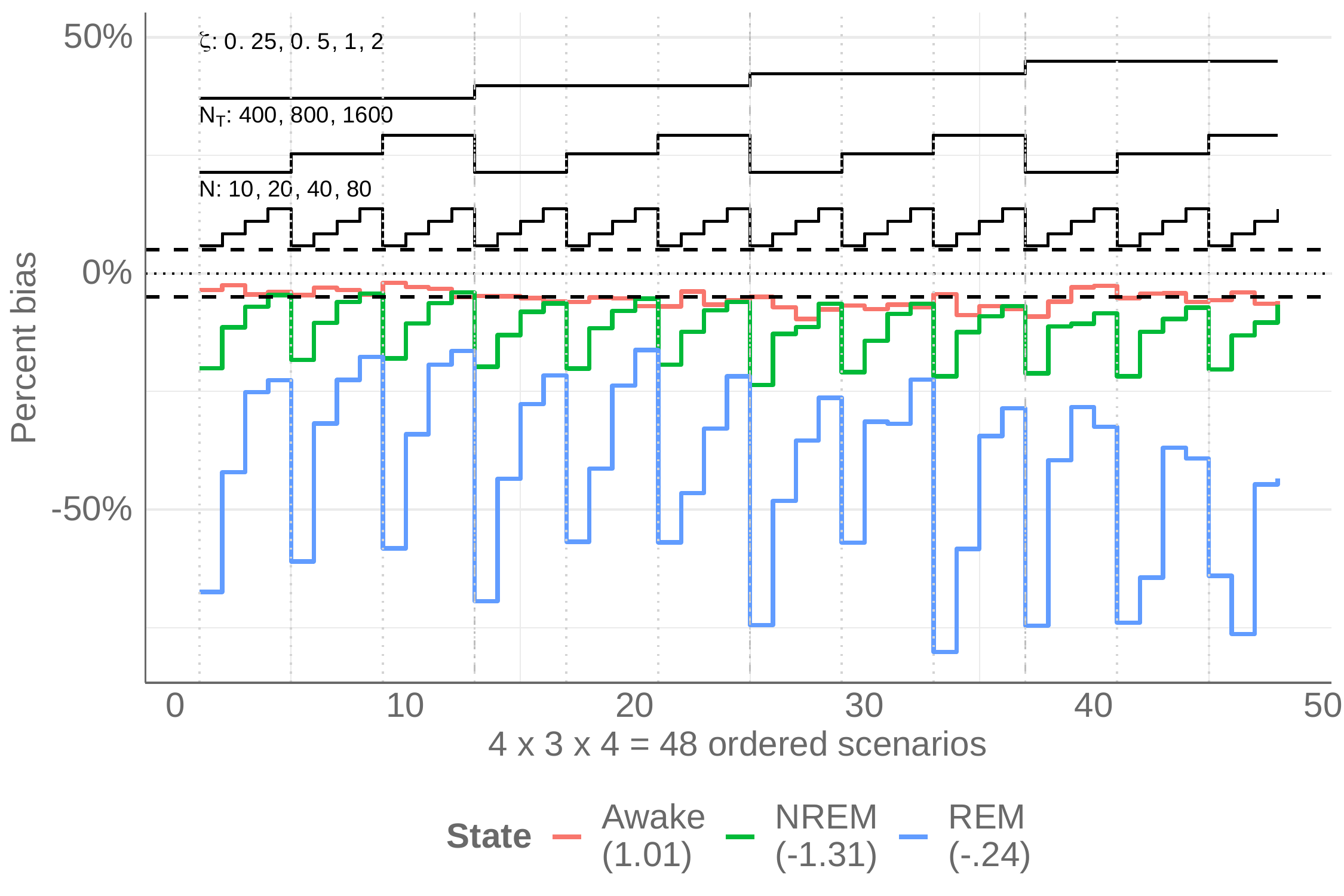}
  \caption{Nested loop plot (percent bias), EOG median theta}
  \label{fig:pbias-emiss-2}
\end{subfigure}
\begin{subfigure}{.5\textwidth}
    \centering
    \includegraphics[width=0.85\linewidth]{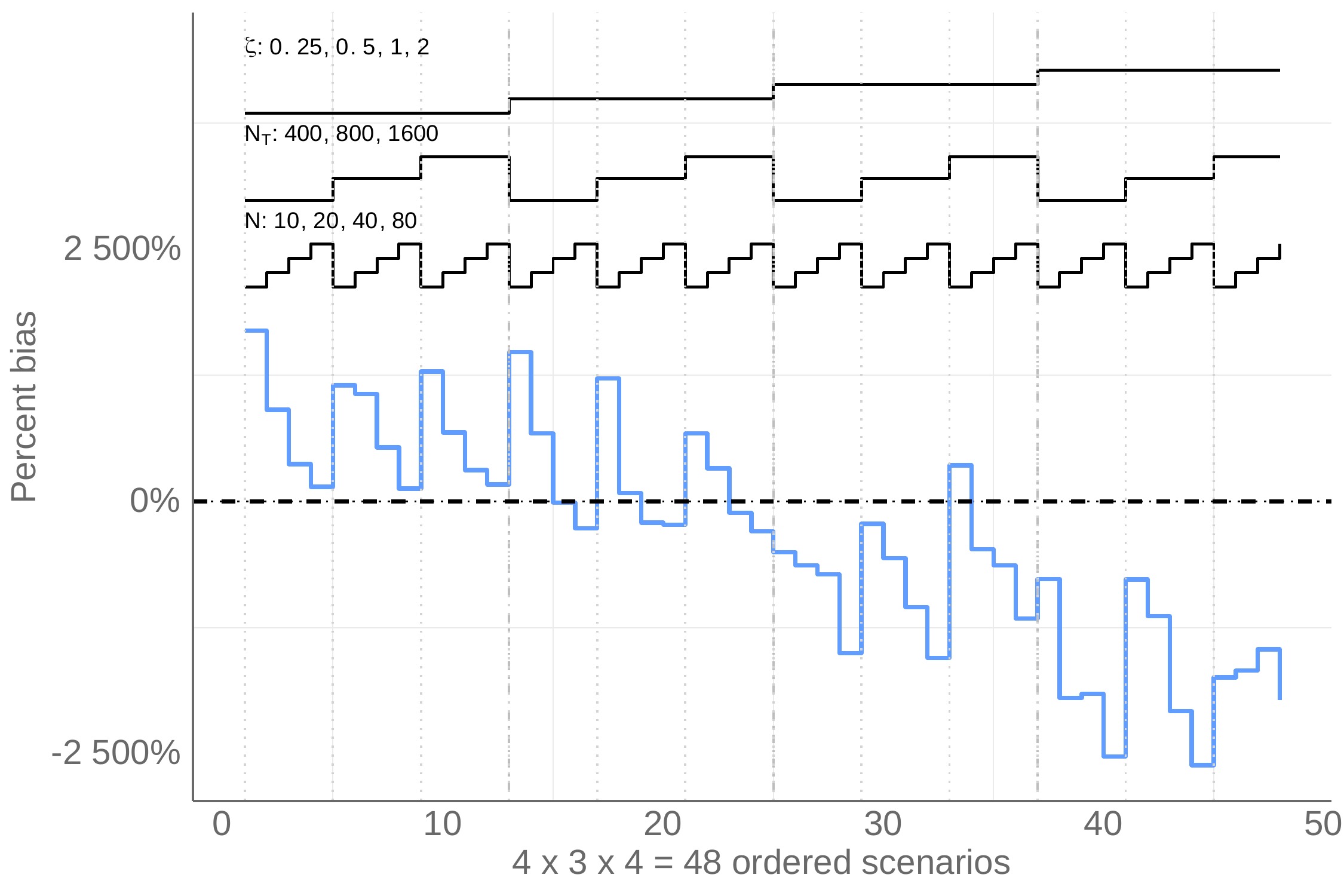}
    \caption{Nested loop plot (percent bias), EOG min beta (REM only)}
    \label{fig:nlp-emiss3-bias-state3}
\end{subfigure}%
\begin{subfigure}{.5\textwidth}
    \centering
    \includegraphics[width=0.85\linewidth]{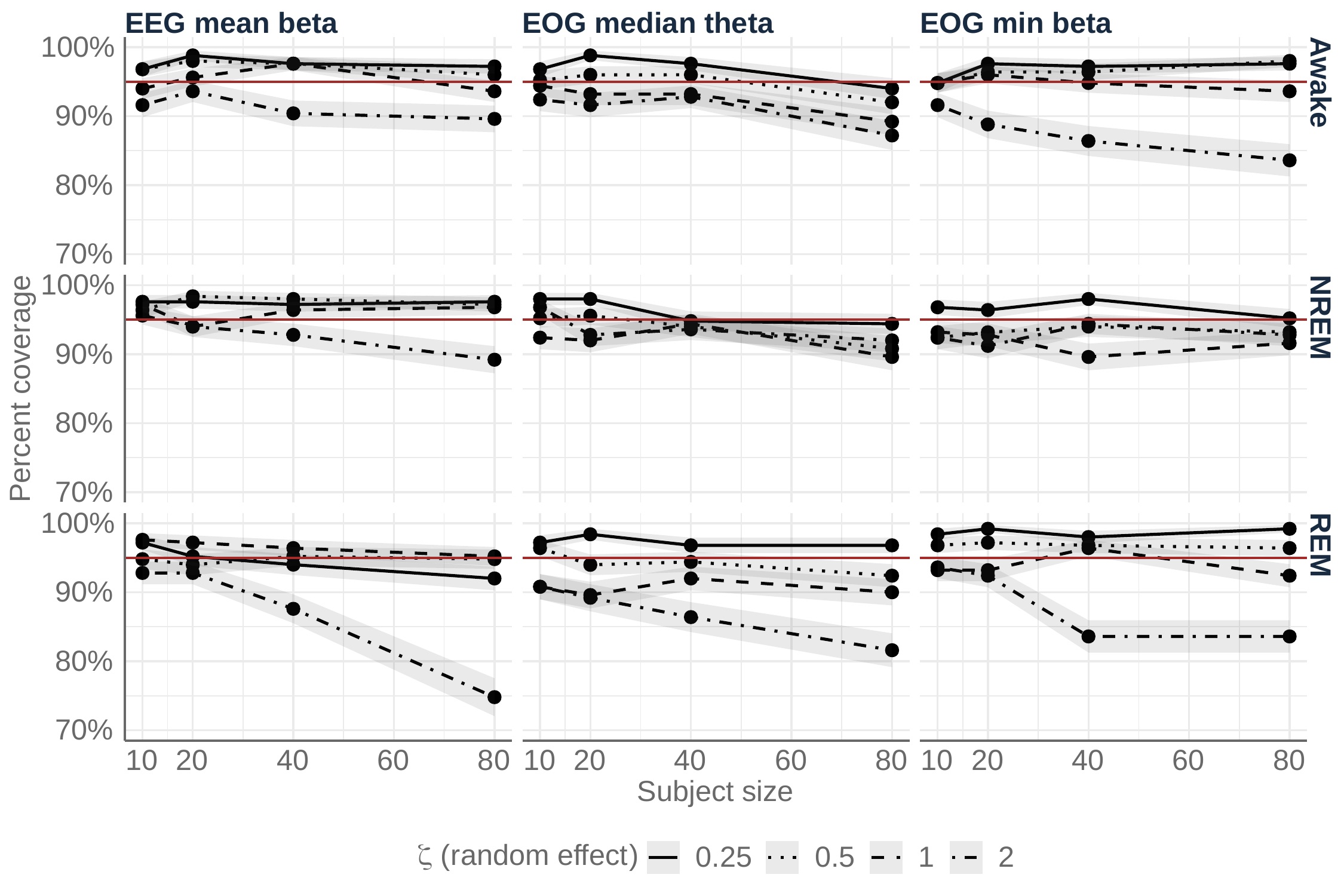}
    \caption{Trellis plot (coverage), all outcome variables}
    \label{fig:emission-means-coverage}
\end{subfigure}
\caption{The nested loop plot (panels a, b, and c) summarizes the percent bias in the parameter estimates of the component distribution group-level means across the simulation settings. The conditions are given at the top of each figure. The colored lines indicate the value of the parameter bias as we vary the simulation settings. The horizontal dashed lines indicate the region of acceptable parameter bias. The conditions are fixed across $Q=0.2$. In panels (a) and (b), the population values for the component distribution group-level means are provided between parentheses in the legend. In panel (c), the population value for the REM state is $.005$. Panel (d) shows the coverage of the component distribution group-level means. Conditions are fixed across $N_t = 1.600$ and $Q=0.2$. MC SEs are denoted by the shaded areas.}
\label{fig:nlp-emiss-bias-fixed}
\end{figure*}

Within the sleep data simulations, the state "Awake" is estimated with the least amount of bias, and the group-level means associated with this state are typically decent when the number of subjects is large ($N=80$). The NREM state is generally estimated well if the value of the random effect is low and the number of subjects is high. In both cases, the parameter estimates are biased downward across the conditions. These trends are clearly visible in the nested loop plot (NLP) shown panels (a) and (b) in figure \ref{fig:nlp-emiss-bias-fixed} for the outcome variables \textit{EEG mean beta} and \textit{EOG median theta}. The NLP is a convenient (although dense) figure that displays the change in an evaluation metric as we vary the simulation settings \citep{rucker2014presenting, gasparini2018rsimsum}.

The REM state exhibits biased estimates that are non-negligible regardless of the value of the random effect, although this bias becomes much lower when the subject sample size is large. Figures \ref{fig:pbias-emiss-1}, \ref{fig:pbias-emiss-2} and \ref{fig:nlp-emiss3-bias-state3} further indicate that, across the outcome variables, the parameter estimates of the REM state are typically best (though still at least 7\% short of the acceptable 5\% mark) when the sample size is large, and the value of the component distribution random effect is low.

Coverage of the estimates typically hovers around 95\% regardless of sample or occasion size, although it does seem to deteriorate as we increase the component distribution random effect on the (see figure \ref{fig:emission-means-coverage}). At low settings of the random effect and low subject sample sizes, we further observe over-coverage on some of the parameter estimates. This happens because the model SE exceeds the empirical SE, indicating that the model is over-estimating the model 95\% CCIs and hence covers the population parameter too often. Across the conditions, empirical and model SEs of the group-level mean estimates decrease as the number of subjects increases, indicating that the parameter estimates become more precise regardless of the value of the component distribution random effect.

\subsection{Component distribution, random effects} \label{sec:results-randomeffects}

\begin{figure*}[t]
\centering
\begin{subfigure}{.5\textwidth}
    \centering
    \includegraphics[width=0.85\linewidth]{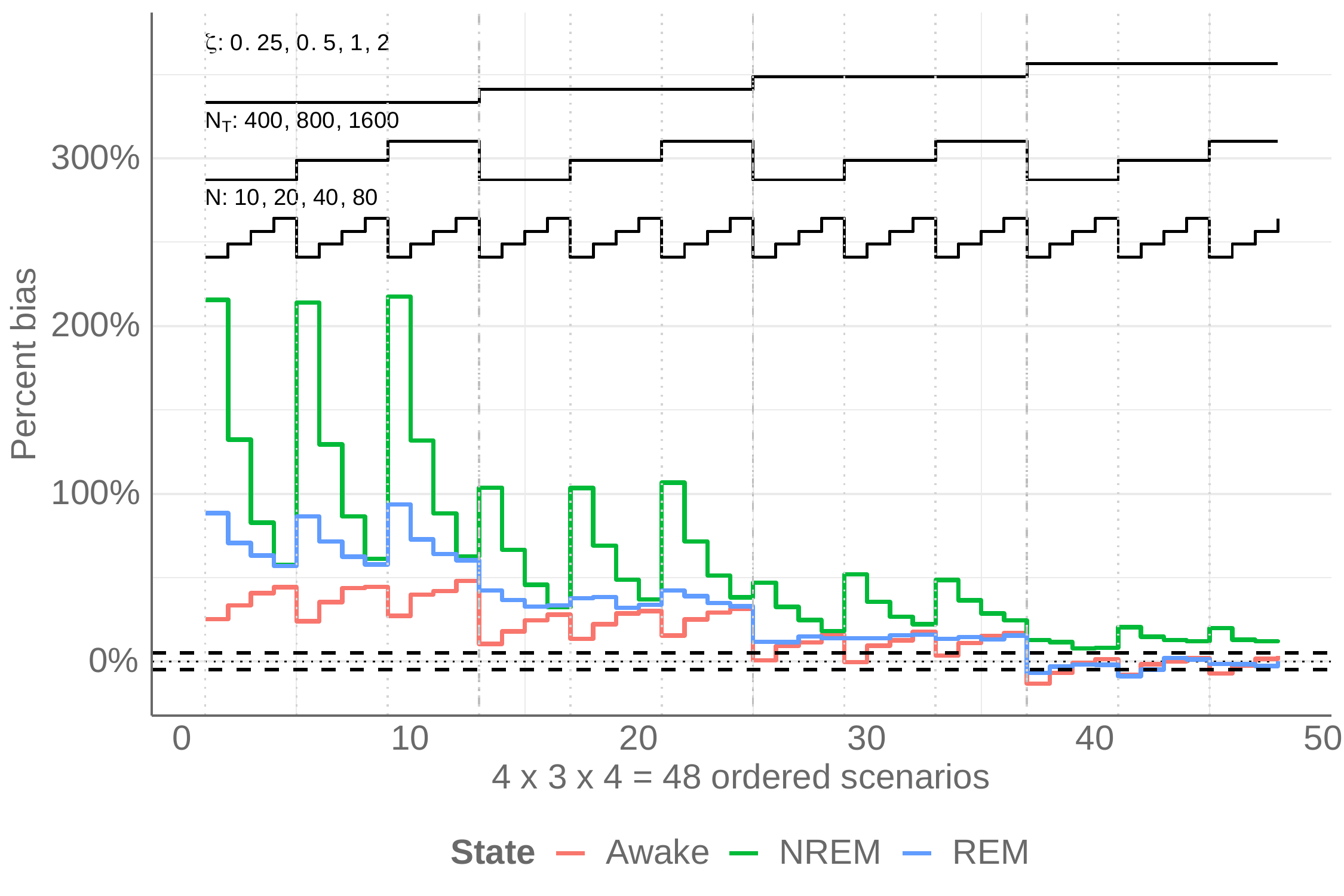}
    \caption{Nested loop plot (percent bias), EOG median theta}
    \label{fig:emission-varmu-pbias}
\end{subfigure}%
\begin{subfigure}{.5\textwidth}
    \centering
    \includegraphics[width=0.85\linewidth]{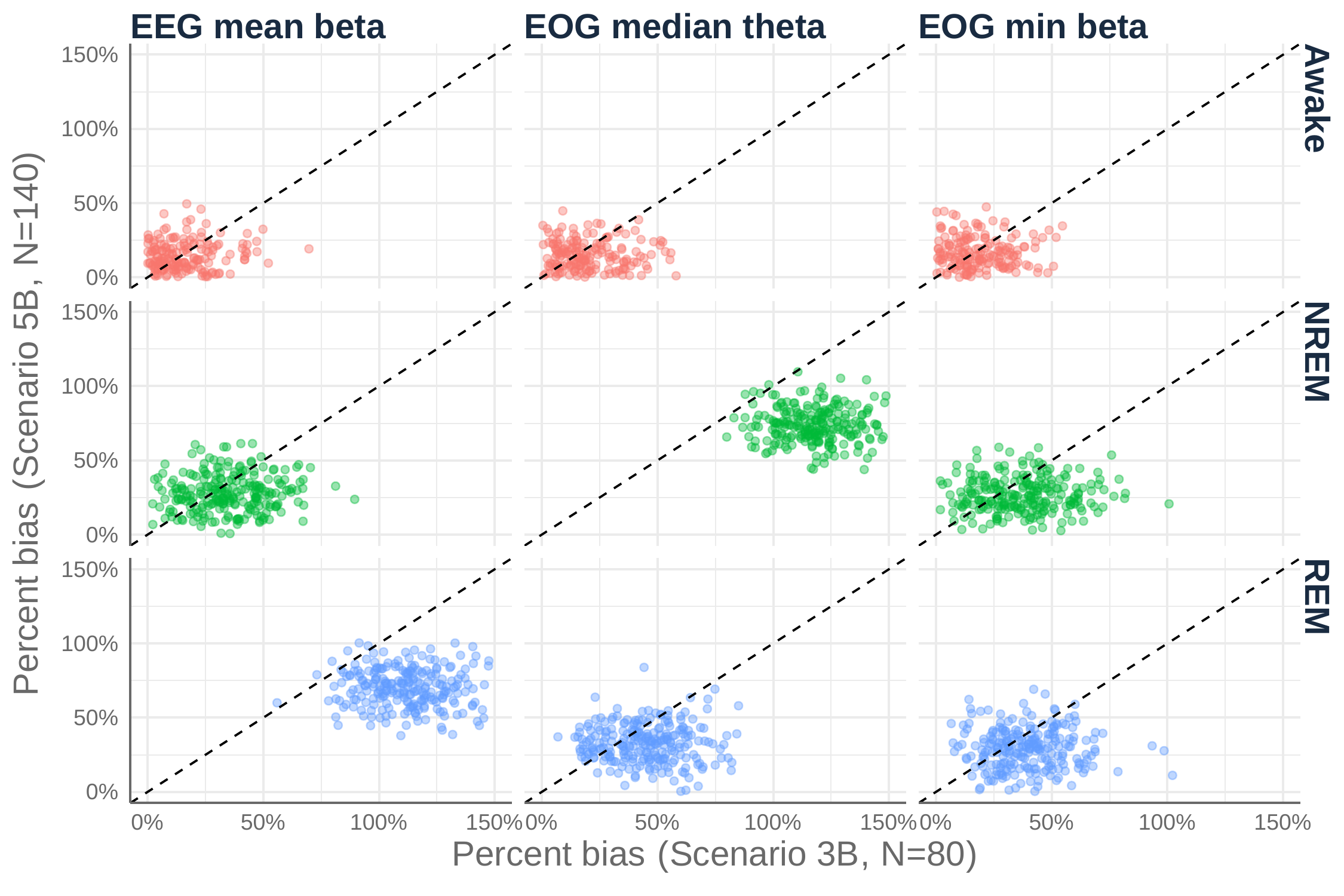}
    \caption{Scatter plot, percent bias of estimates in baseline scenarios $3B$ versus $5B$}
    \label{fig:emission-betvar-baseline-comparison}
\end{subfigure}
\caption{Panel (a) shows a nested loop plot of the percent bias of the component distribution random effect for the variable \textit{EOG median theta} across the simulation settings. The simulation settings are fixed across $Q=0.2$. The population component distribution random effects are displayed as the top-most condition ($\zeta$) in the graph. In panel (b), we show a scatter plot of the percent bias on the component distribution random effect for each simulation iteration in baseline scenario $3B$ (x-axis) versus the percent bias for each simulation iteration in baseline scenario $5B$ (y-axis). The dashed line indicates a one-to-one mapping between the scenarios. For example, if the data points are clustered to the right of the line, this indicates that the percent bias on scenario $3B$ is larger than in scenario $5B$. The population value of the random effect in both scenarios is $\zeta=0.5$.}
\label{fig:nlp-emiss-bias-random}
\end{figure*}

The results of the parameter bias on the component distribution random effects are similar across the baseline and sleep data simulations. Between-subject variance is typically biased upward in scenarios where the between-subject variance is small, and for these cases bias improves with a larger subject size (see figure \ref{fig:emission-varmu-pbias}). Conversely, the bias in the component that is estimated best tends to increase as the number of subjects goes up. 

In figure \ref{fig:emission-betvar-baseline-comparison} we plot the percent bias on the random effects obtained from the simulation iterations in baseline scenario $3B$ ($\zeta = 0.50, N = 80, N_T = 800$) to those obtained in baseline scenario $5B$ ($\zeta = 0.50, N = 140, N_T = 800$). From this plot, it appears that parameter bias seems to decrease mainly in the most extreme cases. The baseline results further indicate that, as the number of subjects grows large, the parameter estimates of the random effects become less biased.  However, there appears to be an interaction with respect to the size of the random effect. From tables \ref{tab:baseline-results-z025} and \ref{tab:baseline-results-z05} in appendix \ref{sec:appendix2}, we observe that the bias on the random effects is much higher at $\zeta=0.25$ than at $\zeta=0.5$. Hence, small values of the random effect display more extreme bias both in the simulation results as well as the baseline results.

Across the outcome variables, the parameter estimates of the random effects become more precise as $N$ grows large. However, this effect only occurs when the random effect is large; at lower values of the random effect, the gains in efficiency are much lower. 

In general, coverage of the parameter estimates declines as $N$ grows large. This effect is more pronounced for settings in which the random effect is small ($\zeta < 1$). The source of under-coverage, however, differs across the values of the random effect. When the random effect is small, there are two sources of poor coverage. On the one hand, the biased estimates lead to under-coverage. On the other hand, the model SE always exceeds the empirical SE, which leads to 95\% CCI that are too wide. When the random effect is large, poor coverage occurs because the empirical SE is generally larger than the model SE, the implication of which is that the 95\% CCI are too narrow and hence do not include the population parameter. These results persist across the baseline scenarios.

\subsection{Transition probabilities}

\begin{figure*}[t]
\centering
\begin{subfigure}{.5\textwidth}
    \centering
    \includegraphics[width=.9\linewidth]{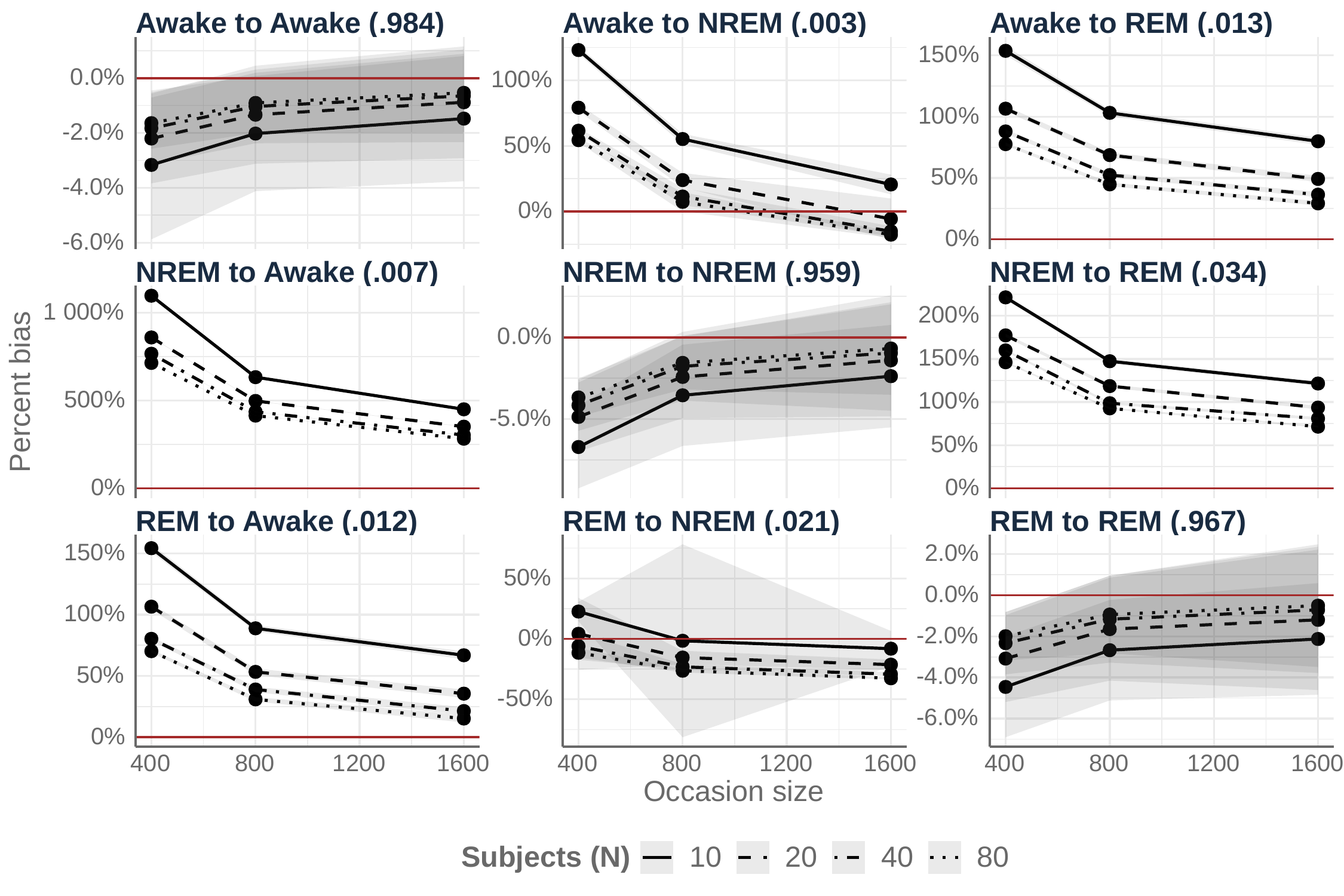}
    \caption{Trellis plot (percent bias), group-level TPM}
    \label{fig:sim-results-TPM-bias}
\end{subfigure}%
\begin{subfigure}{.5\textwidth}
    \centering
    \includegraphics[width=.9\linewidth]{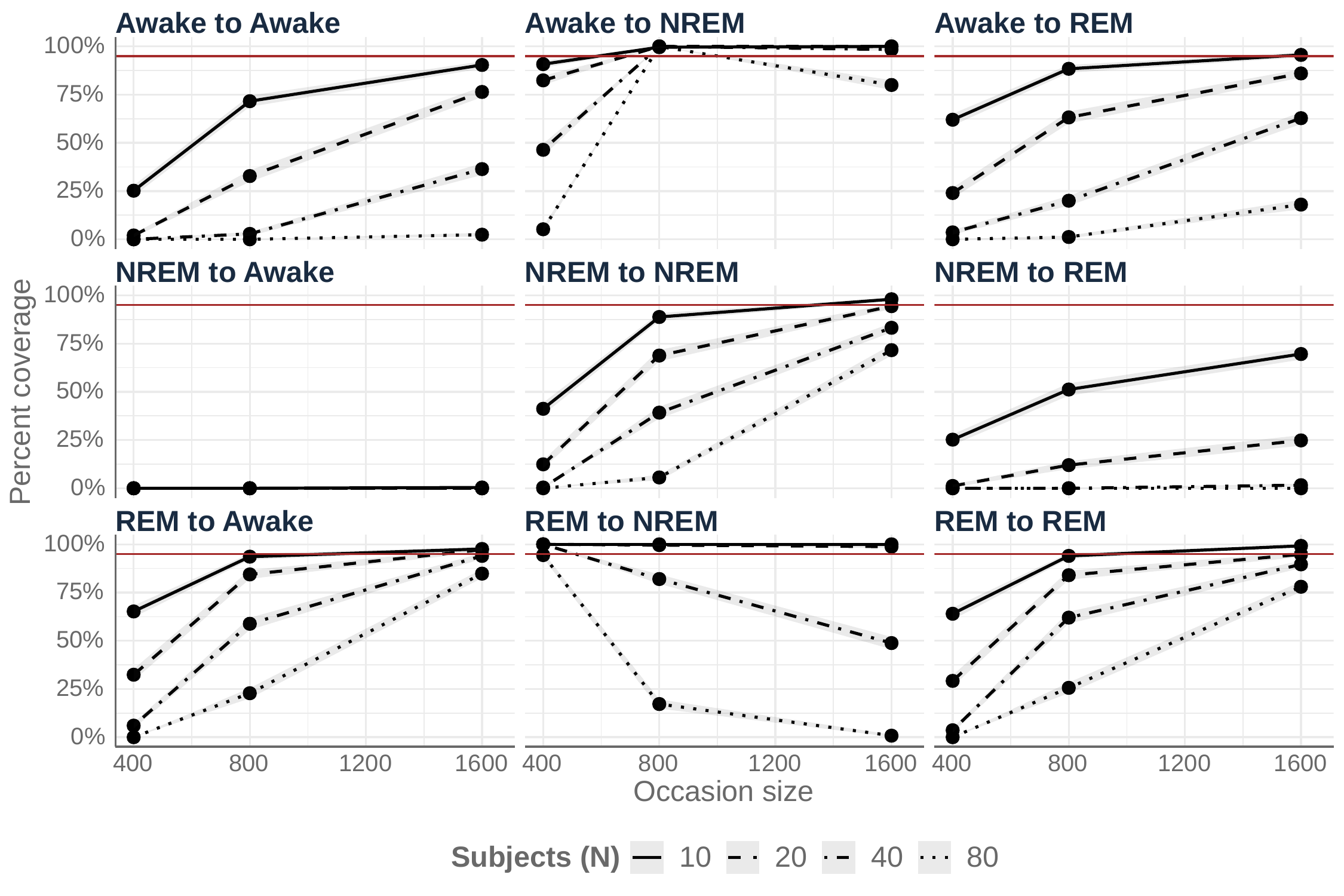}
    \caption{Trellis plot (coverage), group-level TPM}
    \label{fig:sim-results-TPM-cov}
\end{subfigure}
\caption{Panel (a) displays a trellis plot of the bias of the group-level TPM. We fix the TPM random effect to $Q=0.2$ and the component distribution random effect $\zeta = 0.25$. Results are similar for other values of $Q$ and $\zeta$. Monte Carlo SEs are denoted by the shaded areas. The population values for the state transitions are provided between brackets in the transition titles. Panel (b) displays the coverage of the group-level TPM. Results are fixed across the same values of $Q$ and $\zeta$ as in figure \ref{fig:sim-results-TPM-bias}. Monte Carlo SEs are denoted by the shaded area.}
\label{fig:nlp-TPM-bias}
\end{figure*}

In general, the simulation results indicate three effects with respect to the impact of the occasion sample size, the subject sample size, and the TPM random effect on parameter bias . Firstly, the bias decreases substantially as the occasion size grows large (see Figure \ref{fig:sim-results-TPM-bias}). Secondly, increasing the number of subjects also helps to decrease bias, although its effect appears to have less impact beyond a subject sample size of $N=40$ than increasing the number of occasions. Thirdly, these results hold across the values of the TPM random effect, although higher values on the component distribution random effects and TPM tend to result in higher MC SE values.

The baseline results indicate that, when the self-transition probabilities remain high (scenarios $1A$ and $1B$) but the component distributions are separated well across the outcome variables, the parameter estimates for both the self-transitions and off-diagonal values of the TPM will generally be acceptable. Scenarios $4A$ and $4B$ (in which $N_t = 3.200$ and $N=80$) indicate that the bias shrinks further as the number of occasions grows large and that the bias of the transition probabilities is lower than in scenarios $5A$ and $5B$ (in which $N_t = 800$ and $N = 140$). This provides further evidence that that collecting more occasions rather than more subjects leads to more accurate parameter estimates on the transition probabilities. Coverage for scenarios with lowered self-transitions (i.e., scenarios $2A$/$2B$ to $5A$/$5B$) is generally $> 0.90$. However, for scenarios $1A$/$1B$ in which the TPM equals that of the sleep data simulations, coverage of both the off-diagonal and diagonal entries are insufficient. 

For the sleep data simulations (in which state dependent component distributions display a high amount of overlap), the results are as follows. The self-transition probabilities (diagonal entries) exhibit low bias and are accurate even when the number of subjects and occasions are very low. At the higher end of the simulation settings, the bias is negligible. The results with respect to the off-diagonal transition probabilities, however, show an extreme upward bias. Bias decreases for these parameters as the number of subjects and occasions grow large, but do not reach acceptable levels even in the upper range of the simulation settings. 
Moreover, the Monte Carlo SE (MC SE) for the off-diagonal entries is much larger than that of the diagonal entries. In particular, this is the case for the transition from REM to NREM sleep. In section \ref{sec:check-convergence}, we saw that this state transition often fails to converge. In general, high MC SE values on the off-diagonal entries are not unexpected given that the population values of the transition probabilities are small. In turn, this means that they occur much less frequently in the simulated data and hence it is harder to obtain a good estimate for these values, especially when the occasion size is low and the component distributions overlap significantly.

The empirical and model SEs indicate that the precision of the TPM parameter estimates generally improves as the number of occasions increases. However, the improvement in precision across occasion size becomes less pronounced as the number of subjects increases.

In general, the coverage of both the off-diagonal and diagonal entries are insufficient. Firstly, the coverage of the diagonal entries tends to increase as the occasion sample size grows large. Although this effect is visible across each of the subject sample sizes, a small subject sample size typically leads to over-coverage whereas a large number of subjects (i.e. $N=80$) leads to under-coverage. The former effect occurs because the model 95\% CCIs (measured by the model SE) is generally larger than the empirical SE. In other words, the CCIs are estimated too wide and hence include the population parameter too often. The latter effect is caused by the constant bias in the estimates, as the bias-corrected coverage shows that coverage improves when bias is removed as a source of poor coverage.


\section{Empirical application} \label{sec:emp-application}

In this section, we apply the mHMM on an empirical dataset to detect latent sleep states. In the original study, $78$ subjects were asked to wear measurement devices that collected data for roughly $20$ hours on two separate days \citep{kemp2000analysis, goldberger2000physiobank}. The observed data for each subject are derived from electroencephalogram (EEG) and electrooculography (EOG) measurements. An EEG records brain wave patterns through the use of multiple electrodes that are placed on a subject's scalp, and an EOG uses electrodes placed above and below the eyes of a subject to record ocular movements \citep{malhotra2013sleep, aboalayon2016sleep}. The observed EEG and EOG time-series data are then split into epochs of 30 seconds, after which these epochs are labeled with a sleep state by a human scorer. Such data are often used to detect sleep states in individuals, and a wide variety of methods can be used to find useful variables that discriminate between the latent sleep states \citep{flexerand2002automatic, aboalayon2016sleep, li2017hyclasss, tzimourta2018eeg, kishi2018markov, leos2018introduction, humayun2019end}. However, few of these models utilize a multilevel approach. 

\subsection{Motivation}

The main interest in choosing this dataset is to investigate whether the sleep patterns of individuals differ from one another in their observed data and transition probabilities. The latter is especially interesting from the point of view of sleep research, for example because researchers may be interested how various drugs affect sleep patterns. Using a Markov model is a reasonable approach to analyze sleep data, because the outcome variables are uniquely determined by unobserved, latent sleep states that influence the activity in both the brain as well as ocular movements, and transitions between these states occur throughout a night of sleep. Previous studies that use HMMs and variants thereof are \cite{flexerand2002automatic}, \cite{langrock2013combining} and \cite{kishi2018markov}. Using an mHMM is an appropriate model for this type of data because it allows for the simultaneous analysis of multiple subjects while accommodating heterogeneity among them. 

\subsection{Data preprocessing}

We select a subset $41$ nights of sleep of $N=28$ subjects between the ages of $20$ and $50$ years old. The data are truncated such that each subject has observed data on $N_t=1.440$ occasions. Hence, we use a subset of the $20$ hours of collected data that pertains only to the period in which the subject exhibited sleeping behavior. There were no missing entries in the collected dataset. While the original data contained five sleep states that are classified using the Rechtschaffen \& Kales scheme (see e.g. \cite{moser2009sleep} for more details), we follow \cite{flexerand2002automatic} in reclassifying the sleep states into three categories: Awake, Non-REM (NREM) sleep and REM sleep by merging several states together. 

We preprocess the data by applying spectral density decomposition using the multitaper method to decompose the EOG and EEG signals into alpha, beta, gamma, theta and delta channels \citep{thomson1982spectrum, gramfort2014mne}. Such channels indicate different levels of brain activity, and are often used in machine learning applications that extract sleep states automatically (see e.g. \cite{aboalayon2016sleep}). The final preprocessing step consists of extracting summary statistics (e.g. min, median, mean, max, variance) across all channels and epochs. After preprocessing the data, we apply a logit transformation to the extracted channels and center the data. We then select several variables based on their expected ability to discriminate between sleep states. That is, the empirical application uses a different set of variables compared to the simulation study above due to issues with model convergence. Please see appendix \ref{sec:appendix1} for details.

\subsection{Fitting the model}

We use the model described in section \ref{sec:model} to fit the mHMM to the data. When using Bayesian estimation, it is important to check the convergence of the posterior distributions by using multiple chains \citep{lynch2007introduction, gelman2013bayesian}. Hence, we fit the model twice using different starting values. We run $20.000$ iterations for each model, discard the first $10.000$ iterations of each chain as burn-in samples, and thin the posterior chains by selecting every fifth sample.

As mentioned in section \ref{sec:mHMM-in-lit}, choosing good starting values can mitigate issues with label switching. In some preliminary analyses of the data, we observe label switching when the starting values are very different from the sample values. When using the sample values for the emission means and variances, label switching does not appear to be a problem. Note that it is not possible to determine starting values in this way for all applications. In the case of the sleep dataset, it is possible because it contains annotated sleep states across the occasions. If such annotations are unavailable, one can follow the procedure outlined in \cite{shirley2010hidden}.

Given that the values of the outcome variables are hard to interpret (they are logit-transformed EOG and EEG channels), there is little we can say about our prior knowledge of these values. Hence, we only use uninformative hyper-priors for the purposes of this analysis except for the hyper-priors of the component distribution group-level means, for which we use the sample means. An overview of the hyper-prior specifications is given in appendix \ref{sec:appendixI-hyperpriors}.

\subsubsection{Convergence checks}

We inspect the convergence of the posterior parameter distributions by examining the Gelman-Rubin (GR) statistic and by manual inspection of the trace plots, posterior density plots and auto-correlation plots (see appendix \ref{sec:appendix3}). The GR statistic shows satisfactory grounds for convergence for all parameters with $\hat{R} \leq 1.02$. The density plots look uni-modal and the trace plots do not show a trend. Most parameters display some amount of auto-correlation, which indicates slow mixing of the posterior distributions. We considered drawing a higher number of posterior samples in combination with using a larger thinning value. However, hardware limitations did not allow us to do so. Given that the chains have likely converged to the same posterior distribution, we merge the posterior distributions of both chains for further analysis.

\subsection{Results} \label{sec:results}

The MAP parameter estimates obtained by the model are shown in table \ref{tab:map-estimates}. For the purposes of this analysis, we use medians of the posterior distributions as MAP parameters. Furthermore, we use the standard deviations and 95\% Central Credible Intervals (CCI) as uncertainty estimates.

\begin{table}[!b]
\centering
\begin{adjustbox}{width=0.38\textwidth}
\begin{tabular}{ | m{11em} | m{5em} | R{5em} | R{6em} | }
\hline
\textbf{Variable}                          & \textbf{Parameter} & \textbf{MAP (SD)} & \textbf{95\% CCI}  \\ \hline
\multirow{6}{*}{EEG Fpz Cz mean theta} & $\beta_{001,\text{Awake}}$          & -0.93 (0.10)       & {[}-1.13, -0.73{]} \\
                                           & $\beta_{001,\text{NREM}}$           & 0.72 (0.09)       & {[}0.54, 0.90{]}    \\
                                           & $\beta_{001,\text{REM}}$            & 0.47 (0.12)       & {[}0.22, 0.70{]}    \\
                                           & $\sigma^2_{u_01,\text{Awake}}$         & 0.24 (0.10)        & {[}0.12, 0.50{]}    \\
                                           & $\sigma^2_{u_01,\text{NREM}}$          & 0.23 (0.07)       & {[}0.14, 0.40{]}    \\
                                           & $\sigma^2_{u_01,\text{REM}}$           & 0.38 (0.13)       & {[}0.22, 0.73{]}   \\ \hline
\multirow{6}{*}{EOG median theta}        & $\beta_{002,\text{Awake}}$          & 0.98 (0.47)       & {[}0.80, 1.10{]}     \\
                                           & $\beta_{002,\text{NREM}}$          & -0.09 (0.68)      & {[}-1.11, -0.60{]}  \\
                                           & $\beta_{002,\text{REM}}$            & -1.11 (0.33)      & {[}-0.79, -0.12{]} \\
                                           & $\sigma^2_{u_02,\text{Awake}}$         & 0.16 (0.06)       & {[}0.08, 0.30{]}    \\
                                           & $\sigma^2_{u_02,\text{NREM}}$          & 0.48 (0.15)       & {[}0.29, 0.83{]}   \\
                                           & $\sigma^2_{u_02,\text{REM}}$           & 0.81 (0.23)       & {[}0.50, 1.40{]}     \\ \hline
\multirow{6}{*}{EOG min beta}            & $\beta_{003,\text{Awake}}$          & 0.80 (0.06)        & {[}0.67, 0.92{]}   \\
                                           & $\beta_{003,\text{NREM}}$          & -0.80 (0.16)       & {[}-1.10, -0.48{]}  \\
                                           & $\beta_{003,\text{REM}}$            & -0.38 (0.17)      & {[}-0.72, -0.05{]} \\
                                           & $\sigma^2_{u_03,\text{Awake}}$         & 0.09 (0.04)       & {[}0.05, 0.18{]}  \\
                                           & $\sigma^2_{u_03,\text{NREM}}$          & 0.59 (0.18)       & {[}0.34, 1.10{]}    \\
                                           & $\sigma^2_{u_03,\text{REM}}$           & 0.77 (0.23)       & {[}0.46, 1.35{]}   \\ \hline
\multirow{6}{*}{TPM intercepts}            & $\alpha_{12}$           & -5.02 (0.32)      & {[}-5.67, -4.44{]} \\
                                           & $\alpha_{13}$           & -4.70 (0.26)       & {[}-5.22, -4.21{]} \\
                                           & $\alpha_{22}$           & 4.57 (0.26)       & {[}4.1, 5.12{]}    \\
                                           & $\alpha_{23}$           & 0.87 (0.30)        & {[}0.31, 1.50{]}    \\
                                           & $\alpha_{32}$           & 0.62 (0.24)       & {[}0.17, 1.11{]}   \\
                                           & $\alpha_{33}$           & 4.23 (0.22)       & {[}3.82, 4.69{]}   \\ \hline
\end{tabular}
\end{adjustbox}
\caption{Point estimates (medians) for the group-level parameters of the mHMM. Each dependent variable has three component distribution group-level means and random effects. The transition probabilities are represented by their TPM overall intercepts (see section \ref{sec:model-tpm}).}
\label{tab:map-estimates}
\end{table}

By the 95\% CCI of the outcome variables, we can see that the uncertainty is largest around the fixed and random effect estimates of the REM state. This is not unexpected given that it is the hardest state to estimate because it is always "wedged" in between two other state distributions (see also figure \ref{fig:emiss-densities-densplot-empapp} in appendix \ref{sec:appendix1}). Hence, it is associated with the largest measure of uncertainty.

Table \ref{tab:TPM-EMP} displays the MAP values of the group-level transition probabilities. 

\begin{table}[!t]
\centering
\begin{adjustbox}{width=0.45\textwidth}
\begin{tabular}{ C{1em} C{1em}  L{4em}  R{4em}  R{4em}  R{4em}  }
&&& \multicolumn{3}{c }{To state} \\
&&& \multicolumn{3}{c }{\small (occassion $t$ + 1)} \\
\parbox[t]{2mm}{\multirow{4}{*}{\rotatebox[origin=c]{90}{From state}}} &
\parbox[t]{2mm}{\multirow{4}{*}{\rotatebox[origin=c]{90}{(occassion $t$)}}}                             
& & \textbf{Awake} & \textbf{NREM} & \textbf{REM}                                     \\ \cline{4-6}

&&  \textbf{Awake}     & 0.984          & 0.007        & 0.009 \\ \cline{4-6}
& & \textbf{NREM}     & 0.010          & 0.966        & 0.024                                           \\ \cline{4-6}
& & \textbf{REM}        & 0.014          & 0.026        & 0.960                                            \\ \cline{4-6} 

                 \end{tabular}
\end{adjustbox}
\caption{Point estimates for the average transition probabilities. The rows indicate the state at time $t$, the columns indicate the state at time $t+1$.}
\label{tab:TPM-EMP}
\end{table}

\noindent As expected, the self-transition probabilities on the diagonal entries of the TPM are very high ($> 0.95$). Of the three states, the state ‘Awake’ is the most persistent. On average, if an individual is awake at occasion $t$, the probability that they will be awake at occasion $t+1$ is roughly $0.98$. The second most persistent state is NREM sleep. The least persistent state is REM sleep. This is consistent with the literature on sleep state analysis as REM sleep occurs in shorter sequences than the other states. 

By inspection of the subject-specific transition probabilities, we observe that the subjects are very similar in their sleep state transitions. The similarity of the subjects with respect to the transition probabilities is not unexpected given that the data was conducted on healthy individuals. If, as was shown by \cite{langrock2013combining}, the analysis is repeated on sleep data for more heterogeneous individuals - for example due to sleep disorders - the effects on the transition probabilities may be more pronounced.

\begin{figure}[!b]
\centering
\includegraphics[width=0.9\linewidth]{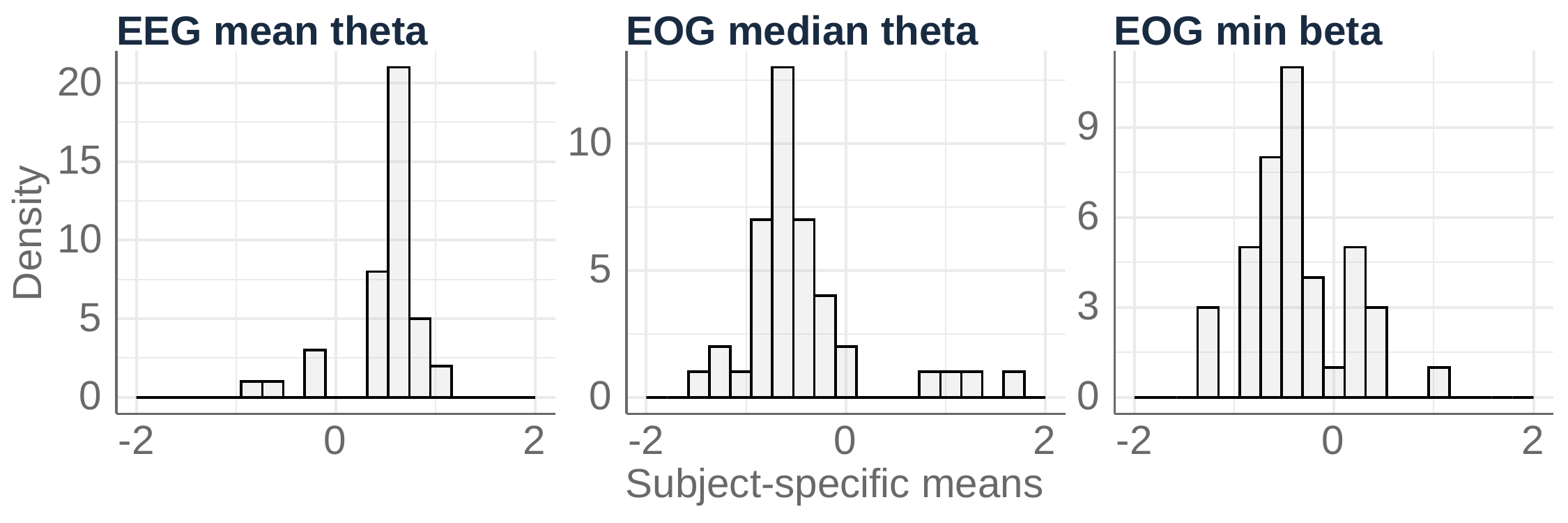}
\caption{Distribution of subject-specific mean values on the REM state of each of the outcome variables.}
\label{fig:bet-subj-means}
\end{figure}

The subjects display more heterogeneity in their subject-specific means of the outcome variables. In particular, the component distribution random effects are large on the REM state (see also figure \ref{fig:bet-subj-means}), which suggests that it may be useful to incorporate covariates to explain differences between subjects.

\subsection{Goodness of fit}

It is common to assess the model fit of Bayesian models through the use of posterior predictive checks (PPC, \cite{lynch2007introduction}; \cite{gelman2013bayesian}). The goal of a PPC is to assess whether it is appropriate to apply a model to the data at hand. This is done by repeatedly sampling the posterior distribution parameters and generating new datasets using these parameters under the assumption that the model is true. Hence, if this assumption is met, the proportion $P_{\text{posterior}}$ of generated samples must not appear extreme on a given test statistic when compared to the observed, empirical dataset (i.e. we expect $P_{\text{posterior}}$ to be approximately equal to $0.5$). 

\cite{shirley2010hidden} use several PPC for their implementation of the Bayesian mHMMs, which were adopted (and slightly modified) by \cite{de2017use}. In turn, we adopt these PPC to assess the model fit of the mHMM on the sleep dataset. Note that the sleep dataset is convenient in that it contains annotated latent states, and hence we can use this information in the design of the PPC. However, not all datasets contain annotated latent states. In those cases, the reader may find suitable PPC in e.g. \cite{de2017use} and \cite{shirley2010hidden}.

To conduct the analysis, we draw $2.000$ samples from the posterior distributions for each parameter. These parameter estimates are then used to generate new samples using the R package \textit{mHMMbayes} \cite{aartsusing2019}. The data simulation procedure requires that the between-subject variance of the component distributions within an outcome variable are equal. Hence, we take the average of the component distribution MAP values for each outcome variable to simulate the dataset. The data simulation procedure also requires a value for the group-level transition probabilities (a parameter which is not returned by the model). Given the small variance between subjects on the transition probability matrix, we set this value to a situation corresponding to little variance between subjects (i.e. $Q = 0.1$). 

\subsubsection{Component distributions}

\begin{figure}[!b]
\centering
\includegraphics[width=1\linewidth]{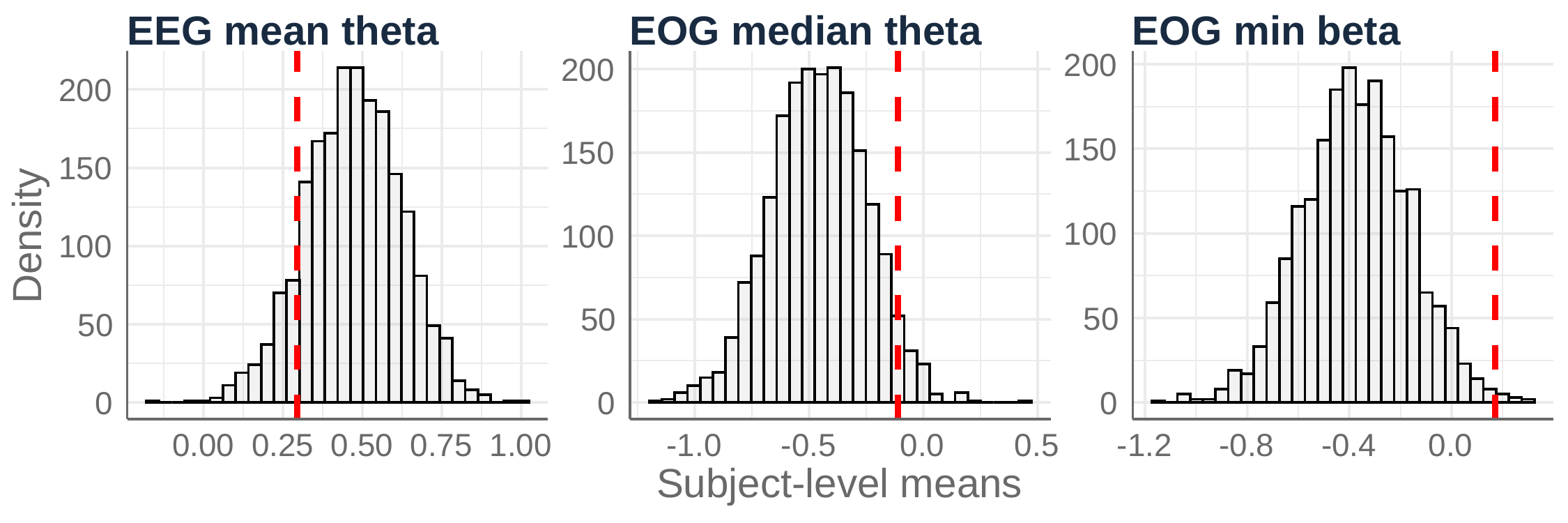}
\caption{Results of PPC 1 for the REM state of the three emission variables. The histograms plots represent the model predictions. The dashed red lines represent the component distribution means in the empirical data.}
\label{fig:PPC-means}
\end{figure}

We first check whether the model produces extreme results on the component distribution means and variances. In the case of the component distribution means, the p-values indicate that the simulated values are not extreme compared to the observed dataset for states "Awake" and "NREM" across all three outcome variables ($0.29 \leq P_{\text{posterior}} \leq 0.56$). In the case of the REM state, however, the p-values indicate that the results in the simulated data are extreme when compared to the observed dataset, as can be seen from figure \ref{fig:PPC-means}. 

The red vertical line in figure \ref{fig:PPC-means} indicates the value of the component distribution mean in the observed dataset. The distribution of the component distribution means in the generated samples are plotted as grey histograms. From the figure, we see that the generated samples produce higher-than-expected mean values on the first outcome variable, while producing lower-than-expected mean values on the second and third outcome variables. This result is not surprising given that the component distribution of the REM state overlaps significantly with the other states on all three outcome variables, and hence may be difficult to separate for the model. 

Due to the specific nature of the data simulation procedure, it is not meaningful to conduct a PPC for the component distribution random effects. Hence, we conduct the PPC for the total variance. The PPC for the total variance indicates that the variance in the simulated data is much larger than the variance in the observed data (all $P_{\text{posterior}} \leq 0.02$). This indicates that the model likely overestimates the diversity between subjects, and is a result that is consistent with the simulation study (see section \ref{sec:results-randomeffects}), in which it was shown that small component distribution random effects result in an upward bias of the random effect parameter estimate. 

\subsubsection{Transition probabilities}

\begin{figure}[!b]
\centering
\includegraphics[width=1\linewidth]{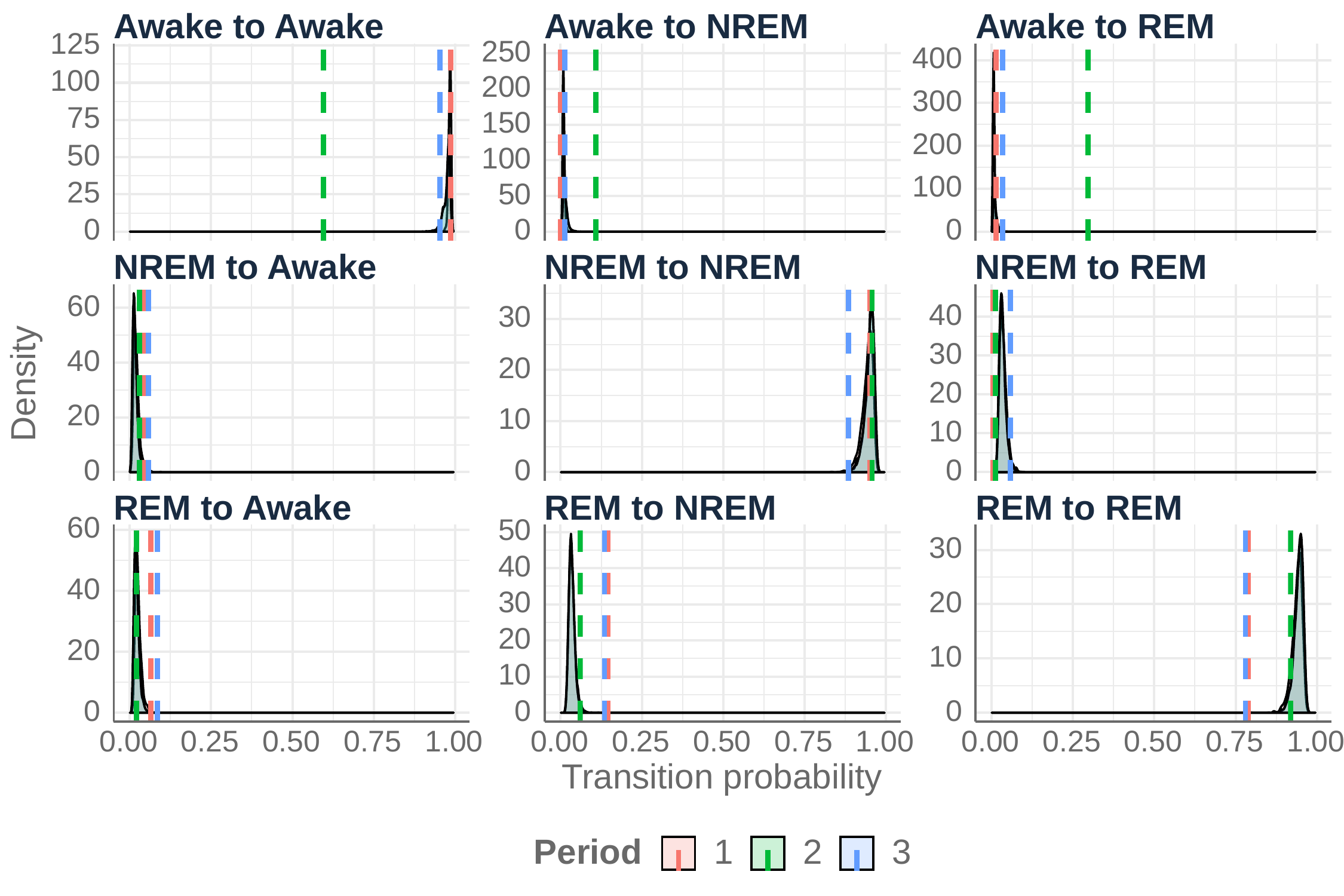}
\caption{Results of PPC 3 for the transition probabilities. The density plots represent the model predictions. The colored lines represent the transition probability in the empirical data for each period.}
\label{fig:PPC-TPM}
\end{figure}

An important assumption of the mHMM is that the transition probabilities are time-homogeneous (or "stable") across the occasions (see section \ref{sec:HMM}). To test this assumption, we split the dataset into three equal-size periods of $480$ occasions. The first period includes the time that subjects start their sleep cycles, in which we expect to see high self-transitions in the awake state. The second period includes the midpoint of their night's sleep, and the third period includes the end of their night's sleep.  In each period, we compute the transition probabilities from the simulated datasets and compare them to the transition probability computed from the observed dataset. The result is shown in figure \ref{fig:PPC-TPM}. 

Figure \ref{fig:PPC-TPM} reveals that the transition probabilities in the observed dataset differ across the periods for some of the state transitions. The dashed lines indicate the value of the transition probability computed from the observed dataset. This is consistent with the academic literature about sleep state transitions (see e.g. \cite{kishi2018markov}), which shows that the REM state in particular changes throughout a night's sleep.

\subsection{Conclusion}

The empirical application demonstrates that it is possible to analyze sleep-state transitions for multiple subjects using the Bayesian mHMM. We restricted this analysis to the modeling of mean and variance structures only, note that it is also possible to include covariates to explain differences between subjects on either the transition probabilities or subject-specific means of the state-dependent normal component distributions.

The results indicate that the mHMM is a useful model to capture the complex nature of the data. By using random effects in the component distributions of the outcome variables, we are able to explicitly account for heterogeneity among the $41$ subjects. Furthermore, the results indicate that the subjects primarily differ on their component distribution means, and in particular on the component distribution of REM sleep. However, they do not differ much in terms of the latent sleep state transitions. This may be a result of data preprocessing, as we only considered those subjects we deemed somewhat similar to ensure that the Gaussian mHMM could be applied to the data. Secondly, all subjects are healthy individuals. 

The PPC indicate that the results of the mHMM on the sleep dataset should be interpreted with some caution. Firstly, the overlap of the component distributions can be considered quite extreme, leading to under-, and over-estimation of the subject-specific means on the REM state. As was shown in section \ref{sec:results-comp-fixed-effects}, this problem should be less pronounced when the component distributions are more separated. Another issue is the over-estimation of the total variance. Most likely, this is the result of relatively small between-subject variance in the subject-specific component distribution means. Finally, the PPC show that the sleep dataset likely violates the assumption of time-homogeneity. To this end, it could be interesting to extend the model to include auto-regressive elements \citep{kishi2018markov, bazzi2017time}.


\section{Summary and discussion}
In this paper, we perform a simulation study in which we vary the number of subjects $N$, the number of occasions $N_t$ and the between-subject variance of subject-specific component distribution means and transition probabilities. Our aim is to investigate the effects of varying these quantities on model performance of the mHMM. We frame this simulation study in the context of modeling latent sleep states based on observed EEG and EOG outcome variables. This dataset is characterized by high self-transition probabilities and component distributions that overlap significantly across all variables. Hence, we also establish $10$ baseline scenarios in which the data is less extreme such that we can compare the results obtained in the simulation study to the results obtained from the baseline scenarios. This allows us to gauge the extent to which the results obtained in the simulation study are generalizable. Finally, we apply the mHMM to an empirical dataset to model latent sleep states of multiple subjects. In the subsections that follow, we discuss the required sample size needed to fit the mHMM, offer recommendations to applied researchers, and discuss limitations and future directions for research.

\subsection{Required sample size}

A previous (small) simulation study conducted by \cite{altman2007mixed} using a Frequentist mHMM with Poisson-distributed outcome variables indicates that $60$ subjects is generally sufficient in the context of her study. However, we note that the model that \citeauthor{altman2007mixed} uses is simpler than the model used here.
Our simulation study suggests that, taken on the whole, increasing the number of subjects has the largest effect on the quality of the parameter estimates. In this sense, our findings are similar to the results obtained by \cite{schultzberg2018number} for DSEM models. However, in the case of the transition probabilities, we find that larger occasion sample sizes are the most important determinant of parameter quality beyond a subject sample size of $N=40$. 
When the component distributions overlap significantly, as is the case in the simulation study, small sample sizes ($N \leq 20$) will lead to extreme bias on the component distribution fixed and random effects and off-diagonal entries in the group-level transition probabilities irrespective of the occasion sample size, and hence we suggest that such small sample sizes should not be used with the mHMM.

With respect to the component distribution group-level means, baseline scenarios $5A$ and $5B$, in which the number of subjects $N = 140$, are the only cases in which the parameter estimates are within the acceptable bounds of 5\% deviation from the population parameter. However, if a higher amount of bias on the group-level means is acceptable and the component distributions are separated relatively well, a sample size of $N=80$ should be sufficient to achieve a parameter bias $< 12$\% on all component distribution group-level means irrespective of occasion sample size. 
When the component distributions show more overlap, a sample size of $N=80$ or larger should yield decent estimates, although this depends on the severity of the overlap as well as the amount of between-subject variance. To bring the parameter bias in this situation to acceptable levels requires sample sizes greater than those examined in this paper.

Typically, random effects are the most difficult parameters to estimate in multilevel models, and our results corroborate those obtained by \cite{schultzberg2018number} in the case of DSEM models in that the random effects show a constant, upward bias. Beyond that, our results for the component distribution random effects indicate two findings. Firstly, the most important determinant of the parameter quality is the amount of between-subject variance.
Secondly, when the between-subject variance is small, the parameter bias of the random effect is extreme even when the component distributions are separated well and the number of subjects is very large ($N=140$, baseline scenarios $5A$ and $5B$). 
However, the simulation results indicate that, when the between-subject variance is very large ($\zeta=2$), this number of subjects may be sufficient. Nonetheless, the findings in this study suggest that much larger sample sizes than are used here are needed to give reliable advice on the required sample size.  

Three factors dominate the quality of the transition probabilities. Firstly, the extent to which the component distributions overlap heavily influences the parameter bias of the off-diagonal entries in the group-level TPM when these off-diagonal entries are small. This can be seen when comparing the simulation results to the results of baseline scenarios $1A$ and $1B$. The bias on the off-diagonal entries is acceptable (i.e. less than 5\%) when the component distributions are separated well. 
In either case, both larger subject and occasion sample sizes tend to improve the quality of the parameter estimates. However, increasing the number of subjects yields diminishing marginal returns beyond $N > 40$. When the self-transition probabilities are high, larger occasion sizes tend to yield less biased estimates, although (as mentioned previously) this interacts with the amount of overlap in the component distributions. Additionally, it is not always possible to increase the occasion sample size. For example, it is rare to find individuals that sleep more than $10$ hours per night. 
Scenarios $1A$ and $1B$ indicate that, when the component distributions show little overlap, an occasion sample size of $N_t = 800$ is sufficiently large to obtain good parameter estimates. When the overlap is more pronounced, larger occasion sample sizes are warranted in combination with larger subject sample sizes to obtain accurate estimates on the off-diagonal entries of the TPM. The findings of the simulation study did not vary much across the between-subject variances of the TPM.

\subsection{Recommendations for researchers}

The goal of this study is to help applied researchers who want to use the mHMM in their own research to choose appropriate sample sizes. In practice, the outcome variables that will be used by applied researchers will probably lie in between the extremes of the baseline scenarios (ideal situation) and the sleep dataset (high overlap in component distributions, high self-transition probabilities). That said, researchers may have different goals and hence different needs of the model. Below, we outline various recommendations that should suit those needs.

In general, researchers may expect over-coverage to occur on all parameter estimates because the simulation results show that the model SEs are typically over-estimated when compared to the empirical SEs, a finding which is consistent with the results of \cite{altman2007mixed}. This means that the 95\% CCIs are generally estimated too wide and hence cover the population parameter too often. In cases where the parameter estimates can reasonably be expected to be biased, coverage deteriorates as the subject sample size increases because of parameter bias. 

Moreover, researchers may encounter issues with label switching (see section \ref{sec:mHMM-in-lit}) and model convergence when the component distributions overlap significantly. When such issues arise, choosing good starting values tends to help (see also \cite{scott2002bayesian, shirley2010hidden}). We note in passing that we also observe that, when component distributions overlap significantly, using multiple outcome variables in which the component distributions overlap in different ways tends to alleviate issues with label switching. For example, in the context of the sleep dataset, we observe that the component distribution group-level means of the REM state are estimated more accurately across when we add an artificially generated outcome variable in which the REM component distribution is clearly separated from the other component distributions.

If component distribution group-level means are the primary research interest, researchers should typically aim for a minimum sample size of $N=80$ subjects. When the component distribution show a more pronounced overlap or the between-subject variance is large, researchers may expect this number of subjects to still provide biased estimates.
If the primary research interest lies in the component distribution random effects of the outcome variables, we note that, in cases where the random effects are expected to be small, the bias in the parameter estimates probably persists even when $N$ grows very large. Based on the outcomes of our study, we cannot recommend a setting that will yield unbiased estimates. However, we note that the component distribution random effects tend to exhibit much lower bias when the differences between subjects are large.  

When the transition probabilities are the primary research interest, we suggest the following. Firstly, if the component distributions show low to medium overlap, a minimum subject sample size of $N=40$ yields sufficient results. In this case, a minimum occasion size of $N_t = 800$ should yield acceptable parameters estimates on the transition probabilities (scenarios $2A$ and $2B$). When the self-transition probabilities are high, these settings should still yield good estimates, and the coverage of the transition probabilities will improve as the occasion sample size increases. However, they are generally not trustworthy because the 95\% CCI will hone in on the biased estimates. 
Secondly, when the overlap in component distributions is more pronounced and the self-transition probabilities are high, researchers can expect the off-diagonal entries in the TPM to show an extreme upward bias. In such cases, collecting data on as many subjects and occasions will help in obtaining less biased estimates, although this should probably be balanced against the financial and operational costs involved with obtaining more data.

Finally, we suggest conducting posterior predictive checks to evaluate model fit. If annotated latent states are available, then these can be incorporated in the PPC. If these are not available, we refer the user to \cite{de2017use} for PPC. 

\subsection{Limitations and future research}

While the situation in which self-transition probabilities are high and the component distributions overlap quite extensively is not uncommon in applications in behavioral research and other areas, the dataset on which the simulations in this paper are based can be considered extreme on both counts. In that respect, the dataset used in this study forms a limitation. For example, it is very likely the case that, if researchers use outcome variables on which the component distributions show considerably less overlap, the extreme bias that is observed in some of model parameters (e.g. the REM states) is considerably lower. However, the baseline results indicate that these are unlikely to disappear entirely in the current implementation of the model unless the number of subjects grows very large. Between the extremes of the the baseline scenarios and the sleep dataset lies a lot of nuance that this research does not capture. Furthermore, we also note that the results obtained in this study may not necessarily hold for outcome data that is not normally distributed, but follows e.g. a Poisson or categorical distribution. 

Secondly, we mainly use only uninformative hyper-priors in this study. One of the most appealing features of Bayesian modeling is the ability to incorporate and weigh prior information in addition to the information that exists in the data. In cases where researchers can specify informative hyper-prior values, our recommendations may be too conservative. 


Finally, this research is limited by the practical choices we made in the design of the simulation study. That is, we considered simulation scenarios and settings such as to balance the computational burden of the model with the large number of iterations that needed to be executed. This constraint prohibited us from choosing larger subject and occasion sample sizes. However, the simulation results indicate that investigating larger sample sizes is warranted. We also note that, in our simulation, the size of the TPM random effect does not appear to influence the estimates of the group-level transition probabilities. This indicates that the values that we use in this study are perhaps too conservative. Furthermore, we do not incorporate covariates for either the outcome variables or the TPM in our study design. Doing so would make the model much more complicated, and would no doubt have an effect on the recommendations we give in this paper.


The mHMM is a relatively "young" model, and provides ample opportunities for future research. First and foremost, this study indicates that a sample size of $80$ subjects is insufficient if one wants to adequately estimate the component distribution random effects. It would be informative to repeat this study with larger subject sample sizes such as to investigate at what sample size the component distribution random effects achieve sufficient quality. In the same vein, future research should compare simulation results across different MCMC samplers and hyper-prior specifications, as there exists some evidence that suggests this may affect the parameter estimates in the mHMM \citep{rueda2013bayesian}. As the results in this study show, this is particularly relevant for the component distribution random effects as they exhibit extreme upward bias if the between-subject variance on the outcome variables is small. 
Moreover, we currently know very little about the effect of the hyper-prior specification in the mHMM. This is relevant because \cite{mcneish2019two} shows that, in the case of DSEM models, using uninformative hyper-priors may in fact \textit{hurt} model results when sample sizes are small (i.e. $N \leq 100$ \citep{mcneish2019two}). More research is needed to investigate whether this observation holds in the case of the mHMM.

As mentioned previously, this study does not investigate the use of covariates in the TPM or the component distributions. However, this is one of the features that makes the mHMM so powerful and should be investigated e.g. with respect to statistical power. Other avenues of research include investigating two characteristics that are common to research data on which the mHMM is often applied. Firstly, an effort should be made to investigate the extent to which the overlap of the component distributions affects the quality of the parameter estimates. Secondly, future research should vary the values of the self-transition probabilities to increase our understanding of how such settings influence the model parameters. \vspace{4.5mm}

\noindent \large{\textbf{Article information}} \vspace{4.5mm}

\small
\noindent \textbf{Declarations of interest:}  none. \vspace{4.5mm}

\noindent \textbf{Ethical clearance: } The authors affirm to have followed professional ethical guidelines in preparing this work. The procedures in this research project were reviewed and granted approval by the Ethics Review Board of the Faculty of Social and Behavioural Sciences at Utrecht University. \vspace{4.5mm}

\noindent \textbf{Data files, software and research archive} In the course of this research project, a variety of software was developed to conduct the collection of data, the execution of the simulation study and the analysis of results. The main repository that contains documentation that can be used to replicate the results can be found here: https://github.com/JasperHG90/sleepsimR-documentation. \vspace{4.5mm}

\noindent \textbf{Acknowledgments: } The authors would like to thank Vera Oosterveen, Goran Ilic, Boaz Manger, and Rens van de Schoot for their feedback on prior versions of this manuscript. All remaining errors are ours.

\bibliography{bib.bib}

\appendix
\newpage
\section{Further details about the simulation study} \label{sec:appendix1}
\subsection{Emission distribution density plot and summary statistics}

\begin{figure}[H]
\centering
\includegraphics[scale=0.5]{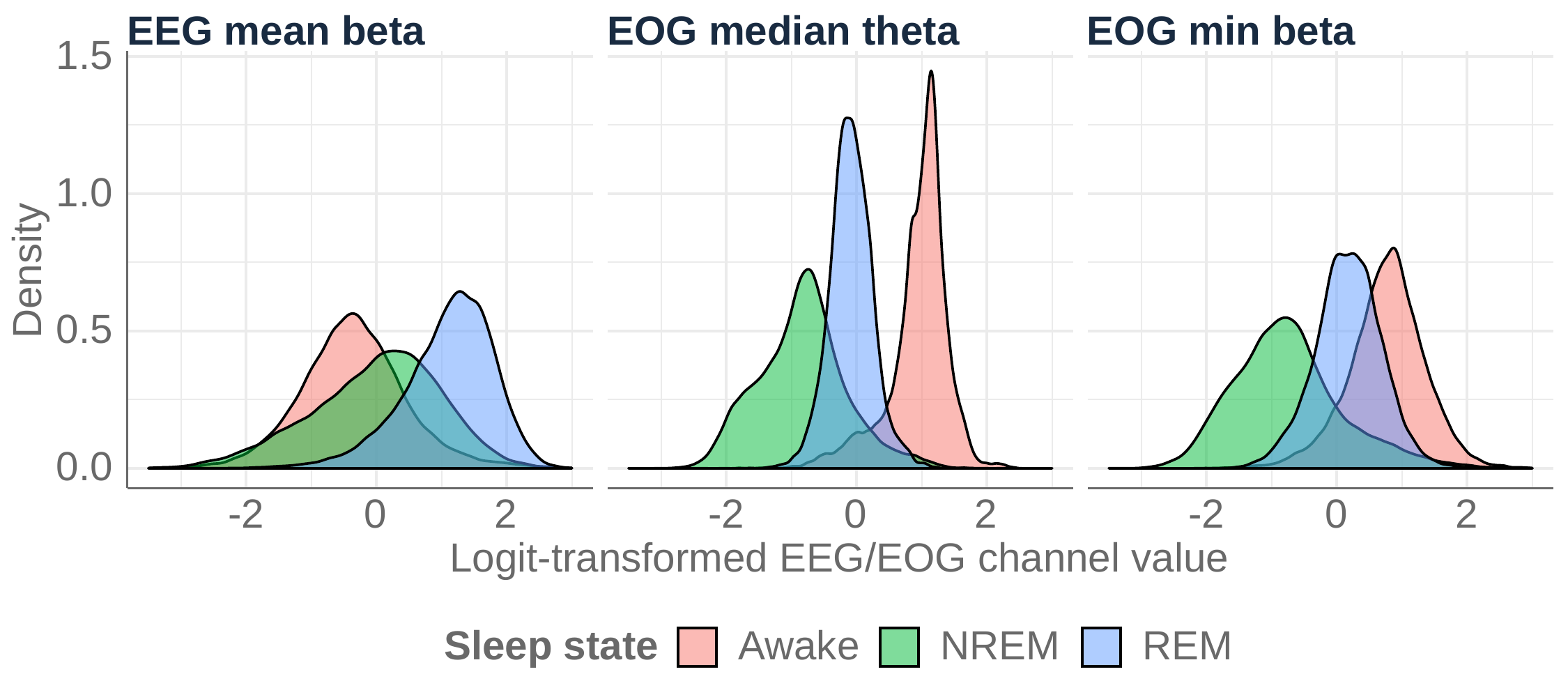}
\caption{Density plot of the emission distributions used to obtain the population parameters described in table \ref{tab:popval-means}.}
\label{fig:emiss-densities-densplot}
\end{figure}

The population parameters used in the simulation study are obtained by fitting the mHMM on the variables that are listed in table \ref{tab:popval-means}. A density plot of these emission variables is shown in figure \ref{fig:emiss-densities-densplot}. We executed two chains, each using different starting values, with $2.000$ iterations each and a burn-in size of $1.000$ samples. At the time, the results obtained from these two models seemed sufficient. However, when we re-ran the models using $20.000$ iterations on specialized hardware, the results indicated that the model suffered from label switching. (see section \ref{sec:mHMM-in-lit}). Hence, the empirical application uses a different set of variables. These variables are described in table \ref{tab:summary-stats-application}.

 \begin{table}[b]
 \centering
\begin{adjustbox}{width=0.35\textwidth}
 \begin{tabular}{ | m{10em} | m{4em} | R{5em} | }
\hline
 \textbf{Parameter}                         & \textbf{State} & \textbf{Mean (SD)} \\ \hline
 \multirow{3}{*}{EEG mean theta} & Awake          & -0.87 (0.93)       \\
                                           & NREM           & 0.70 (0.52)         \\
                                           & REM            & 0.30 (0.42)         \\ \hline
 \multirow{3}{*}{EOG median theta}        & Awake          & 0.98 (0.47)        \\
                                           & NREM           & -0.09 (0.68)       \\
                                           & REM            & -1.11 (0.33)       \\ \hline
\multirow{3}{*}{EOG min beta}            & Awake          & 0.78 (0.59)        \\ 
                                           & NREM           & -0.78 (0.84)       \\
                                           & REM            & 0.17 (0.52)        \\ \cline{1-3} 
 \end{tabular}
 \end{adjustbox}
 \caption{Summary statistics for the three emission variables used in the empirical application.}
 \label{tab:summary-stats-application}
 \end{table}

\begin{figure}[H]
\centering
\includegraphics[scale=0.5]{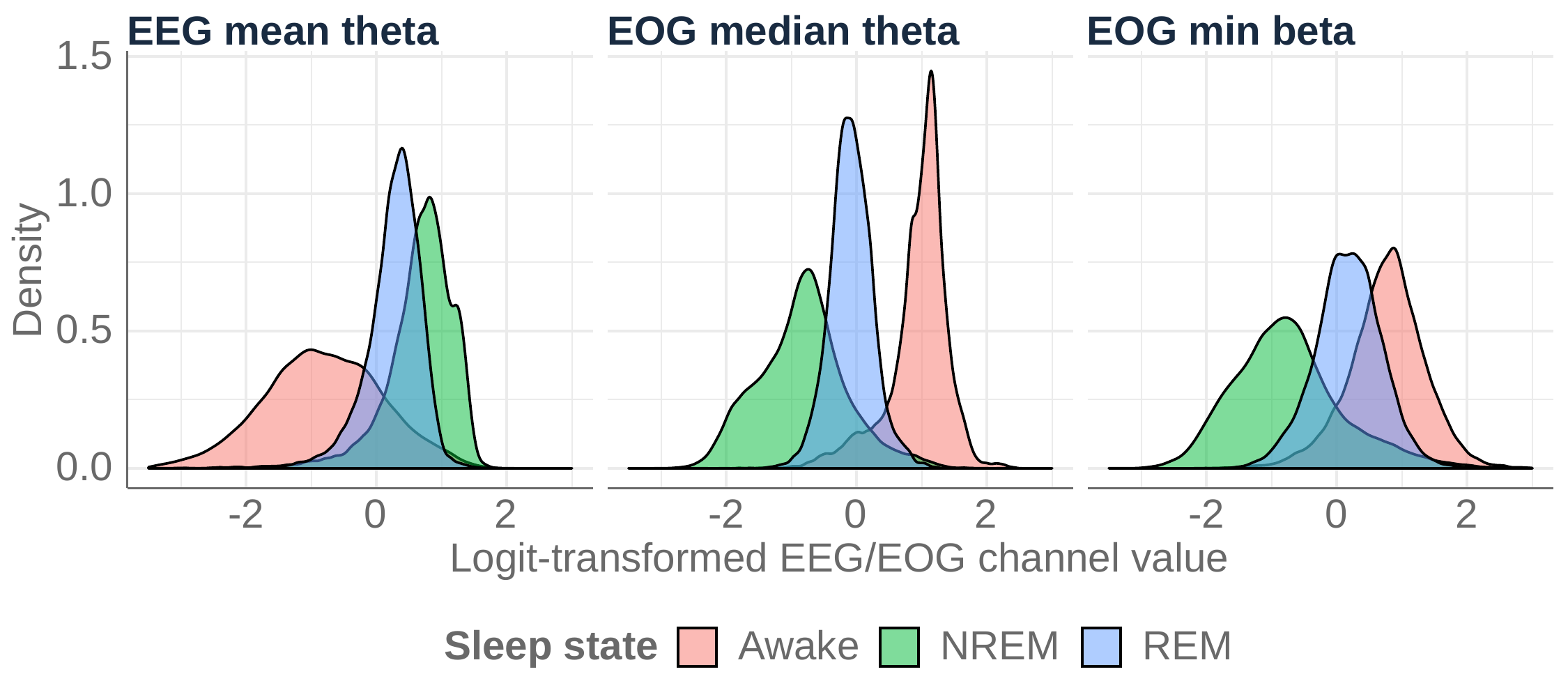}
\caption{Density plot for the emission distributions used in the empirical application.}
\label{fig:emiss-densities-densplot-empapp}
\end{figure}

Figure \ref{fig:emiss-densities-densplot-empapp} shows the variables that were used in the empirical application.

\subsection{Detailed description of the data simulation procedure} \label{sec:appendix1-datasim}

The process by which the simulated datasets are created proceeds as follows:

\begin{enumerate}
    \item{Estimate the Bayesian mHMM on the observed EEG dataset described in section \ref{sec:data}. Collect the parameter estimates at the group level (parameters of the group-level TPM and the group-level component distribution parameters)}.
    \item{For each simulation scenario $h \in 1, 2, \dots, 144$ and iteration $r \in 1, 2, \dots, 250$, use the parameters for the group-level TPM collected in step (1) to generate a dataset of subject-specific TPMs with between-subject variance $Q_h$ for $N_{h}$ subjects. Moreover, use the parameters for the group-level component distributions to generate a dataset of subject-specific component distributions for $N_h$ subjects such that the between-subject variance is scaled by $\zeta_h$.}
    \item{For each of the subjects $n \in N_{h}$, use the subject-specific TPM $\pmb{\Gamma}_{n}$ to simulate a sequence of latent states of length $T_h$. Then, use the subject-specific parameters of the component distributions obtained in step (2) to simulate multivariate outcome data for each state in the sequence of states for subject $n$ according to the model specified in section \ref{sec:model}.}
\end{enumerate}

\subsection{MC SE and number of simulation iterations for each scenario} \label{sec:mcmcse}

Simulation studies involve random processes (e.g. data-generating mechanisms, starting values for the model). As such, the metrics by which the simulations are evaluated have some uncertainty associated with them. This is captured by the Monte Carlo Standard Error (MC SE). Ideally, the MC SE formulas are used to compute the required number of iterations given some acceptable degree of uncertainty in the estimates. In practice, however, researchers also need to take into account considerations such as the computational complexity and the run time of the model, as well as the number number of scenarios. 

Using the number of iterations described above allows the computation of the expected uncertainty in the parameter estimates. My primary evaluation metric is parameter bias. The formula for the MCMC SE for this evaluation metric is given in equation \ref{eq:MCMCSE}.

\begin{equation} \label{eq:MCMCSE}
    \text{MCSE}_{\text{bias}, h} = \sqrt{\text{Var}(\hat{\theta}_h) / n_{\text{iterations}}}
\end{equation}

Where $\text{Var}(\hat{\theta}_h)$ is the variance term of parameter estimate $h$ in scenario $r$.\footnote{Note that the term 'parameter estimate' here refers to state-dependent emission distribution means and variances, as well as the multinomial regression intercept values.} We estimate this term for each parameter estimate by running an initial small simulation run of $48$ iterations. The expected MC SE is largest for the emission distribution \textit{EOG min beta}, and in particular the REM state. This indicates that there is a lot of uncertainty around the bias metric for this emission distribution.  

If the coverage of all parameters is 95\%, the implication of using $n_\text{iterations} = 250$ is that:

\begin{equation}
    \text{MCSE}_\text{coverage} = \sqrt{(95 \times 5) / 250} = 1.38\%
\end{equation}

With 50\% coverage, the MC SE is maximized at 3.16\%. Careful consideration will be given to those estimates with large expected MC SE. In practice, these are settings with a low number of subjects and occasions per subject, as well as scenarios in which the between-subject variance is very large. It is possible that the size of the MC SE will require additional iterations to reduce the variance in the parameter estimates. 

\subsection{Hyper-prior specification} \label{sec:appendixI-hyperpriors}

Table \ref{tab:hyperpriors} gives an overview of the hyper-priors that need to be specified to run the model using the \textit{mHMMbayes} R library. The value for each hyper-prior is chosen such that they are 'uninformative'. That is, they carry no prior information about the problem context.

\begin{table}[H]
\centering
\begin{tabular}{C{4em} L{15em} L{10em} L{10em}}
\hline
\textbf{Symbol} & \textbf{Description}                                                                                           & \textbf{Parameter}                                      & \textbf{Value}                                                       \\ \hline
$\mu_0$         & Hypothesized (prior) mean of the emission distribution.                                                        & Emission distribution means                             & For each state-dependent emission distribution, use the sample mean. \\ \hline
$K_0$          & Hypothesized (prior) number of subjects on which the prior means are based.                                    & Emission distribution means                             & Set to $1$ for all emission distributions.                           \\ \hline
$\nu$           & Degrees of freedom of the inverse Gamma hyper-prior distribution connected to the emission distribution means. & Emission distribution means                             & Set to $1$ for all emission distributions.                           \\ \hline
$V$             & Hypothesized prior variances between the hypothetical prior subjects.                                          & Between-subject variances of the emission distributions & Set to $1$ for all emission distributions.                           \\ \hline
$\alpha_0$      & Shape parameter of the inverse Gamma hyper-prior used for the residual error.                                  & Residual error.                                         & Set to $0.1$ for all emission distributions.                         \\ \hline
$\beta_0$       & Scale parameter of the inverse Gamma hyper-prior used for the residual error.                                  & Residual error.                                         & Set to $0.1$ for all emission distributions.                         \\ \hline
\end{tabular}
\caption{Overview of the hyper-priors used in the model.}
\label{tab:hyperpriors}
\end{table}

\newpage
\section{Selected results from the simulation study} \label{sec:appendix2}
\subsection{Baseline scenarios} \label{sec:appendixII-baseline}

Tables \ref{tab:baseline-results-z025} and \ref{tab:baseline-results-z05} contain the results of the baseline scenarios mentioned in section \ref{sec:methods}. The scenarios in table \ref{tab:baseline-results-z05} differ only from those in table \ref{tab:baseline-results-z025} in terms of the value of $\zeta$. Scenarios $1A$ and $1B$ are characterized by component distributions that are clearly separated and by high self-transition probabilities. In scenarios $2A$ and $2B$, we also lower the self-transition probabilities (see column "pop. value" in tables \ref{tab:baseline-results-z025} and \ref{tab:baseline-results-z05}). Scenarios $3A$-$5A$ and $3B$-$5B$ use the same population values as scenarios $2A$ and $2B$.

\newpage

\begin{table}[H]
\centering
\captionsetup{width=.98\textwidth}
\begin{adjustbox}{width=1.2\textwidth, angle=90}
\begin{tabular}{ m{7em}  m{5em} | R{2.5em} | R{3em} | R{3em} | R{3em} | R{3em} | m{3.5em} | m{3.5em} | R{2.5em} | R{3em} | R{3em} | R{3em} | R{3em} | m{3.5em} | m{3.5em} | R{3em} | R{3em} | R{3em} | R{3em} | m{3.5em} | m{3.5em} | R{3em} | R{3em} | R{3em} | R{3em} | m{3.5em} | m{3.5em} | R{3em} | R{3em} | R{3em} | R{3em} | m{3.5em} | m{3.5em} }
\cline{3-34}
 & \multicolumn{1}{l|}{} & \multicolumn{7}{l|}{\begin{tabular}[c]{@{}l@{}}\large Scenario 1A - emission distributions not overlapping\\ \large ($N=40$, $N_t=800$, $\zeta=0.25$, $Q=0.1$, $n_{\text{simulations}} = 250$)\end{tabular}} & \multicolumn{7}{l|}{\begin{tabular}[c]{@{}l@{}}\large Scenario 2A - lower self-transitions probabilities\\ \large ($N=40$, $N_t=800$, $\zeta=0.25$, $Q=0.1$, $n_{\text{simulations}} = 250$)\end{tabular}} & \multicolumn{6}{l|}{\begin{tabular}[c]{@{}l@{}}\large Scenario 3A - $N$ large\\ \large ($N=80$, $N_t=800$, $\zeta=0.25$, $Q=0.1$, $n_{\text{simulations}} = 250$)\end{tabular}} & \multicolumn{6}{l|}{\begin{tabular}[c]{@{}l@{}}\large Scenario 4A - $N_t$ large\\ \large ($N=80$, $N_t=3.200$, $\zeta=0.25$, $Q=0.1$, $n_{\text{simulations}} = 86$)\end{tabular}} & \multicolumn{6}{l}{\begin{tabular}[c]{@{}l@{}}\large Scenario 5A - $N$ very large\\ \large ($N=140$, $N_t=800$, $\zeta=0.25$, $Q=0.1$, $n_{\text{simulations}} = 250$)\end{tabular}} \\ \hline
\textbf{Variable} & \textbf{Parameter} & \textbf{Pop. value} & \textbf{Percent bias} & \textbf{Emp. SE} & \textbf{Model SE} & \textbf{MSE} & \textbf{Coverage} & \textbf{Bias Corr. Coverage} & \textbf{Pop. value} & \textbf{Percent bias} & \textbf{Emp. SE} & \textbf{Model SE} & \textbf{MSE} & \textbf{Coverage} & \textbf{Bias Corr. Coverage} & \textbf{Percent bias} & \textbf{Emp. SE} & \textbf{Model SE} & \textbf{MSE} & \textbf{Coverage} & \textbf{Bias Corr. Coverage} & \textbf{Percent bias} & \textbf{Emp. SE} & \textbf{Model SE} & \textbf{MSE} & \textbf{Coverage} & \textbf{Bias Corr. Coverage} & \textbf{Percent bias} & \textbf{Emp. SE} & \textbf{Model SE} & \textbf{MSE} & \textbf{Coverage} & \textbf{Bias Corr. Coverage} \\ \hline
\multirow{6}{*}{\begin{tabular}[c]{@{}l@{}}EEG mean\\ theta\end{tabular}} & $\beta_{001,\text{Awake}}$ & -3.90 & -0.01 (-0.13) & 0.079 (0.004) & 0.097 (0.001) & 0.006 (0.001) & 0.98 (0.01) & 0.98 (0.01) & - & -0.01 (-0.13) & 0.080 (0.004) & 0.097 (0.001) & 0.006 (0.001) & 0.99 (0.01) & 0.99 (0.01) & -0.06 (-0.09) & 0.056 (0.002) & 0.068 (0.000) & 0.003 (0.000) & 0.99 (0.01) & 0.99 (0.01) & 0.08 (-0.15) & 0.053 (0.004) & 0.069 (0.000) & 0.003 (0.000) & 1.00 (0.00) & 1.00 (0.00) & -0.06 (-0.06) & 0.037 (0.002) & 0.052 (0.000) & 0.001 (0.000) & 0.99 (0.01) & 0.99 (0.01) \\ \cline{2-34} 
 & $\beta_{001,\text{NREM}}$ & -1.00 & 8.11 (-0.47) & 0.074 (0.003) & 0.121 (0.000) & 0.012 (0.001) & 0.97 (0.01) & 1.00 (0.00) & - & 8.04 (-0.47) & 0.074 (0.003) & 0.121 (0.000) & 0.012 (0.001) & 0.98 (0.01) & 1.00 (0.00) & 3.26 (-0.36) & 0.057 (0.003) & 0.078 (0.000) & 0.004 (0.000) & 0.99 (0.01) & 0.99 (0.01) & 2.60 (-0.55) & 0.051 (0.004) & 0.079 (0.000) & 0.003 (0.000) & 1.00 (0.00) & 1.00 (0.00) & 2.01 (-0.25) & 0.040 (0.002) & 0.056 (0.000) & 0.002 (0.000) & 0.99 (0.01) & 0.99 (0.01) \\ \cline{2-34} 
 & $\beta_{001,\text{REM}}$ & 2.40 & -6.72 ( 0.20) & 0.077 (0.003) & 0.186 (0.000) & 0.032 (0.002) & 1.00 (0.00) & 1.00 (0.00) & - & -6.74 ( 0.21) & 0.078 (0.003) & 0.186 (0.000) & 0.032 (0.002) & 1.00 (0.00) & 1.00 (0.00) & -3.27 ( 0.14) & 0.054 (0.002) & 0.106 (0.000) & 0.009 (0.001) & 0.99 (0.01) & 1.00 (0.00) & -3.34 ( 0.22) & 0.049 (0.004) & 0.106 (0.000) & 0.009 (0.001) & 1.00 (0.00) & 1.00 (0.00) & -1.96 ( 0.12) & 0.044 (0.002) & 0.070 (0.000) & 0.004 (0.000) & 0.97 (0.01) & 0.99 (0.01) \\ \cline{2-34} 
 & $\sigma^2_{u_01,\text{Awake}}$ & 0.25 & 23.62 (1.45) & 0.057 (0.003) & 0.088 (0.001) & 0.007 (0.001) & 0.95 (0.01) & 1.00 (0.00) & - & 23.76 (1.45) & 0.057 (0.003) & 0.089 (0.001) & 0.007 (0.001) & 0.95 (0.01) & 1.00 (0.00) & 22.35 (1.04) & 0.041 (0.002) & 0.059 (0.000) & 0.005 (0.000) & 0.89 (0.02) & 1.00 (0.00) & 24.85 (1.77) & 0.041 (0.003) & 0.060 (0.001) & 0.006 (0.001) & 0.84 (0.04) & 1.00 (0.00) & 25.06 (0.76) & 0.030 (0.001) & 0.045 (0.000) & 0.005 (0.000) & 0.70 (0.03) & 1.00 (0.00) \\ \cline{2-34} 
 & $\sigma^2_{u_01,\text{NREM}}$ & 0.25 & 103.78 (1.48) & 0.059 (0.003) & 0.133 (0.001) & 0.071 (0.002) & 0.02 (0.01) & 1.00 (0.00) & - & 104.40 (1.48) & 0.059 (0.003) & 0.134 (0.001) & 0.072 (0.002) & 0.01 (0.01) & 1.00 (0.00) & 66.27 (1.01) & 0.040 (0.002) & 0.076 (0.000) & 0.029 (0.001) & 0.03 (0.01) & 1.00 (0.00) & 67.90 (1.94) & 0.045 (0.003) & 0.077 (0.001) & 0.031 (0.002) & 0.04 (0.02) & 0.99 (0.01) & 49.27 (0.83) & 0.033 (0.001) & 0.052 (0.000) & 0.016 (0.001) & 0.04 (0.01) & 1.00 (0.00) \\ \cline{2-34} 
 & $\sigma^2_{u_01,\text{REM}}$ & 0.25 & 414.81 (1.68) & 0.066 (0.003) & 0.316 (0.001) & 1.080 (0.009) & 0.00 (0.00) & 1.00 (0.00) & - & 414.02 (1.66) & 0.066 (0.003) & 0.316 (0.001) & 1.076 (0.009) & 0.00 (0.00) & 1.00 (0.00) & 224.67 (1.01) & 0.040 (0.002) & 0.138 (0.000) & 0.317 (0.003) & 0.00 (0.00) & 1.00 (0.00) & 225.82 (1.65) & 0.038 (0.003) & 0.139 (0.001) & 0.320 (0.005) & 0.00 (0.00) & 1.00 (0.00) & 140.97 (0.80) & 0.032 (0.001) & 0.078 (0.000) & 0.125 (0.001) & 0.00 (0.00) & 1.00 (0.00) \\ \hline
\multirow{6}{*}{\begin{tabular}[c]{@{}l@{}}EOG median\\ theta\end{tabular}} & $\beta_{002,\text{Awake}}$ & 3.05 & -0.13 ( 0.16) & 0.078 (0.003) & 0.095 (0.001) & 0.006 (0.001) & 0.99 (0.01) & 0.99 (0.01) & - & -0.12 ( 0.16) & 0.077 (0.003) & 0.096 (0.001) & 0.006 (0.001) & 0.98 (0.01) & 0.99 (0.01) & 0.01 ( 0.11) & 0.052 (0.002) & 0.069 (0.000) & 0.003 (0.000) & 0.99 (0.01) & 0.99 (0.01) & 0.01 ( 0.21) & 0.058 (0.004) & 0.069 (0.000) & 0.003 (0.000) & 0.99 (0.01) & 0.99 (0.01) & 0.02 ( 0.09) & 0.043 (0.002) & 0.052 (0.000) & 0.002 (0.000) & 0.99 (0.01) & 0.99 (0.01) \\ \cline{2-34} 
 & $\beta_{002,\text{NREM}}$ & -3.40 & -4.72 (-0.15) & 0.079 (0.004) & 0.189 (0.000) & 0.032 (0.002) & 1.00 (0.00) & 1.00 (0.00) & - & -4.71 (-0.15) & 0.079 (0.004) & 0.189 (0.000) & 0.032 (0.002) & 1.00 (0.00) & 1.00 (0.00) & -2.36 (-0.11) & 0.059 (0.003) & 0.108 (0.000) & 0.010 (0.001) & 0.98 (0.01) & 1.00 (0.00) & -2.33 (-0.18) & 0.057 (0.004) & 0.107 (0.000) & 0.010 (0.001) & 0.98 (0.02) & 1.00 (0.00) & -1.43 (-0.07) & 0.040 (0.002) & 0.071 (0.000) & 0.004 (0.000) & 0.99 (0.01) & 1.00 (0.00) \\ \cline{2-34} 
 & $\beta_{002,\text{REM}}$ & -0.50 & -16.31 (-0.87) & 0.069 (0.003) & 0.133 (0.000) & 0.011 (0.001) & 0.99 (0.01) & 1.00 (0.00) & - & -16.36 (-0.89) & 0.070 (0.003) & 0.132 (0.000) & 0.012 (0.001) & 1.00 (0.00) & 1.00 (0.00) & -10.91 (-0.69) & 0.054 (0.002) & 0.082 (0.000) & 0.006 (0.000) & 0.97 (0.01) & 1.00 (0.00) & -7.64 (-1.22) & 0.056 (0.004) & 0.082 (0.000) & 0.005 (0.001) & 1.00 (0.00) & 1.00 (0.00) & -4.38 (-0.52) & 0.041 (0.002) & 0.059 (0.000) & 0.002 (0.000) & 0.97 (0.01) & 0.98 (0.01) \\ \cline{2-34} 
 & $\sigma^2_{u_02,\text{Awake}}$ & 0.25 & 19.65 (1.45) & 0.057 (0.003) & 0.086 (0.001) & 0.006 (0.001) & 0.94 (0.02) & 0.99 (0.01) & - & 19.87 (1.46) & 0.058 (0.003) & 0.086 (0.001) & 0.006 (0.001) & 0.94 (0.02) & 0.99 (0.01) & 24.92 (1.04) & 0.041 (0.002) & 0.061 (0.000) & 0.006 (0.000) & 0.86 (0.02) & 1.00 (0.00) & 25.69 (1.93) & 0.044 (0.003) & 0.061 (0.001) & 0.006 (0.001) & 0.82 (0.04) & 0.98 (0.02) & 24.78 (0.76) & 0.030 (0.001) & 0.045 (0.000) & 0.005 (0.000) & 0.71 (0.03) & 1.00 (0.00) \\ \cline{2-34} 
 & $\sigma^2_{u_02,\text{NREM}}$ & 0.25 & 433.22 (1.79) & 0.071 (0.003) & 0.326 (0.001) & 1.178 (0.010) & 0.00 (0.00) & 1.00 (0.00) & - & 433.18 (1.75) & 0.069 (0.003) & 0.326 (0.001) & 1.178 (0.010) & 0.00 (0.00) & 1.00 (0.00) & 237.40 (1.10) & 0.043 (0.002) & 0.143 (0.000) & 0.354 (0.003) & 0.00 (0.00) & 1.00 (0.00) & 234.82 (1.78) & 0.041 (0.003) & 0.142 (0.001) & 0.346 (0.005) & 0.00 (0.00) & 1.00 (0.00) & 148.20 (0.76) & 0.030 (0.001) & 0.080 (0.000) & 0.138 (0.001) & 0.00 (0.00) & 1.00 (0.00) \\ \cline{2-34} 
 & $\sigma^2_{u_02,\text{REM}}$ & 0.25 & 149.32 (1.46) & 0.058 (0.003) & 0.159 (0.001) & 0.143 (0.003) & 0.00 (0.00) & 1.00 (0.00) & - & 149.43 (1.45) & 0.057 (0.003) & 0.159 (0.001) & 0.143 (0.003) & 0.00 (0.00) & 1.00 (0.00) & 87.87 (1.09) & 0.043 (0.002) & 0.084 (0.000) & 0.050 (0.001) & 0.00 (0.00) & 1.00 (0.00) & 87.68 (1.92) & 0.044 (0.003) & 0.084 (0.001) & 0.050 (0.002) & 0.00 (0.00) & 1.00 (0.00) & 63.09 (0.76) & 0.030 (0.001) & 0.056 (0.000) & 0.026 (0.001) & 0.00 (0.00) & 1.00 (0.00) \\ \hline
\multirow{6}{*}{\begin{tabular}[c]{@{}l@{}}EOG min \\ beta\end{tabular}} & $\beta_{003,\text{Awake}}$ & 0.40 & 1.66 ( 1.16) & 0.073 (0.003) & 0.095 (0.001) & 0.005 (0.000) & 0.99 (0.01) & 1.00 (0.00) & - & 1.55 ( 1.16) & 0.074 (0.003) & 0.095 (0.001) & 0.005 (0.000) & 0.99 (0.01) & 1.00 (0.00) & 0.44 ( 0.98) & 0.062 (0.003) & 0.068 (0.000) & 0.004 (0.000) & 0.96 (0.01) & 0.96 (0.01) & -0.26 ( 1.44) & 0.053 (0.004) & 0.069 (0.000) & 0.003 (0.000) & 0.99 (0.01) & 0.99 (0.01) & 0.12 ( 0.69) & 0.044 (0.002) & 0.052 (0.000) & 0.002 (0.000) & 0.98 (0.01) & 0.98 (0.01) \\ \cline{2-34} 
 & $\beta_{003,\text{NREM}}$ & 3.50 & -2.38 ( 0.14) & 0.079 (0.004) & 0.124 (0.000) & 0.013 (0.001) & 0.98 (0.01) & 1.00 (0.00) & - & -2.38 ( 0.14) & 0.078 (0.004) & 0.124 (0.000) & 0.013 (0.001) & 0.98 (0.01) & 1.00 (0.00) & -1.12 ( 0.10) & 0.056 (0.002) & 0.079 (0.000) & 0.005 (0.000) & 0.98 (0.01) & 0.99 (0.01) & -0.83 ( 0.17) & 0.056 (0.004) & 0.079 (0.000) & 0.004 (0.001) & 0.99 (0.01) & 1.00 (0.00) & -0.79 ( 0.08) & 0.042 (0.002) & 0.057 (0.000) & 0.003 (0.000) & 0.98 (0.01) & 1.00 (0.00) \\ \cline{2-34} 
 & $\beta_{003,\text{REM}}$ & -2.80 & -2.94 (-0.18) & 0.080 (0.004) & 0.127 (0.000) & 0.013 (0.001) & 0.97 (0.01) & 1.00 (0.00) & - & -2.93 (-0.18) & 0.080 (0.004) & 0.127 (0.000) & 0.013 (0.001) & 0.96 (0.01) & 1.00 (0.00) & -1.32 (-0.12) & 0.055 (0.002) & 0.080 (0.000) & 0.004 (0.000) & 0.99 (0.01) & 1.00 (0.00) & -1.50 (-0.22) & 0.058 (0.004) & 0.080 (0.000) & 0.005 (0.001) & 0.99 (0.01) & 0.99 (0.01) & -0.82 (-0.10) & 0.045 (0.002) & 0.057 (0.000) & 0.003 (0.000) & 0.97 (0.01) & 0.98 (0.01) \\ \cline{2-34} 
 & $\sigma^2_{u_03,\text{Awake}}$ & 0.25 & 19.04 (1.47) & 0.058 (0.003) & 0.086 (0.001) & 0.006 (0.001) & 0.95 (0.01) & 0.99 (0.01) & - & 19.01 (1.48) & 0.058 (0.003) & 0.086 (0.001) & 0.006 (0.001) & 0.95 (0.01) & 0.99 (0.01) & 22.96 (1.00) & 0.040 (0.002) & 0.060 (0.000) & 0.005 (0.000) & 0.87 (0.02) & 1.00 (0.00) & 27.18 (1.75) & 0.040 (0.003) & 0.062 (0.001) & 0.006 (0.001) & 0.81 (0.04) & 0.99 (0.01) & 24.08 (0.81) & 0.032 (0.001) & 0.045 (0.000) & 0.005 (0.000) & 0.70 (0.03) & 0.99 (0.01) \\ \cline{2-34} 
 & $\sigma^2_{u_03,\text{NREM}}$ & 0.25 & 116.52 (1.41) & 0.056 (0.002) & 0.140 (0.001) & 0.088 (0.002) & 0.00 (0.00) & 1.00 (0.00) & - & 116.95 (1.41) & 0.056 (0.003) & 0.140 (0.001) & 0.089 (0.002) & 0.00 (0.00) & 1.00 (0.00) & 73.19 (1.09) & 0.043 (0.002) & 0.079 (0.000) & 0.035 (0.001) & 0.03 (0.01) & 1.00 (0.00) & 73.16 (2.11) & 0.049 (0.004) & 0.079 (0.001) & 0.036 (0.002) & 0.00 (0.00) & 1.00 (0.00) & 52.79 (0.74) & 0.029 (0.001) & 0.053 (0.000) & 0.018 (0.001) & 0.00 (0.00) & 1.00 (0.00) \\ \cline{2-34} 
 & $\sigma^2_{u_03,\text{REM}}$ & 0.25 & 127.13 (1.49) & 0.059 (0.003) & 0.146 (0.001) & 0.104 (0.002) & 0.00 (0.00) & 1.00 (0.00) & - & 126.82 (1.47) & 0.058 (0.003) & 0.146 (0.001) & 0.104 (0.002) & 0.00 (0.00) & 1.00 (0.00) & 77.40 (1.12) & 0.044 (0.002) & 0.080 (0.000) & 0.039 (0.001) & 0.01 (0.01) & 1.00 (0.00) & 77.54 (1.75) & 0.040 (0.003) & 0.081 (0.001) & 0.039 (0.002) & 0.02 (0.02) & 1.00 (0.00) & 55.84 (0.76) & 0.030 (0.001) & 0.054 (0.000) & 0.020 (0.001) & 0.02 (0.01) & 1.00 (0.00) \\ \hline
\multirow{9}{*}{\begin{tabular}[c]{@{}l@{}}Transition \\ probabilities\end{tabular}} & $\gamma_{\text{ Awake,Awake}}$ & 0.984 & -3.41 (0.05) & 0.002 (0.000) & 0.004 (0.000) & 0.000 (0.000) & 0.06 (0.02) & 1.00 (0.00) & 0.80 & -5.03 (0.17) & 0.007 (0.000) & 0.013 (0.000) & 0.000 (0.000) & 0.97 (0.01) & 1.00 (0.00) & -3.13 (0.13) & 0.005 (0.000) & 0.007 (0.000) & 0.000 (0.000) & 0.92 (0.02) & 1.00 (0.00) & -2.17 (0.19) & 0.004 (0.000) & 0.007 (0.000) & 0.000 (0.000) & 0.96 (0.02) & 1.00 (0.00) & -2.41 (0.08) & 0.003 (0.000) & 0.005 (0.000) & 0.000 (0.000) & 0.88 (0.02) & 0.99 (0.01) \\ \cline{2-34} 
 & $\gamma_{\text{ Awake,NREM}}$ & 0.007 & -0.14 (0.02) & 0.001 (0.000) & 0.001 (0.000) & 0.000 (0.000) & 1.00 (0.00) & 1.00 (0.00) & 0.10 & 2.46 (0.14) & 0.005 (0.000) & 0.009 (0.000) & 0.000 (0.000) & 0.96 (0.01) & 1.00 (0.00) & 1.46 (0.10) & 0.004 (0.000) & 0.005 (0.000) & 0.000 (0.000) & 0.94 (0.02) & 1.00 (0.00) & 0.99 (0.15) & 0.003 (0.000) & 0.005 (0.000) & 0.000 (0.000) & 0.98 (0.02) & 0.99 (0.01) & 1.26 (0.07) & 0.003 (0.000) & 0.004 (0.000) & 0.000 (0.000) & 0.92 (0.02) & 0.98 (0.01) \\ \cline{2-34} 
 & $\gamma_{\text{ Awake,REM}}$ & 0.012 & 2.33 (0.04) & 0.002 (0.000) & 0.003 (0.000) & 0.000 (0.000) & 0.20 (0.02) & 1.00 (0.00) & 0.10 & 2.44 (0.14) & 0.005 (0.000) & 0.009 (0.000) & 0.000 (0.000) & 0.98 (0.01) & 1.00 (0.00) & 1.62 (0.10) & 0.004 (0.000) & 0.005 (0.000) & 0.000 (0.000) & 0.94 (0.02) & 0.98 (0.01) & 1.14 (0.15) & 0.003 (0.000) & 0.005 (0.000) & 0.000 (0.000) & 0.96 (0.02) & 1.00 (0.00) & 1.14 (0.07) & 0.003 (0.000) & 0.004 (0.000) & 0.000 (0.000) & 0.94 (0.02) & 1.00 (0.00) \\ \cline{2-34} 
 & $\gamma_{\text{ NREM,Awake}}$ & 0.003 & 4.54 (0.04) & 0.002 (0.000) & 0.003 (0.000) & 0.000 (0.000) & 0.00 (0.00) & 1.00 (0.00) & 0.15 & 2.37 (0.16) & 0.006 (0.000) & 0.010 (0.000) & 0.000 (0.000) & 0.98 (0.01) & 1.00 (0.00) & 1.96 (0.12) & 0.005 (0.000) & 0.006 (0.000) & 0.000 (0.000) & 0.95 (0.01) & 1.00 (0.00) & 0.86 (0.16) & 0.004 (0.000) & 0.005 (0.000) & 0.000 (0.000) & 0.98 (0.02) & 1.00 (0.00) & 1.42 (0.09) & 0.003 (0.000) & 0.004 (0.000) & 0.000 (0.000) & 0.94 (0.02) & 0.98 (0.01) \\ \cline{2-34} 
 & $\gamma_{\text{ NREM,NREM}}$ & 0.959 & -6.69 (0.12) & 0.005 (0.000) & 0.008 (0.000) & 0.000 (0.000) & 0.24 (0.03) & 1.00 (0.00) & 0.70 & -4.47 (0.32) & 0.013 (0.001) & 0.020 (0.000) & 0.000 (0.000) & 0.99 (0.01) & 1.00 (0.00) & -3.78 (0.23) & 0.009 (0.000) & 0.012 (0.000) & 0.000 (0.000) & 0.96 (0.01) & 0.98 (0.01) & -2.04 (0.37) & 0.009 (0.001) & 0.011 (0.000) & 0.000 (0.000) & 0.98 (0.02) & 0.99 (0.01) & -2.74 (0.17) & 0.007 (0.000) & 0.008 (0.000) & 0.000 (0.000) & 0.93 (0.02) & 0.98 (0.01) \\ \cline{2-34} 
 & $\gamma_{\text{ NREM,REM}}$ & 0.021 & 8.90 (0.10) & 0.004 (0.000) & 0.007 (0.000) & 0.001 (0.000) & 0.00 (0.00) & 1.00 (0.00) & 0.15 & 1.99 (0.22) & 0.009 (0.000) & 0.014 (0.000) & 0.000 (0.000) & 1.00 (0.00) & 1.00 (0.00) & 1.76 (0.16) & 0.006 (0.000) & 0.008 (0.000) & 0.000 (0.000) & 0.97 (0.01) & 0.98 (0.01) & 1.14 (0.26) & 0.006 (0.000) & 0.008 (0.000) & 0.000 (0.000) & 0.98 (0.02) & 0.99 (0.01) & 1.29 (0.12) & 0.005 (0.000) & 0.006 (0.000) & 0.000 (0.000) & 0.96 (0.01) & 0.99 (0.01) \\ \cline{2-34} 
 & $\gamma_{\text{ REM,Awake}}$ & 0.013 & 1.54 (0.04) & 0.001 (0.000) & 0.003 (0.000) & 0.000 (0.000) & 0.73 (0.03) & 1.00 (0.00) & 0.18 & 2.06 (0.18) & 0.007 (0.000) & 0.011 (0.000) & 0.000 (0.000) & 0.99 (0.01) & 1.00 (0.00) & 1.91 (0.13) & 0.005 (0.000) & 0.007 (0.000) & 0.000 (0.000) & 0.94 (0.02) & 0.99 (0.01) & 1.19 (0.18) & 0.004 (0.000) & 0.006 (0.000) & 0.000 (0.000) & 0.96 (0.02) & 0.99 (0.01) & 1.20 (0.09) & 0.004 (0.000) & 0.005 (0.000) & 0.000 (0.000) & 0.96 (0.01) & 0.98 (0.01) \\ \cline{2-34} 
 & $\gamma_{\text{ REM,NREM}}$ & 0.034 & -3.14 (0.06) & 0.002 (0.000) & 0.004 (0.000) & 0.000 (0.000) & 0.64 (0.03) & 1.00 (0.00) & 0.18 & 2.60 (0.28) & 0.011 (0.000) & 0.015 (0.000) & 0.000 (0.000) & 0.98 (0.01) & 0.99 (0.01) & 1.60 (0.20) & 0.008 (0.000) & 0.009 (0.000) & 0.000 (0.000) & 0.96 (0.01) & 0.97 (0.01) & 1.02 (0.24) & 0.006 (0.000) & 0.009 (0.000) & 0.000 (0.000) & 1.00 (0.00) & 1.00 (0.00) & 1.41 (0.13) & 0.005 (0.000) & 0.007 (0.000) & 0.000 (0.000) & 0.96 (0.01) & 0.99 (0.01) \\ \cline{2-34} 
 & $\gamma_{\text{ REM,REM}}$ & 0.967 & -4.03 (0.08) & 0.003 (0.000) & 0.006 (0.000) & 0.000 (0.000) & 0.57 (0.03) & 1.00 (0.00) & 0.64 & -4.79 (0.36) & 0.014 (0.001) & 0.021 (0.000) & 0.000 (0.000) & 0.99 (0.01) & 0.99 (0.01) & -3.56 (0.26) & 0.010 (0.000) & 0.013 (0.000) & 0.000 (0.000) & 0.94 (0.02) & 0.99 (0.01) & -2.26 (0.32) & 0.007 (0.001) & 0.012 (0.000) & 0.000 (0.000) & 0.98 (0.02) & 1.00 (0.00) & -2.64 (0.18) & 0.007 (0.000) & 0.009 (0.000) & 0.000 (0.000) & 0.94 (0.02) & 0.98 (0.01) \\ \hline
\end{tabular}
\end{adjustbox}
\caption{Simulation results on the baseline scenarios. The component distribution random effect is set to $\zeta=0.25$ in all scenarios. The TPM random effect is set to $Q=0.1$. Scenario-specific information can be found at the top of the table. MC standard errors are given in parentheses.}
\label{tab:baseline-results-z025}
\end{table}

\newpage

Table \ref{tab:baseline-results-z05} shows the same results as table \ref{tab:baseline-results-z025}. Here, however, we use a between-subject variance of $\zeta = 0.5$.

\begin{table}[H]
\centering
\captionsetup{width=.98\textwidth}
\begin{adjustbox}{width=1.2\textwidth, angle=90}
\begin{tabular}{ m{7em}  m{5em} | R{2.5em} | R{3em} | R{3em} | R{3em} | R{3em} | m{3.5em} | m{3.5em} | R{2.5em} | R{3em} | R{3em} | R{3em} | R{3em} | m{3.5em} | m{3.5em} | R{3em} | R{3em} | R{3em} | R{3em} | m{3.5em} | m{3.5em} | R{3em} | R{3em} | R{3em} | R{3em} | m{3.5em} | m{3.5em} | R{3em} | R{3em} | R{3em} | R{3em} | m{3.5em} | m{3.5em} }
\cline{3-34}
 & \multicolumn{1}{l|}{} & \multicolumn{7}{l|}{\begin{tabular}[c]{@{}l@{}}\large Scenario 1B - emission distributions not overlapping\\\large ($N=40$, $N_t=800$, $\zeta=0.5$, $Q=0.1$, $n_{\text{simulations}} = 250$)\end{tabular}} & \multicolumn{7}{l|}{\begin{tabular}[c]{@{}l@{}}\large Scenario 2B - lower self-transitions probabilities\\ \large ($N=40$, $N_t=800$, $\zeta=0.5$, $Q=0.1$, $n_{\text{simulations}} = 250$)\end{tabular}} & \multicolumn{6}{l|}{\begin{tabular}[c]{@{}l@{}}\large Scenario 3B - $N$ large\\ \large ($N=80$, $N_t=800$, $\zeta=0.5$, $Q=0.1$, $n_{\text{simulations}} = 250$)\end{tabular}} & \multicolumn{6}{l|}{\begin{tabular}[c]{@{}l@{}}\large Scenario 4B - $N_t$ large\\ \large ($N=80$, $N_t=3.200$, $\zeta=0.5$, $Q=0.1$, $n_{\text{simulations}} = 46$)\end{tabular}} & \multicolumn{6}{l}{\begin{tabular}[c]{@{}l@{}}\large Scenario 5B - $N$ very large\\ \large ($N=140$, $N_t=800$, $\zeta=0.5$, $Q=0.1$, $n_{\text{simulations}} = 250$)\end{tabular}} \\ \hline
\textbf{Variable} & \textbf{Parameter} & \textbf{Pop. value} & \textbf{Percent bias} & \textbf{Emp. SE} & \textbf{Model SE} & \textbf{MSE} & \textbf{Coverage} & \textbf{Bias Corr. Coverage} & \textbf{Pop. value} & \textbf{Percent bias} & \textbf{Emp. SE} & \textbf{Model SE} & \textbf{MSE} & \textbf{Coverage} & \textbf{Bias Corr. Coverage} & \textbf{Percent bias} & \textbf{Emp. SE} & \textbf{Model SE} & \textbf{MSE} & \textbf{Coverage} & \textbf{Bias Corr. Coverage} & \textbf{Percent bias} & \textbf{Emp. SE} & \textbf{Model SE} & \textbf{MSE} & \textbf{Coverage} & \textbf{Bias Corr. Coverage} & \textbf{Percent bias} & \textbf{Emp. SE} & \textbf{Model SE} & \textbf{MSE} & \textbf{Coverage} & \textbf{Bias Corr. Coverage} \\ \hline
\multirow{6}{*}{\begin{tabular}[c]{@{}l@{}}EEG mean\\ theta\end{tabular}} & $\beta_{001,\text{Awake}}$ & -3.90 & 0.18 (-0.18) & 0.109 (0.005) & 0.126 (0.001) & 0.012 (0.001) & 0.98 (0.01) & 0.98 (0.01) & - & 0.18 (-0.18) & 0.109 (0.005) & 0.126 (0.001) & 0.012 (0.001) & 0.98 (0.01) & 0.98 (0.01) & 0.11 (-0.13) & 0.078 (0.003) & 0.089 (0.000) & 0.006 (0.001) & 0.98 (0.01) & 0.99 (0.01) & 0.27 (-0.30) & 0.078 (0.008) & 0.090 (0.001) & 0.006 (0.001) & 0.98 (0.02) & 0.98 (0.02) & -0.03 (-0.10) & 0.058 (0.003) & 0.068 (0.000) & 0.003 (0.000) & 0.98 (0.01) & 0.98 (0.01) \\ \cline{2-34} 
 & $\beta_{001,\text{NREM}}$ & -1.00 & 7.76 (-0.66) & 0.104 (0.005) & 0.146 (0.001) & 0.017 (0.001) & 0.98 (0.01) & 1.00 (0.00) & - & 7.60 (-0.65) & 0.103 (0.005) & 0.147 (0.001) & 0.016 (0.001) & 0.98 (0.01) & 1.00 (0.00) & 3.59 (-0.51) & 0.080 (0.004) & 0.097 (0.000) & 0.008 (0.001) & 0.97 (0.01) & 0.98 (0.01) & 3.68 (-1.37) & 0.092 (0.010) & 0.097 (0.001) & 0.010 (0.002) & 0.98 (0.02) & 0.98 (0.02) & 2.19 (-0.34) & 0.053 (0.002) & 0.072 (0.000) & 0.003 (0.000) & 1.00 (0.00) & 1.00 (0.00) \\ \cline{2-34} 
 & $\beta_{001,\text{REM}}$ & 2.40 & -6.68 ( 0.29) & 0.108 (0.005) & 0.201 (0.001) & 0.037 (0.003) & 0.97 (0.01) & 1.00 (0.00) & - & -6.67 ( 0.28) & 0.108 (0.005) & 0.201 (0.000) & 0.037 (0.002) & 0.97 (0.01) & 1.00 (0.00) & -3.43 ( 0.21) & 0.081 (0.004) & 0.120 (0.000) & 0.013 (0.001) & 0.97 (0.01) & 1.00 (0.00) & -3.89 ( 0.41) & 0.066 (0.007) & 0.119 (0.001) & 0.013 (0.002) & 0.98 (0.02) & 1.00 (0.00) & -1.77 ( 0.15) & 0.057 (0.003) & 0.082 (0.000) & 0.005 (0.000) & 0.97 (0.01) & 1.00 (0.00) \\ \cline{2-34} 
 & $\sigma^2_{u_01,\text{Awake}}$ & 0.25 & 10.43 (1.35) & 0.107 (0.005) & 0.151 (0.002) & 0.014 (0.001) & 0.96 (0.01) & 0.99 (0.01) & - & 10.59 (1.36) & 0.107 (0.005) & 0.151 (0.002) & 0.014 (0.001) & 0.98 (0.01) & 0.99 (0.01) & 11.99 (0.96) & 0.075 (0.003) & 0.104 (0.001) & 0.009 (0.001) & 0.94 (0.02) & 0.99 (0.01) & 14.68 (2.52) & 0.084 (0.009) & 0.106 (0.002) & 0.012 (0.003) & 0.91 (0.04) & 1.00 (0.00) & 12.56 (0.75) & 0.058 (0.003) & 0.077 (0.000) & 0.007 (0.001) & 0.91 (0.02) & 1.00 (0.00) \\ \cline{2-34} 
 & $\sigma^2_{u_01,\text{NREM}}$ & 0.25 & 54.10 (1.39) & 0.110 (0.005) & 0.200 (0.002) & 0.085 (0.004) & 0.53 (0.03) & 1.00 (0.00) & - & 54.62 (1.40) & 0.111 (0.005) & 0.201 (0.002) & 0.087 (0.004) & 0.54 (0.03) & 1.00 (0.00) & 35.10 (1.03) & 0.081 (0.004) & 0.121 (0.001) & 0.037 (0.002) & 0.59 (0.03) & 1.00 (0.00) & 35.52 (2.98) & 0.100 (0.011) & 0.122 (0.003) & 0.041 (0.006) & 0.60 (0.07) & 0.98 (0.02) & 27.66 (0.80) & 0.062 (0.003) & 0.086 (0.001) & 0.023 (0.001) & 0.54 (0.03) & 1.00 (0.00) \\ \cline{2-34} 
 & $\sigma^2_{u_01,\text{REM}}$ & 0.25 & 203.73 (1.53) & 0.121 (0.005) & 0.374 (0.002) & 1.052 (0.016) & 0.00 (0.00) & 1.00 (0.00) & - & 203.75 (1.51) & 0.119 (0.005) & 0.374 (0.002) & 1.052 (0.016) & 0.00 (0.00) & 1.00 (0.00) & 112.58 (1.12) & 0.088 (0.004) & 0.182 (0.001) & 0.325 (0.006) & 0.00 (0.00) & 1.00 (0.00) & 107.94 (2.20) & 0.074 (0.008) & 0.179 (0.002) & 0.297 (0.012) & 0.00 (0.00) & 1.00 (0.00) & 69.94 (0.82) & 0.064 (0.003) & 0.110 (0.001) & 0.126 (0.003) & 0.00 (0.00) & 1.00 (0.00) \\ \hline
\multirow{6}{*}{\begin{tabular}[c]{@{}l@{}}EOG median\\ theta\end{tabular}} & $\beta_{002,\text{Awake}}$ & 3.05 & 0.02 ( 0.23) & 0.110 (0.005) & 0.126 (0.001) & 0.012 (0.001) & 0.98 (0.01) & 0.98 (0.01) & - & 0.02 ( 0.23) & 0.110 (0.005) & 0.126 (0.001) & 0.012 (0.001) & 0.98 (0.01) & 0.98 (0.01) & -0.08 ( 0.16) & 0.078 (0.004) & 0.090 (0.000) & 0.006 (0.001) & 0.98 (0.01) & 0.98 (0.01) & 0.07 ( 0.44) & 0.090 (0.010) & 0.089 (0.001) & 0.008 (0.001) & 0.98 (0.02) & 0.98 (0.02) & 0.08 ( 0.12) & 0.058 (0.003) & 0.068 (0.000) & 0.003 (0.000) & 0.97 (0.01) & 0.97 (0.01) \\ \cline{2-34} 
 & $\beta_{002,\text{NREM}}$ & -3.40 & -4.93 (-0.23) & 0.122 (0.005) & 0.204 (0.001) & 0.043 (0.003) & 0.96 (0.01) & 1.00 (0.00) & - & -4.90 (-0.23) & 0.123 (0.005) & 0.204 (0.001) & 0.043 (0.003) & 0.96 (0.01) & 1.00 (0.00) & -2.39 (-0.13) & 0.071 (0.003) & 0.122 (0.000) & 0.012 (0.001) & 0.99 (0.01) & 1.00 (0.00) & -2.04 (-0.35) & 0.080 (0.008) & 0.121 (0.001) & 0.011 (0.002) & 1.00 (0.00) & 1.00 (0.00) & -1.37 (-0.11) & 0.060 (0.003) & 0.083 (0.000) & 0.006 (0.001) & 0.98 (0.01) & 0.99 (0.01) \\ \cline{2-34} 
 & $\beta_{002,\text{REM}}$ & -0.50 & -17.20 (-1.40) & 0.111 (0.005) & 0.155 (0.001) & 0.020 (0.002) & 0.98 (0.01) & 1.00 (0.00) & - & -17.05 (-1.39) & 0.110 (0.005) & 0.155 (0.001) & 0.019 (0.001) & 0.98 (0.01) & 1.00 (0.00) & -9.33 (-1.04) & 0.082 (0.004) & 0.101 (0.000) & 0.009 (0.001) & 0.96 (0.01) & 0.98 (0.01) & -9.40 (-2.45) & 0.082 (0.009) & 0.101 (0.001) & 0.009 (0.002) & 0.96 (0.03) & 0.98 (0.02) & -4.29 (-0.83) & 0.065 (0.003) & 0.073 (0.000) & 0.005 (0.000) & 0.97 (0.01) & 0.97 (0.01) \\ \cline{2-34} 
 & $\sigma^2_{u_02,\text{Awake}}$ & 0.25 & 11.59 (1.36) & 0.107 (0.005) & 0.153 (0.002) & 0.015 (0.001) & 0.98 (0.01) & 0.99 (0.01) & - & 11.71 (1.36) & 0.107 (0.005) & 0.153 (0.002) & 0.015 (0.001) & 0.98 (0.01) & 1.00 (0.00) & 14.24 (1.05) & 0.083 (0.004) & 0.106 (0.001) & 0.012 (0.001) & 0.92 (0.02) & 1.00 (0.00) & 11.46 (2.35) & 0.079 (0.008) & 0.103 (0.002) & 0.009 (0.002) & 0.93 (0.04) & 1.00 (0.00) & 13.62 (0.70) & 0.055 (0.002) & 0.078 (0.000) & 0.008 (0.001) & 0.89 (0.02) & 0.99 (0.01) \\ \cline{2-34} 
 & $\sigma^2_{u_02,\text{NREM}}$ & 0.25 & 213.50 (1.51) & 0.119 (0.005) & 0.385 (0.002) & 1.154 (0.016) & 0.00 (0.00) & 1.00 (0.00) & - & 213.62 (1.49) & 0.118 (0.005) & 0.386 (0.002) & 1.155 (0.016) & 0.00 (0.00) & 1.00 (0.00) & 119.09 (1.03) & 0.082 (0.004) & 0.187 (0.001) & 0.361 (0.006) & 0.00 (0.00) & 1.00 (0.00) & 116.30 (2.28) & 0.076 (0.008) & 0.184 (0.002) & 0.344 (0.013) & 0.00 (0.00) & 1.00 (0.00) & 73.84 (0.76) & 0.059 (0.003) & 0.112 (0.001) & 0.140 (0.003) & 0.00 (0.00) & 1.00 (0.00) \\ \cline{2-34} 
 & $\sigma^2_{u_02,\text{REM}}$ & 0.25 & 74.46 (1.51) & 0.120 (0.005) & 0.224 (0.002) & 0.153 (0.006) & 0.19 (0.02) & 1.00 (0.00) & - & 74.50 (1.51) & 0.119 (0.005) & 0.224 (0.002) & 0.153 (0.006) & 0.21 (0.03) & 1.00 (0.00) & 44.97 (1.03) & 0.082 (0.004) & 0.129 (0.001) & 0.057 (0.002) & 0.32 (0.03) & 1.00 (0.00) & 47.52 (2.35) & 0.079 (0.008) & 0.132 (0.002) & 0.063 (0.006) & 0.29 (0.07) & 1.00 (0.00) & 32.46 (0.83) & 0.065 (0.003) & 0.089 (0.001) & 0.031 (0.001) & 0.38 (0.03) & 0.98 (0.01) \\ \hline
\multirow{6}{*}{\begin{tabular}[c]{@{}l@{}}EOG min \\ beta\end{tabular}} & $\beta_{003,\text{Awake}}$ & 0.40 & 0.94 ( 1.83) & 0.116 (0.005) & 0.126 (0.001) & 0.013 (0.001) & 0.95 (0.01) & 0.95 (0.01) & - & 0.90 ( 1.84) & 0.116 (0.005) & 0.126 (0.001) & 0.013 (0.001) & 0.94 (0.02) & 0.94 (0.02) & -2.70 ( 1.28) & 0.081 (0.004) & 0.090 (0.000) & 0.007 (0.001) & 0.98 (0.01) & 0.98 (0.01) & 1.44 ( 3.47) & 0.093 (0.010) & 0.090 (0.001) & 0.009 (0.002) & 0.93 (0.04) & 0.96 (0.03) & -0.68 ( 1.01) & 0.063 (0.003) & 0.068 (0.000) & 0.004 (0.000) & 0.98 (0.01) & 0.97 (0.01) \\ \cline{2-34} 
 & $\beta_{003,\text{NREM}}$ & 3.50 & -1.88 ( 0.19) & 0.106 (0.005) & 0.148 (0.001) & 0.016 (0.001) & 0.97 (0.01) & 0.99 (0.01) & - & -1.88 ( 0.19) & 0.105 (0.005) & 0.148 (0.001) & 0.015 (0.001) & 0.98 (0.01) & 1.00 (0.00) & -1.03 ( 0.14) & 0.079 (0.004) & 0.098 (0.000) & 0.008 (0.001) & 0.96 (0.01) & 0.97 (0.01) & -0.99 ( 0.33) & 0.077 (0.008) & 0.099 (0.001) & 0.007 (0.001) & 1.00 (0.00) & 1.00 (0.00) & -0.65 ( 0.11) & 0.059 (0.003) & 0.071 (0.000) & 0.004 (0.000) & 0.98 (0.01) & 0.98 (0.01) \\ \cline{2-34} 
 & $\beta_{003,\text{REM}}$ & -2.80 & -2.90 (-0.25) & 0.109 (0.005) & 0.150 (0.001) & 0.018 (0.001) & 0.97 (0.01) & 0.99 (0.01) & - & -2.90 (-0.25) & 0.109 (0.005) & 0.150 (0.001) & 0.018 (0.001) & 0.97 (0.01) & 0.99 (0.01) & -1.57 (-0.17) & 0.076 (0.003) & 0.098 (0.000) & 0.008 (0.001) & 0.98 (0.01) & 0.99 (0.01) & -1.40 (-0.45) & 0.084 (0.009) & 0.098 (0.001) & 0.008 (0.002) & 0.96 (0.03) & 0.98 (0.02) & -0.88 (-0.14) & 0.064 (0.003) & 0.072 (0.000) & 0.005 (0.000) & 0.96 (0.01) & 0.96 (0.01) \\ \cline{2-34} 
 & $\sigma^2_{u_03,\text{Awake}}$ & 0.25 & 10.26 (1.47) & 0.116 (0.005) & 0.151 (0.002) & 0.016 (0.002) & 0.96 (0.01) & 0.96 (0.01) & - & 10.46 (1.46) & 0.116 (0.005) & 0.152 (0.002) & 0.016 (0.002) & 0.96 (0.01) & 0.97 (0.01) & 12.59 (0.99) & 0.078 (0.004) & 0.104 (0.001) & 0.010 (0.001) & 0.95 (0.01) & 1.00 (0.00) & 14.24 (2.88) & 0.097 (0.010) & 0.106 (0.003) & 0.014 (0.004) & 0.91 (0.04) & 0.96 (0.03) & 13.89 (0.78) & 0.061 (0.003) & 0.078 (0.000) & 0.009 (0.001) & 0.87 (0.02) & 0.98 (0.01) \\ \cline{2-34} 
 & $\sigma^2_{u_03,\text{NREM}}$ & 0.25 & 57.26 (1.41) & 0.112 (0.005) & 0.203 (0.002) & 0.094 (0.005) & 0.48 (0.03) & 1.00 (0.00) & - & 57.49 (1.39) & 0.110 (0.005) & 0.204 (0.002) & 0.095 (0.004) & 0.50 (0.03) & 1.00 (0.00) & 36.61 (1.14) & 0.090 (0.004) & 0.122 (0.001) & 0.042 (0.002) & 0.52 (0.03) & 1.00 (0.00) & 41.00 (2.29) & 0.077 (0.008) & 0.126 (0.002) & 0.048 (0.005) & 0.44 (0.07) & 1.00 (0.00) & 26.38 (0.75) & 0.059 (0.003) & 0.085 (0.000) & 0.021 (0.001) & 0.59 (0.03) & 1.00 (0.00) \\ \cline{2-34} 
 & $\sigma^2_{u_03,\text{REM}}$ & 0.25 & 63.00 (1.60) & 0.127 (0.006) & 0.211 (0.002) & 0.115 (0.006) & 0.40 (0.03) & 1.00 (0.00) & - & 63.00 (1.62) & 0.128 (0.006) & 0.211 (0.002) & 0.116 (0.006) & 0.40 (0.03) & 0.99 (0.01) & 38.96 (1.04) & 0.082 (0.004) & 0.124 (0.001) & 0.045 (0.002) & 0.49 (0.03) & 0.98 (0.01) & 39.66 (2.56) & 0.086 (0.009) & 0.125 (0.002) & 0.047 (0.007) & 0.56 (0.07) & 0.98 (0.02) & 28.58 (0.89) & 0.069 (0.003) & 0.087 (0.001) & 0.025 (0.001) & 0.47 (0.03) & 0.99 (0.01) \\ \hline
\multirow{9}{*}{\begin{tabular}[c]{@{}l@{}}Transition \\ probabilities\end{tabular}} & $\gamma_{\text{ Awake,Awake}}$ & 0.984 & -1.65 (0.02) & 0.002 (0.000) & 0.004 (0.000) & 0.000 (0.000) & 0.07 (0.02) & 1.00 (0.00) & 0.80 & -2.27 (0.09) & 0.007 (0.000) & 0.013 (0.000) & 0.000 (0.000) & 0.97 (0.01) & 1.00 (0.00) & -1.52 (0.06) & 0.005 (0.000) & 0.007 (0.000) & 0.000 (0.000) & 0.90 (0.02) & 1.00 (0.00) & -0.65 (0.12) & 0.004 (0.000) & 0.007 (0.000) & 0.000 (0.000) & 0.98 (0.02) & 1.00 (0.00) & -1.20 (0.05) & 0.004 (0.000) & 0.005 (0.000) & 0.000 (0.000) & 0.85 (0.02) & 0.98 (0.01) \\ \cline{2-34} 
 & $\gamma_{\text{ Awake,NREM}}$ & 0.007 & -0.11 (0.01) & 0.001 (0.000) & 0.001 (0.000) & 0.000 (0.000) & 1.00 (0.00) & 1.00 (0.00) & 0.10 & 1.12 (0.07) & 0.006 (0.000) & 0.009 (0.000) & 0.000 (0.000) & 0.98 (0.01) & 1.00 (0.00) & 0.76 (0.05) & 0.004 (0.000) & 0.005 (0.000) & 0.000 (0.000) & 0.95 (0.01) & 1.00 (0.00) & 0.46 (0.11) & 0.004 (0.000) & 0.005 (0.000) & 0.000 (0.000) & 0.96 (0.03) & 0.98 (0.02) & 0.61 (0.04) & 0.003 (0.000) & 0.004 (0.000) & 0.000 (0.000) & 0.92 (0.02) & 0.98 (0.01) \\ \cline{2-34} 
 & $\gamma_{\text{ Awake,REM}}$ & 0.012 & 1.14 (0.02) & 0.002 (0.000) & 0.003 (0.000) & 0.000 (0.000) & 0.20 (0.02) & 1.00 (0.00) & 0.10 & 1.08 (0.07) & 0.005 (0.000) & 0.009 (0.000) & 0.000 (0.000) & 0.97 (0.01) & 1.00 (0.00) & 0.74 (0.05) & 0.004 (0.000) & 0.005 (0.000) & 0.000 (0.000) & 0.94 (0.02) & 0.99 (0.01) & 0.16 (0.09) & 0.003 (0.000) & 0.005 (0.000) & 0.000 (0.000) & 1.00 (0.00) & 1.00 (0.00) & 0.58 (0.04) & 0.003 (0.000) & 0.004 (0.000) & 0.000 (0.000) & 0.90 (0.02) & 0.98 (0.01) \\ \cline{2-34} 
 & $\gamma_{\text{ NREM,Awake}}$ & 0.003 & 2.32 (0.02) & 0.002 (0.000) & 0.003 (0.000) & 0.000 (0.000) & 0.00 (0.00) & 1.00 (0.00) & 0.15 & 1.07 (0.08) & 0.006 (0.000) & 0.010 (0.000) & 0.000 (0.000) & 0.99 (0.01) & 0.99 (0.01) & 0.88 (0.06) & 0.004 (0.000) & 0.006 (0.000) & 0.000 (0.000) & 0.96 (0.01) & 1.00 (0.00) & 0.65 (0.14) & 0.005 (0.000) & 0.005 (0.000) & 0.000 (0.000) & 0.93 (0.04) & 1.00 (0.00) & 0.65 (0.04) & 0.004 (0.000) & 0.004 (0.000) & 0.000 (0.000) & 0.91 (0.02) & 0.98 (0.01) \\ \cline{2-34} 
 & $\gamma_{\text{ NREM,NREM}}$ & 0.959 & -3.45 (0.07) & 0.005 (0.000) & 0.008 (0.000) & 0.000 (0.000) & 0.25 (0.03) & 1.00 (0.00) & 0.70 & -2.19 (0.16) & 0.013 (0.001) & 0.020 (0.000) & 0.000 (0.000) & 0.98 (0.01) & 1.00 (0.00) & -1.76 (0.12) & 0.009 (0.000) & 0.012 (0.000) & 0.000 (0.000) & 0.94 (0.02) & 1.00 (0.00) & -1.29 (0.29) & 0.010 (0.001) & 0.011 (0.000) & 0.000 (0.000) & 0.93 (0.04) & 1.00 (0.00) & -1.38 (0.09) & 0.007 (0.000) & 0.008 (0.000) & 0.000 (0.000) & 0.89 (0.02) & 0.98 (0.01) \\ \cline{2-34} 
 & $\gamma_{\text{ NREM,REM}}$ & 0.021 & 4.51 (0.06) & 0.004 (0.000) & 0.007 (0.000) & 0.001 (0.000) & 0.00 (0.00) & 1.00 (0.00) & 0.15 & 1.05 (0.11) & 0.009 (0.000) & 0.014 (0.000) & 0.000 (0.000) & 1.00 (0.00) & 1.00 (0.00) & 0.86 (0.08) & 0.007 (0.000) & 0.008 (0.000) & 0.000 (0.000) & 0.95 (0.01) & 0.98 (0.01) & 0.63 (0.19) & 0.006 (0.001) & 0.008 (0.000) & 0.000 (0.000) & 1.00 (0.00) & 1.00 (0.00) & 0.72 (0.06) & 0.005 (0.000) & 0.006 (0.000) & 0.000 (0.000) & 0.93 (0.02) & 0.97 (0.01) \\ \cline{2-34} 
 & $\gamma_{\text{ REM,Awake}}$ & 0.013 & 0.76 (0.02) & 0.001 (0.000) & 0.003 (0.000) & 0.000 (0.000) & 0.75 (0.03) & 1.00 (0.00) & 0.18 & 1.29 (0.09) & 0.007 (0.000) & 0.011 (0.000) & 0.000 (0.000) & 0.98 (0.01) & 1.00 (0.00) & 0.84 (0.07) & 0.005 (0.000) & 0.007 (0.000) & 0.000 (0.000) & 0.96 (0.01) & 0.99 (0.01) & 0.29 (0.15) & 0.005 (0.001) & 0.006 (0.000) & 0.000 (0.000) & 0.98 (0.02) & 0.98 (0.02) & 0.75 (0.05) & 0.004 (0.000) & 0.005 (0.000) & 0.000 (0.000) & 0.93 (0.02) & 0.98 (0.01) \\ \cline{2-34} 
 & $\gamma_{\text{ REM,NREM}}$ & 0.034 & -1.58 (0.03) & 0.002 (0.000) & 0.004 (0.000) & 0.000 (0.000) & 0.60 (0.03) & 1.00 (0.00) & 0.18 & 1.20 (0.13) & 0.010 (0.000) & 0.015 (0.000) & 0.000 (0.000) & 1.00 (0.00) & 1.00 (0.00) & 0.70 (0.10) & 0.008 (0.000) & 0.009 (0.000) & 0.000 (0.000) & 0.96 (0.01) & 0.98 (0.01) & 0.47 (0.18) & 0.006 (0.001) & 0.009 (0.000) & 0.000 (0.000) & 1.00 (0.00) & 1.00 (0.00) & 0.70 (0.06) & 0.005 (0.000) & 0.007 (0.000) & 0.000 (0.000) & 0.98 (0.01) & 0.99 (0.01) \\ \cline{2-34} 
 & $\gamma_{\text{ REM,REM}}$ & 0.967 & -1.99 (0.04) & 0.003 (0.000) & 0.006 (0.000) & 0.000 (0.000) & 0.56 (0.03) & 1.00 (0.00) & 0.64 & -2.57 (0.17) & 0.013 (0.001) & 0.021 (0.000) & 0.000 (0.000) & 1.00 (0.00) & 1.00 (0.00) & -1.55 (0.13) & 0.010 (0.000) & 0.013 (0.000) & 0.000 (0.000) & 0.95 (0.01) & 0.99 (0.01) & -0.79 (0.24) & 0.008 (0.001) & 0.012 (0.000) & 0.000 (0.000) & 1.00 (0.00) & 1.00 (0.00) & -1.46 (0.09) & 0.007 (0.000) & 0.009 (0.000) & 0.000 (0.000) & 0.93 (0.02) & 0.99 (0.01) \\ \hline
\end{tabular}
\end{adjustbox}
\caption{Simulation results on the baseline scenarios. The component distribution random effect is set to $\zeta=0.5$ in all scenarios. The TPM random effect is set to $Q=0.1$. Scenario-specific information can be found at the top of the table. MC standard errors are given in parentheses.}
\label{tab:baseline-results-z05}
\end{table}

\newpage
\section{Selected results from the empirical application} \label{sec:appendix3}


\begin{figure}[H]
\centering
\begin{adjustbox}{width=1.1\textwidth, angle=90}
\includegraphics[scale=1]{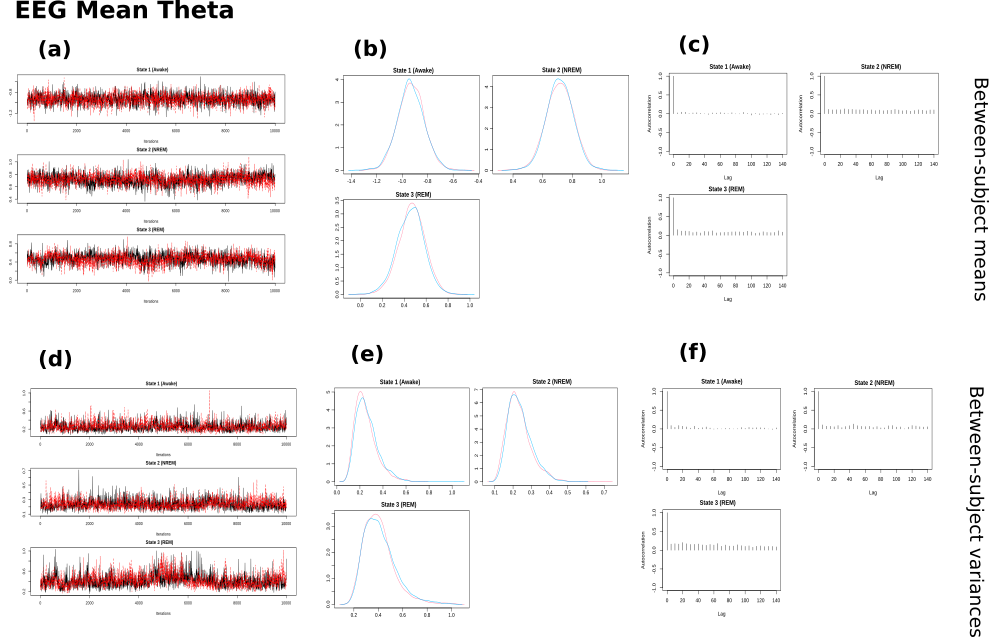}
\end{adjustbox}
\caption{Convergence plots for the outcome variable \textit{EEG mean theta} used in the empirical application. Panels (a), (b) and (c) show the trace plots, density plots and auto-correlation plots for the component distribution group-level means. Panels (d), (e) and (f) show the trace plots, density plots and auto-correlation plots for the component distribution random effects.}
\label{fig:convergence-emiss-empap-I}
\end{figure}

\newpage

\begin{figure}[H]
\centering
\begin{adjustbox}{width=1.1\textwidth, angle=90}
\includegraphics[scale=1]{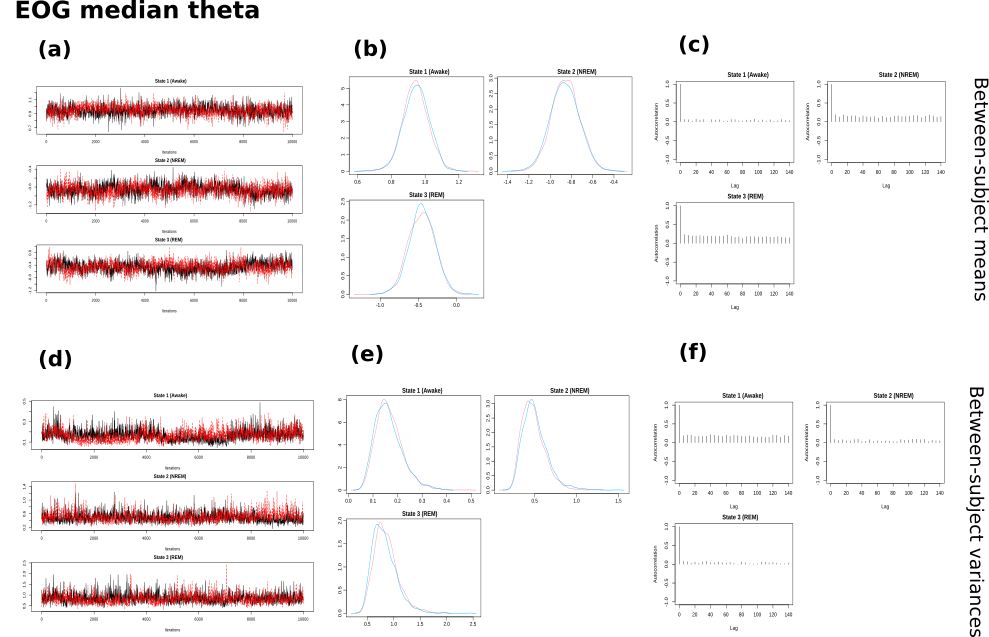}
\end{adjustbox}
\caption{Convergence plots for the outcome variable \textit{EOG median theta} used in the empirical application. Panels (a), (b) and (c) show the trace plots, density plots and auto-correlation plots for the component distribution group-level means. Panels (d), (e) and (f) show the trace plots, density plots and auto-correlation plots for the component distribution random effects.}
\label{fig:convergence-emiss-empap-II}
\end{figure}

\newpage

\begin{figure}[H]
\centering
\begin{adjustbox}{width=1.1\textwidth, angle=90}
\includegraphics[scale=1]{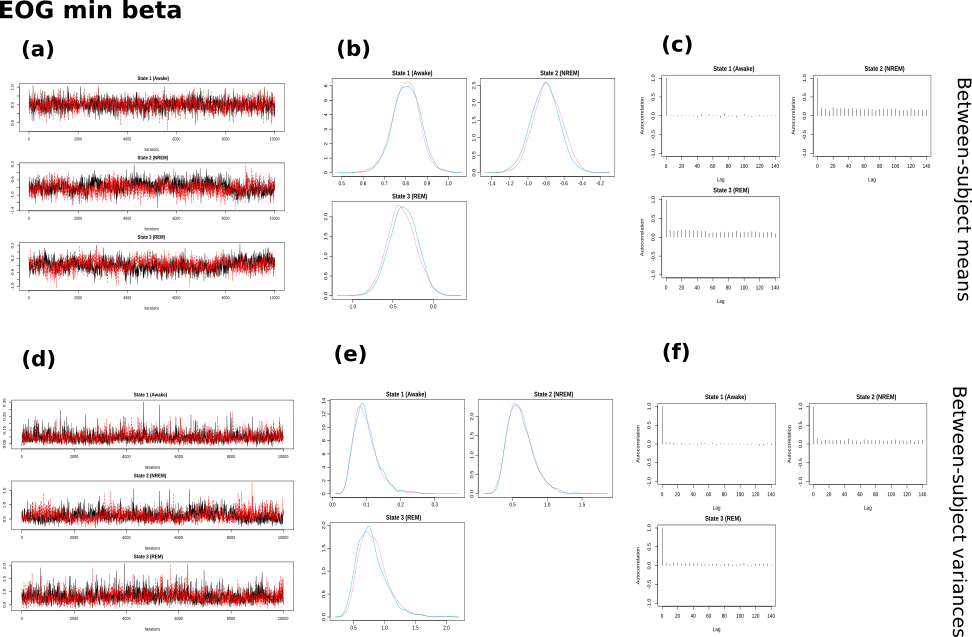}
\end{adjustbox}
\caption{Convergence plots for the outcome variable \textit{EOG min beta} used in the empirical application. Panels (a), (b) and (c) show the trace plots, density plots and auto-correlation plots for the component distribution group-level means. Panels (d), (e) and (f) show the trace plots, density plots and auto-correlation plots for the component distribution random effects.}
\label{fig:convergence-emiss-empap-III}
\end{figure}

\newpage

\begin{figure}[H]
\centering
\begin{adjustbox}{width=1.1\textwidth, angle=90}
\includegraphics[scale=1]{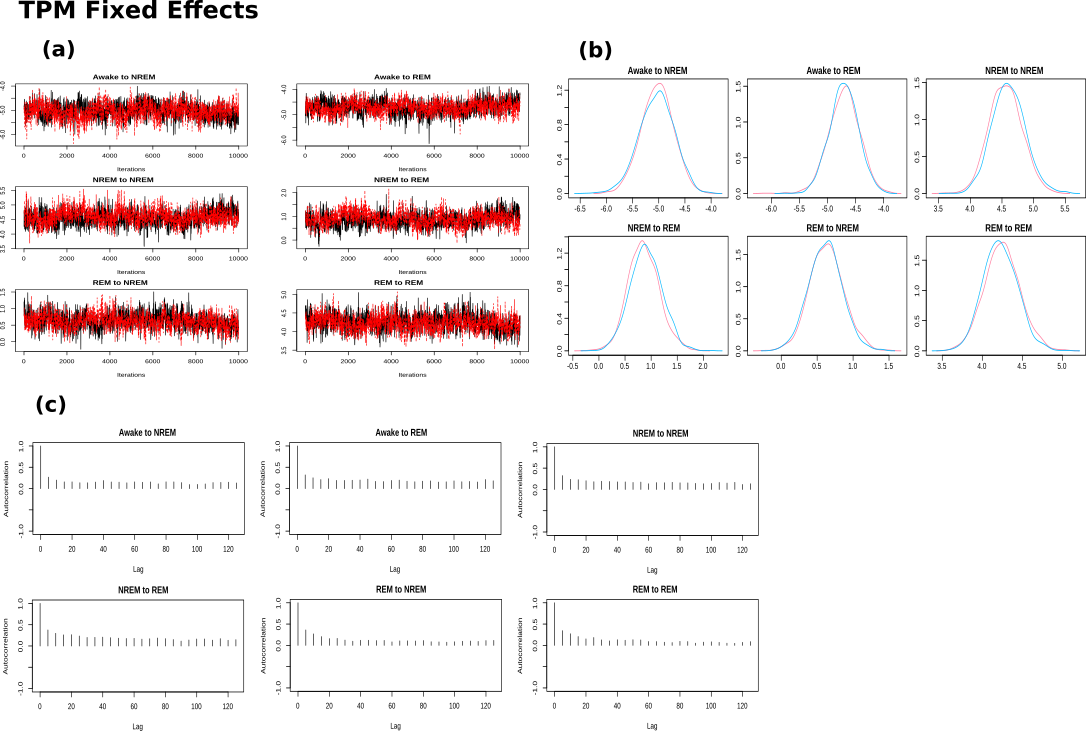}
\end{adjustbox}
\caption{Convergence plots for the TPM group-level intercepts used to compute the group-level TPM. Panel (a) shows the trace plots of the MLR intercepts. Panel (b) shows the density plots for these parameter estimates, and panel (c) shows the auto-correlation plots.}
\label{fig:convergence-MLR-empap}
\end{figure}

\end{document}